\documentclass{article}
\usepackage[margin=2cm]{geometry}
\usepackage{graphicx,bm,amsmath,amssymb,latexsym,psfrag,epsfig,euscript, multirow,color,verbatim,cancel}
\title{A WIMPy Baryogenesis Miracle}
\usepackage{mathrsfs}
\usepackage{float}

\allowdisplaybreaks
\begin{document}
\newcommand{\PL}{\mathcal{P}_{\mathrm L}}
\newcommand{\PR}{\mathcal{P}_{\mathrm R}}
\newcommand{\gs}{g_{\mathrm s}}
\newcommand{\be}{\begin{equation*}}
\newcommand{\ee}{\end{equation*}}
\newcommand{\ben}{\begin{equation}}
\newcommand{\een}{\end{equation}}
\newcommand{\bea}{\begin{eqnarray*}}
\newcommand{\eea}{\end{eqnarray*}}
\newcommand{\bean}{\begin{eqnarray}}
\newcommand{\eean}{\end{eqnarray}}
\newcommand{\MPl}{M_{\mathrm{Pl}}}
\newcommand{\Td}{T_{\mathrm{d}}}
\newcommand{\ptl}{\partial}
\newcommand{\benum}{\begin{enumerate}}
\newcommand{\eenum}{\end{enumerate}}
\newcommand{\bi}{\begin{itemize}}
\newcommand{\ei}{\end{itemize}}
\newcommand{\newfootnotemark}[1]{\addtocounter{footnote}{#1} \footnotemark[\value{footnote}]}
\newcommand{\newfootnotetext}[2]{\addtocounter{footnote}{#1} \footnotetext[\value{footnote}]{#2}}

\begin{titlepage}

\begin{flushright}
\end{flushright}

\vspace{0.2cm}
\begin{center}
\Large\bf
A WIMPy Baryogenesis Miracle
\end{center}

\vspace{0.2cm}
\begin{center}
{\sc Yanou Cui,$^{a,b}$\footnote{E-mail:cuiyo@umd.edu~~~~~~~~~~~~~~~~~~~~~~~~{UMD-PP-011-015}}
 Lisa Randall$^{a}$\footnote{E-mail:randall@physics.harvard.edu}
 and  Brian Shuve$^{a}$\footnote{E-mail:shuve@physics.harvard.edu}}\\
\vspace{0.4cm}
{\sl $^a$ Center for the Fundamental Laws of Nature\\
Jefferson Physical Laboratory\\
Harvard University\\
Cambridge, MA 02138, U.S.A.\\
$^b$ Department of Physics, University of Maryland, College Park, MD 20742, USA.}
\end{center}

\vspace{0.2cm}
\begin{abstract}\vspace{0.2cm}
\noindent
We explore models in which weakly interacting massive particle (WIMP) dark matter annihilation is directly responsible for baryogenesis, thereby connecting dark matter with baryogenesis. We call this process ``WIMPy baryogenesis''. The dark matter relic density in these models, as with conventional WIMP models, is obtained with only order one couplings and TeV-scale masses according to the WIMP miracle. Thus, WIMPy baryogenesis models naturally accommodate weak-scale dark matter. Furthermore, an extension of the WIMP miracle simultaneously explains the observed baryon asymmetry and the correct dark matter abundance. The models we present have the further feature that they  create the baryon number asymmetry at the weak scale, thereby avoiding the problems in some models of baryogenesis associated with high reheat temperatures in supersymmetric theories. Some of these models yield observable consequences in ongoing and future experiments.
\end{abstract}
\vfil

\end{titlepage}

\tableofcontents

\setcounter{footnote}{0}

\section{Introduction}
Generally, baryogenesis and the establishment of the dark matter number density are treated as independent processes.   With the notable exception of one class of models \cite{McDonald:2011zz},  it has largely been overlooked that models with symmetric, weakly interacting massive particle (WIMP) dark matter can connect dark matter physics with baryogenesis. We present a new mechanism that creates such a link and is based on a simple premise: if WIMP annihilation satisfies the Sakharov conditions, a non-zero baryon number asymmetry can be generated from dark matter annihilation, and in some instances, can account for the entire observed baryon asymmetry. We call this process WIMPy baryogenesis. Our models are distinct from models of asymmetric dark matter, which propose that dark matter and baryons have their origins in a common asymmetry. In our models, the energy densities of baryons and dark matter are more loosely linked but can accommodate the observed dark-matter-to-baryon ratio.

\cite{Nussinov:1985xr,Kaplan:2009ag,Cohen:2009fz,Buckley:2010ui,Cui:2011qe,Buckley:2011ye}

We list below the  Sakharov conditions and how they are satisfied in WIMPy baryogenesis:
\begin{enumerate}
\item \emph{Baryon number violation:} WIMP annihilations violate  baryon or lepton number. A preserved $\mathrm U(1)$  symmetry is allowed if the  baryon asymmetry is balanced by a negative asymmetry in a decoupled sector  that restores the net global symmetry. We have such a $\rm U(1)$ symmetry in most models we present.
\item \emph{$CP$ violation:} WIMP couplings to Standard Model fields violate $CP$.
\item \emph{Departure from thermal equilibrium:} The cooling of the universe provides the necessary departure from thermal equilibrium. Net dark matter annihilation begins around temperatures $T\lesssim m_{\rm DM}$, resulting in a small deviation of the dark matter number density from its equilibrium value. The annihilations can generate a baryon asymmetry that depends on the amount of dark matter annihilation occurring during washout freeze-out, which is comparable to the dark matter density at that time.

\end{enumerate}
We present models that satisfy all three Sakharov conditions, and that simultaneously generate the observed baryon asymmetry and  WIMP relic density. In particular, we find that there exist successful models of WIMPy baryogenesis with $\mathcal O(1)$ couplings and CP phases, and weak-scale masses for all new fields. This is an extension of the WIMP miracle to also include baryogenesis, although we show that the size of the generated asymmetry is sensitive to the parameters in the theory and can vary by several orders of magnitude from the observed asymmetry.

\begin{figure}
\begin{center}
\includegraphics[width=7cm]{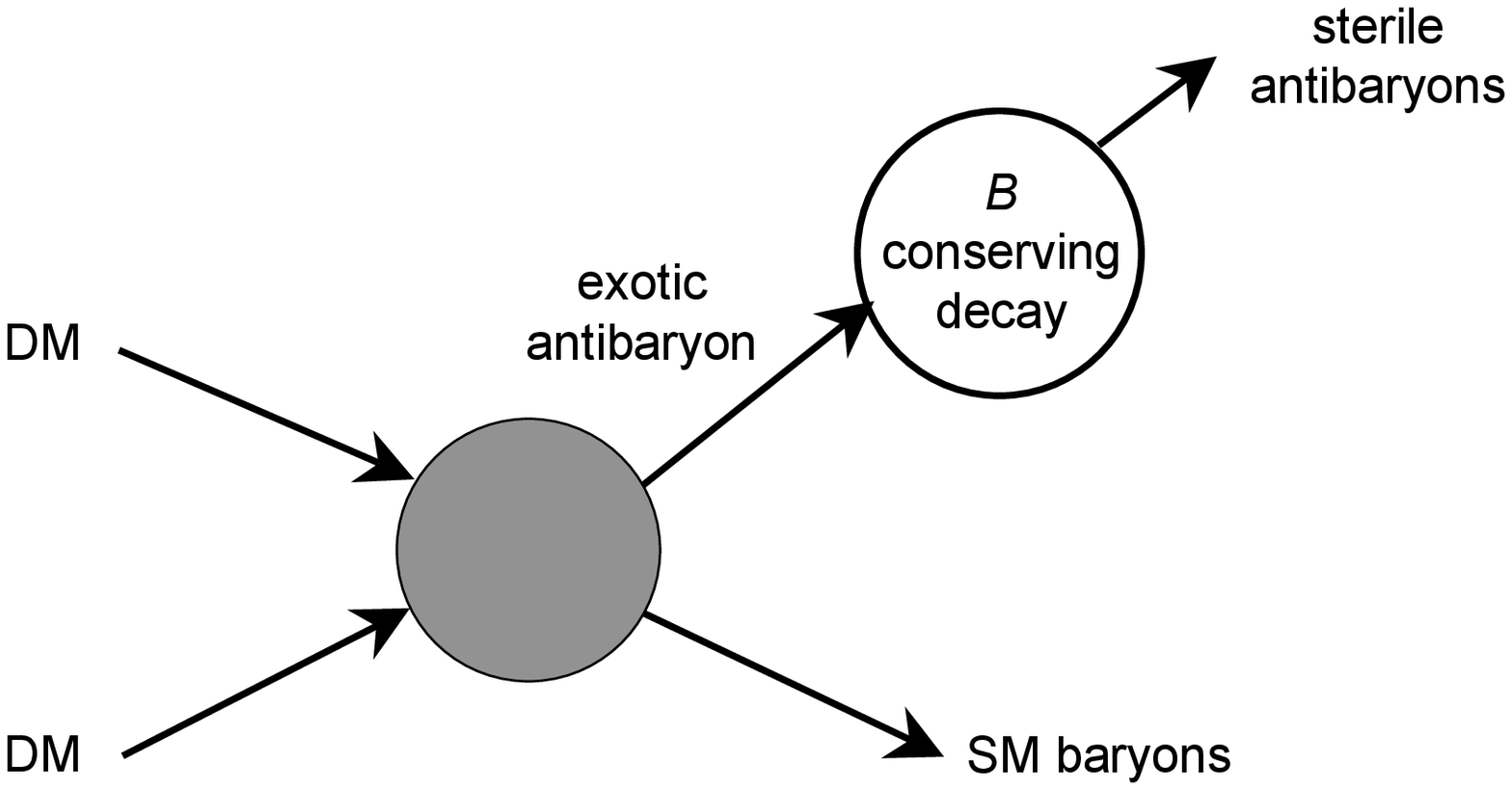}\hspace{1cm}
\includegraphics[width=6cm]{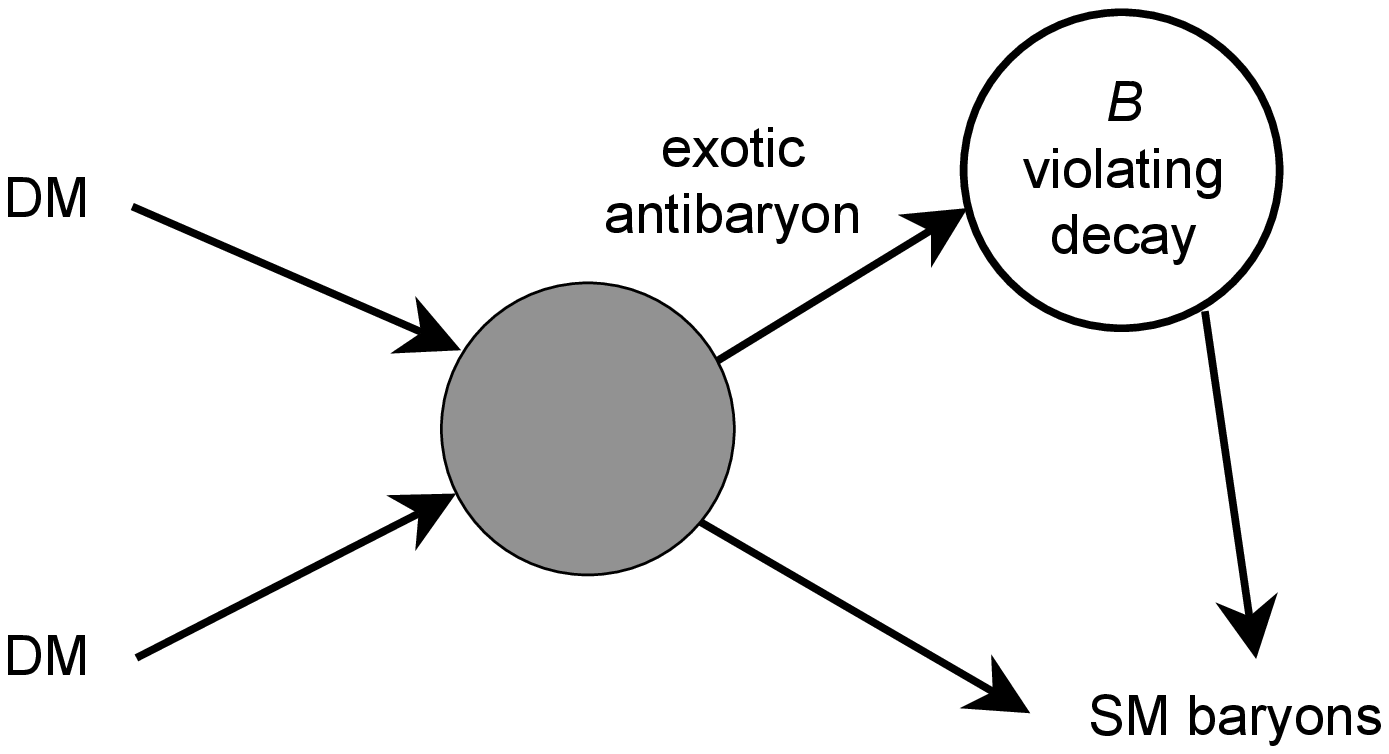}
\caption{Schematic diagrams showing the evolution of the asymmetry created by dark matter annihilation. {\bf (left)} Model where asymmetry created in exotic antibaryons is sequestered in a sterile sector through baryon-number-conserving decays. {\bf (right)} Model where asymmetry created in exotic antibaryons is converted into a Standard Model baryon asymmetry through baryon-number-violating decays.
}\label{fig:flow}
\end{center}
\end{figure}

Although WIMP annihilation can  generate a baryon asymmetry, there are other  processes that have the potential to \emph{wash out} the asymmetry, and their freeze-out is crucial to create the observed baryon asymmetry. In our models, the two leading sources of washout are inverse annihilations of baryons into dark matter and baryon-to-antibaryon processes.
Washout scatterings must be suppressed to generate a sizeable baryon asymmetry because, as we show in Section \ref{sec:wimpybaryogenesis}, any asymmetry generated prior to washout freeze-out\footnote{The time of washout  freeze-out is defined as when the rate of washout processes falls below the Hubble expansion rate. This is analogous to the freeze-out of WIMP annihilation.}
is rapidly damped away. After washout processes freeze out, dark matter annihilations can efficiently create a baryon asymmetry, and the final asymmetry depends on how much dark matter remains when washout scatterings freeze out. Washout freeze-out must occur before that of WIMP annihilation, at which point dark matter annihilation is no longer efficient and no sizeable asymmetry can be created.  Thus, we find our central result: \emph{if washout processes freeze out  before WIMP freeze-out, then a large baryon asymmetry may accumulate, and its final value is proportional to the WIMP abundance at the time that washout becomes inefficient}.

The early freeze-out of washout processes can occur for kinematic reasons. Inverse annihilations will be Boltzmann-suppressed for $T<m_{\rm DM}$ because the thermal baryon fields are no longer energetic enough to annihilate back into dark matter. Baryon-antibaryon scatterings, however, can remain rapid at temperatures well below $m_{\rm DM}$. The only way to suppress baryon-to-antibaryon washout is if all washout processes involve a heavy exotic baryon field in the initial state. We illustrate this scenario in Figure \ref{fig:flow}, showing how dark matter annihilates to Standard Model baryons plus an exotic baryon, as well as the possible decays of the exotic baryon (either through baryon-preserving or baryon-violating interactions). If this exotic field has a mass $\gtrsim m_{\rm DM}$, its abundance is Boltzmann-suppressed at $T<m_{\rm DM}$ and suppresses the washout rate. Meanwhile, dark matter annihilations are not kinematically allowed if the heavy baryon field has mass $\gtrsim2m_{\rm DM}$, so the mass condition  $m_{\rm DM}\lesssim m_{\rm exotic\,baryon}\lesssim 2m_{\rm DM}$ is essential to generate a large baryon asymmetry through WIMPy baryogenesis.

Dark matter annihilations  generate a positive baryon asymmetry stored in Standard Model quarks along with an equal negative  asymmetry stored in the exotic baryon field. It is important that the decays of the exotic baryon do not eliminate the Standard Model baryon asymmetry. In models of WIMPy baryogenesis with a preserved  $\rm U(1)$ baryon symmetry, the exotic baryon-number-carrying field is charged under an additional discrete symmetry, while Standard Model fields are uncharged, preventing the exotic baryon from decaying into Standard Model baryons and destroying the asymmetry. The heavy baryon-number-carrying field decays instead into light gauge singlet fields that are  charged under the discrete symmetry and decoupled from Standard Model fields at temperatures below the scale of WIMPy baryogenesis. We also present a model where the exotic baryon decays to Standard Model quarks through baryon-number-violating couplings, and such models manifestly satisfy the first Sakharov condition.

For many years, the WIMP miracle -- the fact that dark matter fields with weak-scale masses and annihilation cross sections give the correct dark matter thermal relic density -- has been a compelling paradigm for dark matter model-building. WIMPy baryogenesis preserves the WIMP miracle while also offering an explanation for the observed baryon asymmetry. While WIMPy baryogenesis models do not predict the precise relationship between the dark matter and baryon number densities, natural models do restrict the baryon asymmetry to a range of about seven orders of magnitude (see Section \ref{sec:wimpybaryogenesis}), and the observed asymmetry is within this range. Since baryogenesis arises from WIMP annihilations, WIMPy baryogenesis is also necessarily connected to weak-scale physics. While we do not discuss an embedding of WIMPy baryogenesis in a particular solution of the  hierarchy problem, we assume that whatever new physics lies at the weak scale stabilizes any scalar potentials in our theory and gives a natural explanation for their weak-scale masses. A consequence of this is that, with weak-scale masses, some of the new fields necessary for baryogenesis may give signals at future experiments. Additionally, the present-day dark matter is symmetric, leading to the possibility of the indirect detection of dark matter annihilations as in conventional WIMP models. This is in contrast with generic asymmetric dark matter models, in which the majority of dark matter annihilations ceased long before the present day, although it is also noteworthy that there do exist scenarios in which the symmetric component of dark matter is regenerated at late times, giving indirect detection signals for some asymmetric dark matter models \cite{Buckley:2011ye}.

A further advantage of this scenario is that bounds on the reheat temperature in supersymmetric models do not constrain WIMPy baryogenesis. Typical reheat temperature constraints come from overproduction of gravitinos and are in the range $T_{\rm RH}\lesssim10^6-10^9$ GeV \cite{Khlopov:1984pf}, and $T_{\rm RH}$ is consequently below the scale required for conventional leptogenesis through the decay of heavy, Majorana right-handed neutrinos. Although low-scale mechanisms for baryogenesis are known, such as electroweak baryogenesis, WIMPy baryogenesis is a new way of generating the baryon asymmetry at $T\sim$ TeV while satisfying the reheat bound.

We discuss the general conditions  for successful WIMPy baryogenesis in Section \ref{sec:wimpybaryogenesis}, finding that interactions washing out the baryon asymmetry must become ineffective prior to WIMP freeze-out in order to generate the observed asymmetry. In Section \ref{sec:model}, we focus on a particular model where dark matter annihilates through a lepton-number-violating interaction and the asymmetry is subsequently transferred to baryons by sphalerons. Because sphaleron processes are only rapid in the unbroken electroweak phase, such baryogenesis must occur before the electroweak phase transition. We compute the dark matter relic density and baryon asymmetry, and find the range of masses and couplings that agrees with the observed densities of both. We also consider the implications of additional lepton-number-conserving dark matter annihilation channels. In Section \ref{sec:alternatives}, we consider  models with WIMPs annihilating directly to quarks, where baryogenesis can occur over a wider range of temperatures because sphalerons are no longer needed to establish the baryon asymmetry. We discuss experimental constraints and possible signals for  models of WIMPy baryogenesis in Section \ref{sec:signals}. Finally, we summarize in Section \ref{sec:concl}.

\section{General Analysis of WIMPy Baryogenesis} \label{sec:wimpybaryogenesis}
To begin our discussion of WIMPy baryogenesis, we highlight in Section \ref{sec:boltzmann} some of its general features and use an analytic approximation to determine the regimes in which baryogenesis is successful. Our central result is that the final baryon asymmetry from WIMPy baryogenesis is proportional to the dark matter density at the time when washout processes freeze out. This means that  washout scatterings must freeze out at a time when the dark matter density was larger than or comparable to the observed final baryon asymmetry, and that washout freeze-out occurs at such a time when one of the baryon-number-carrying products of WIMP annihilation is heavier than the dark matter mass, as described in the introduction. In Section \ref{sec:estimate}, we estimate the magnitude of the baryon asymmetry for input parameters consistent with the WIMP miracle and show that it can lie within a range of approximately seven orders of magnitude.

\subsection{Boltzmann Equations and Solutions}\label{sec:boltzmann}

We consider a  theory with dark matter species $X$ whose annihilation violates baryon number, creating one quark (or anti-quark) along with other field(s) $\psi_i$. In this section we will not specify the precise interactions mediating dark matter annihilation in order to avoid specific model dependence. $X$ can be either Majorana or Dirac, and the results derived below remain the same up to $\mathcal O(1)$ multiplicative factors. All manifestations of the Boltzmann equation for baryon number evolution in WIMPy baryogenesis models have two important terms: one that describes the annihilation of dark matter and the consequent generation of a baryon asymmetry, and another that drives the asymmetry towards its equilibrium value of zero through baryon-number-violating washout scatterings. The full Boltzmann equations describing the evolution of the various particle abundances are model-dependent and can have many terms, which we give explicitly for concrete models in Sections \ref{sec:model} and \ref{sec:alternatives}. However, in the models of interest to us, namely models where the asymmetry arises predominantly from WIMP annihilations, the overall dynamics are  well-described by the inclusion of only these terms.

Consider the limit where WIMP annihilations are the dominant source of the baryon asymmetry and for which the asymmetry is small as observed.  We derive the Boltzmann equations in terms of dimensionless quantities: the number density per comoving volume of field $i$, $Y_i = n_i/s$ ($s$ is the entropy density), and the temperature, which we express as $x=m_X/T$. The dark matter number density is denoted $Y_X$ and the baryon asymmetry is denoted $Y_{\Delta B}$. The $Y_X$ evolution equation has one term that is important in all models of WIMPy baryogenesis, namely the conventional WIMP annihilation term that is proportional to the annihilation cross section $\sigma_{\rm ann}$ and drives $Y_X$ to its equilibrium value. This term arises from both $XX\rightarrow$ baryon processes and the inverse processes, baryons $\rightarrow XX$. The $Y_X$ Boltzmann equation with this term is
\ben
\label{eq:YXgeneral}\frac{dY_X}{dx} =-\frac{2s(x)}{x\,H(x)}\,\langle\sigma_{\rm ann}v\rangle\left[Y_X^2-(Y_X^{\rm eq})^2\right],
\een
where $H(x)$ is the Hubble scale.

We neglect a back-reaction term in the $Y_X$ Boltzmann equation $\epsilon\,s(x)\,\langle\sigma_{\rm ann}v\rangle\,Y_{\Delta B}(Y_X^{\rm eq})^2/(2Y_\gamma\,x\,H(x))$, where $\epsilon$ is  the net  baryon number created per dark matter annihilation and is a measure of the magnitude of $CP$-violation (it is defined more precisely in Section \ref{sec:model}). This term accounts for the modification of the inverse scattering rate of baryons into $X$ when there is a baryon asymmetry.  This approximation is valid because this term is small when $Y_{\Delta B}\ll1$, as is true in our universe ($Y_{\Delta B}\sim10^{-10}$). This simplification also decouples the equations for $Y_X$ and $Y_{\Delta B}$, which makes it easier to get an approximate analytic solution for $Y_{\Delta B}$. The equation (\ref{eq:YXgeneral}) in this limit is the same as  the familiar Boltzmann equation for conventional WIMPs. $Y_X$ is then obtained from the standard WIMP relic density calculation \cite{Kolb:1990vq} and is approximately inversely proportional to the annihilation cross section.

The Boltzmann equation for the evolution of the baryon asymmetry  has two important terms. In the first term, a baryon asymmetry is generated through $X$ annihilations, and is proportional to $\epsilon/2\times dY_X/dx$, which is the annihilation rate multiplied by the fractional asymmetry generated per annihilation. The factor of $1/2$ arises because the annihilation term in (\ref{eq:YXgeneral}) includes the sum of annihilation into baryons and antibaryons, whereas the term generating the asymmetry includes the difference. The second term in the baryon asymmetry Boltzmann equation reduces the existing baryon asymmetry and is  the \emph{washout} term. It is proportional to $Y_{\Delta B}$ multiplied by the cross section of processes that eliminate the baryon asymmetry $\sigma_{\rm washout}$. The Boltzmann equation is
\ben
\frac{dY_{\Delta B}}{dx} = \frac{\epsilon\,s(x)}{x\,H(x)}\,\langle\sigma_{\rm ann}v\rangle\left[Y_X^2-(Y_X^{\rm eq})^2\right] - \frac{s(x)}{x\,H(x)}\,\langle\sigma_{\rm washout}v\rangle \frac{Y_{\Delta B}}{2Y_\gamma}\prod_i Y^{\rm eq}_i.\label{eq:YBgeneral}
\een
The factor of $Y_{\Delta B}/2Y_\gamma$ comes from the the fact that the chemical potential $\mu_{\Delta B}$ for the baryon asymmetry can be written as $\mu_{\Delta B}/T = Y_{\Delta B}/2Y_\gamma$ \cite{Kolb:1990vq}.  We assume that all species except for $X$ are in equilibrium. There are other terms that we have not included, such as washout terms proportional to $Y_{\mathrm{exotic-}B}^{\rm eq}\,Y_X$ that come from scattering of baryon-number-carrying fields off dark matter fields. Typically, the suppression coming from the small value of $Y_X$ for $x\gg1$ makes this term subdominant to other washout terms. In the models of Section \ref{sec:model} and \ref{sec:alternatives}, the $Y_{\mathrm{exotic-}B}^{\rm eq}\,Y_X$ term can be ignored without substantially affecting the numerical results.

As expected, the Boltzmann equations show that the total baryon number is zero when all fields are in equilibrium and with the initial condition $Y_{\Delta B}=0$.
A solution for $Y_{\Delta B}$ can be written in integral form in terms of the $X$ density:
\bean
Y_{\Delta B}(x)&=&\int_0^x dx'\,\frac{\epsilon\,s(x')}{x'\,H(x')}\,\langle\sigma_{\rm ann}v\rangle\left[Y_X^2-(Y_X^{\rm eq})^2\right](x')\,\exp\left[-\int_{x'}^x\frac{dx''}{x''}\,\frac{s(x'')}{2Y_\gamma\,H(x'')}\,\langle\sigma_{\rm washout}v\rangle \prod_i Y^{\rm eq}_i(x'')\right]\\
&\approx& -\frac{\epsilon}{2}\int_0^x dx'\,\frac{dY_X(x')}{dx'}\,\exp\left[-\int_{x'}^x\frac{dx''}{x''}\,\frac{s(x'')}{2Y_\gamma\,H(x'')}\,\langle\sigma_{\rm washout}v\rangle \prod_i Y^{\rm eq}_i(x'')\right].\label{eq:asymmetryanal}
\eean
Equation (\ref{eq:asymmetryanal})  explicitly shows that $Y_{\Delta B}(x)$ can be expressed in terms of a source term from dark matter annihilations, and an exponential term that attempts to erase any asymmetry  generated by WIMP annihilations. The source term can be written as $dY_X/dx$, as in \cite{Kolb:1979qa}. At $T\gtrsim m_X$, or $x\lesssim1$, WIMP annihilations are balanced by inverse scattering processes and $dY_X/dx\approx0$, meaning no asymmetry is generated according to (\ref{eq:asymmetryanal}). At $x\gtrsim1$, the expansion and cooling of the universe result in net WIMP annihilations ($dY_X/dx\neq0$), providing the departure from equilibrium necessary for baryogenesis.  The net asymmetry at any $x$ is  sensitive to the rate of washout processes during the epoch of WIMP annihilations.

The integrand in the exponent (\ref{eq:asymmetryanal}) is the washout rate $\Gamma_{\rm washout}(x)$ normalized to the Hubble scale $H(x)$,
\ben\label{eq:washoutrate}
\frac{\Gamma_{\rm washout}(x)}{H(x)} = \frac{s(x)}{2Y_\gamma\,H(x)}\,\langle\sigma_{\rm washout}v\rangle \prod_i Y^{\rm eq}_i(x).
\een
Washout freezes out when $\Gamma_{\rm washout}/H<1$. In the limit where $\Gamma_{\rm washout}$ is a rapidly decreasing function of $x$, (\ref{eq:asymmetryanal}) takes a particularly simple form. This is true if, for example, the washout rate freezes out because $m_i/T$ becomes large and yields an exponential suppression of $Y^{\rm eq}_i$. In this case, we can model the exponential in (\ref{eq:asymmetryanal}) as a step function and obtain
\ben\label{eq:DLapprox}
Y_{\Delta B}(\infty) \approx -\frac{\epsilon}{2}\int_{x_{\rm washout}}^\infty dx'\,\frac{dY_X(x')}{dx} = \frac{\epsilon}{2}\left[Y_X(x_{\rm washout})-Y_X(\infty)\right],
\een
where $x_{\rm washout} = m_X/T_{\rm washout}$ is the point at which  washout processes freeze out, and $Y_X(\infty)$ is the late-time dark matter relic density.

Equation (\ref{eq:DLapprox}) has a very clear physical interpretation: after washout scatterings freeze out, all subsequent WIMP annihilations generate a baryon asymmetry with efficiency $\epsilon$. This is why, according to (\ref{eq:DLapprox}), $Y_{\Delta B}$ is proportional to $\epsilon$ times the total number of WIMP annihilations that happen after $x_{\rm washout}$, which is $Y_X(x_{\rm washout})-Y_X(\infty)$. The observed baryon asymmetry is  $Y_{\Delta B}\approx 9\times 10^{-11}$ \cite{Komatsu:2010fb}. Since dark matter at late times satisfies the relation
\ben\label{eq:dmrelation}
Y_X(\infty) \approx \frac{(5\,\,\mathrm{GeV})\,Y_{\Delta B}(\infty)}{m_X},
\een
we require that $Y_X(\infty) < Y_{\Delta B}(\infty)$ for weak-scale dark matter. Along with the requirement  $\epsilon <1$, (\ref{eq:dmrelation}) and (\ref{eq:DLapprox}) imply that $Y_X(x_{\rm washout}) \gg Y_X(\infty)$. In other words, the washout interactions must become ineffective prior to $XX$ annihilation freeze-out in order to generate a sufficiently large baryon asymmetry through WIMPy baryogenesis. As an example of the numerical scales in WIMPy baryogenesis: for a WIMP of mass 1 TeV, $Y_X(\infty)\approx4\times10^{-13}$ and WIMP freeze-out happens at $x_{\rm f.o.}\approx27$, or $T\approx37$ GeV. For $\epsilon=0.1$, washout scatterings must freeze out at $x_{\rm washout}\approx20$ or $T\approx50$ GeV. The final baryon asymmetry is proportional to the WIMP density at the time when washout ceases to be important, with $Y_{\Delta B}(\infty)\approx9\times10^{-11}$.

For what parameters do we expect washout processes to freeze out prior to WIMP annihilation freeze-out? We compare $\Gamma_{\rm washout}$ in (\ref{eq:washoutrate}) to the corresponding rate of WIMP annihilation, which is \cite{Kolb:1990vq}
\ben\label{eq:gammaWIMP}
\frac{\Gamma_{\rm WIMP}(x)}{H(x)}=\frac{2s(x)}{H(x)}\,\langle\sigma_{\rm ann}\,v\rangle\,Y_X(x).
\een
We then find that
\ben
\frac{\Gamma_{\rm washout}(x)}{\Gamma_{\rm WIMP}(x)} \approx \frac{\langle\sigma_{\rm washout}v\rangle \prod_i Y^{\rm eq}_i(x)}{4\langle\sigma_{\rm ann}\,v\rangle\,Y_X^{\rm eq}(x)\,Y_\gamma}.
\een
This ratio must be less than one at the time of washout freeze-out for washout processes to freeze out prior to WIMP freeze-out. This can be realized if either of the following is true \emph{for every process washing out the baryon asymmetry}:
\benum
\item One of the baryon states is heavier than dark matter so $\frac{\prod_iY_i^{\rm eq}(x)}{Y_X^{\rm eq}(x)Y_\gamma}\ll1$.
\item The baryon-number-violating coupling is small so $\langle\sigma_{\rm washout}\,v\rangle \ll \langle\sigma_{\rm ann}\,v\rangle$.
\eenum
The second scenario is challenging to realize, because the same baryon-number-violating couplings appear in both the washout and annihilation cross sections, and $\langle\sigma_{\rm ann}\,v\rangle$ is  fixed by the dark matter relic density. Furthermore, as we show in Section \ref{sec:washout}, suppressing the washout cross section also suppresses the fractional asymmetry generated per annihilation, $\epsilon$, and the resulting baryon asymmetry is typically too small. Therefore, we expect that viable models of WIMPy baryogenesis have at least one baryon-number-carrying field with mass $\gtrsim m_X$.

\subsection{Estimates of Baryon Asymmetry}\label{sec:estimate}
In this section, we derive an  estimate of the baryon asymmetry generated by a WIMP dark matter candidate with mass $m_X\sim$ TeV, and we determine the size of the baryon energy density compared to the WIMP relic density. In the following, we assume for simplicity that a dark matter field $X$ annihilates into a Standard Model quark $Q$ plus an exotic baryon field $\psi$ (see Sections \ref{sec:model} and \ref{sec:alternatives} for specific model details). We find that the baryon asymmetry depends strongly on the mass $m_\psi$ and is constrained to lie within a seven or eight order-of-magnitude window, with the observed baryon asymmetry within an order of magnitude of the upper limit. Therefore, WIMPy baryogenesis does not predict the value of the dark matter-baryon ratio, but neither is the relationship between the two energy densities completely arbitrary.

To determine the range of baryon asymmetries obtained from WIMPy baryogenesis, we use the result from the last section that the final baryon asymmetry is proportional to the number of dark matter annihilations that occur after washout freeze-out, as shown in (\ref{eq:DLapprox}). The largest possible asymmetry is generated when the exotic baryon field is heavy relative to dark matter ($m_\psi\gtrsim m_X$) so that washout processes freeze out while there is still a large dark matter abundance. To determine the upper bound on the asymmetry, we use the fact that $m_\psi < 2m_X$ for WIMP annihilation to be allowed kinematically, and this limits how many dark matter particles can remain when washout freezes out. By contrast, the baryon asymmetry is small when washout processes turn off at a late time ($m_\psi\ll m_X$) after dark matter annihilation has frozen out.  To calculate the lower bound, we determine the rate of residual dark matter annihilation after dark matter freezes out and use this to determine the size of the asymmetry.

For both the upper and lower limits, we first calculate the allowed baryon asymmetry and then determine the corresponding dark matter-baryon ratio. In both scenarios, the baryon asymmetry depends on two time scales: the point of washout freeze-out, $x_{\rm washout} = m_X/T_{\rm washout}$, and the point at which WIMP annihilation freezes out, $x_{\rm ann} = m_X/T_{\rm ann}$.  \\

\noindent{\bf Estimate of upper limit:} We first estimate the upper limit of the baryon asymmetry generated within our framework, which occurs when $m_\psi$ is heavy to suppress washout and is therefore also at the TeV-scale. Kinematically, dark matter annihilation occurs only if $m_\psi < 2m_X$, which bounds how early $x_{\rm washout}$ can be relative to $x_{\rm ann}$.  For a TeV-scale dark matter field, WIMP annihilation freezes out when the temperature is about $1/30$ of its mass. Therefore $x_{\rm washout}\approx x_{\rm ann}(m_X/m_\psi)\gtrsim15$. We also know that, when washout freezes out while WIMP annihilation is still active, $Y_X(x_{\rm washout}) \gg Y_X(\infty)$. We then obtain from (\ref{eq:DLapprox}):
\ben\label{eq:asymmetryestimate}
Y_{\Delta B}(\infty) \approx \frac{\epsilon}{2}\,Y_X(x_{\rm washout})< \frac{\epsilon}{2}\,Y_X^{\rm eq}(15) \approx \epsilon\times10^{-8}.
\een
According to (\ref{eq:asymmetryestimate}), the asymmetry is independent of $m_X$ and depends only on the ratio $x_{\rm washout} \approx m_\psi/m_X$, with a large asymmetry when $m_\psi$ is comparable to or larger than $m_X$.

To compare the baryon density to the dark matter energy density, recall that the WIMP density changes little after annihilation freezes out, and so $Y_X(\infty) \approx Y_X(x_{\rm ann}) \approx Y_X(30)$. We then find that
\ben
\frac{\Omega_B}{\Omega_X} = \frac{m_{\rm proton}\,Y_{\Delta B}(\infty)}{m_X\,Y_X(\infty)}\approx\frac{\epsilon}{2}\,\frac{Y_X(x_{\rm washout})}{Y_X(x_{\rm ann})}\left(\frac{\rm GeV}{m_X}\right)\approx\frac{\epsilon}{2}\,\frac{Y_X^{\rm eq}(15)}{Y_X^{\rm eq}(30)}\left(\frac{\rm GeV}{m_X}\right)\lesssim10\,\left(\frac{\epsilon}{10^{-2}}\right)\left(\frac{\mathrm{TeV}}{m_X}\right)\label{eq:bdratio1}.
\een
Therefore, for a model with weak-scale $m_X$, $\mathcal O(1)$ couplings (in accordance with the WIMP miracle), and the loop-suppressed $\epsilon\sim10^{-2}$ as in (\ref{eq:epsilonexpr}), we find that the energy density of baryons can be at most an order of magnitude larger than the energy density of dark matter.\\

\noindent{\bf Estimate of lower limit:} In deriving the upper bound, we assumed that $m_\psi$ saturated the bound $m_\psi<2 m_X$ and we found that dark matter annihilation could generate the observed baryon asymmetry. When $m_\psi\ll 2m_X$, washout processes remain in equilibrium until after dark matter freeze-out, and the asymmetry from WIMPy baryogenesis is too small to account for the observed asymmetry.  We now estimate the full range of baryon asymmetries achieved in our models when $m_\psi\ll 2m_X$. In this case, the equilibrium number density of $X$ is much smaller than the actual, frozen-out $X$ abundance. As a result of this overabundance of $X$ relative to its equilibrium value, some residual dark matter annihilations continue at late times, even though the annihilation rate is insufficient to appreciably change $Y_X$ after $x_{\rm ann}$. Such annihilations can, however, generate a small baryon asymmetry. According to  (\ref{eq:DLapprox}), this asymmetry can be estimated by calculating $Y_X(\infty)-Y_X(x_{\rm washout})$, where $x_{\rm washout}>x_{\rm ann}$.

To determine the asymmetry, we solve the Boltzmann equation (\ref{eq:YXgeneral}), neglecting the subdominant term $(Y_X^{\rm eq})^2$ in equation (\ref{eq:YXgeneral}).
 Furthermore, if $XX$ annihilation is $s$-wave, then $\langle\sigma_{\rm ann}\,v\rangle$ is approximately constant in the domain $x_{\rm ann} < x < x_{\rm washout}$. The only $x$-dependence comes from the factor
\ben
\frac{s(x)}{x\,H(x)} = \frac{s(x_{\rm ann})}{H(x_{\rm ann})}\,\frac{x_{\rm ann}}{x^2}.
 \een
 Integrating (\ref{eq:YXgeneral}) from $x=x_{\rm washout}$ to $x=\infty$ gives
\ben
 Y_X(x_{\rm washout})-Y_X(\infty) \approx \frac{2s(x_{\rm ann})x_{\rm ann}\,\langle\sigma_{\rm ann}\,v\rangle\,Y_X(x_{\rm ann})^2}{H(x_{\rm ann})}\frac{1}{x_{\rm washout}}.
\een
Using the definition of $\Gamma_{\rm WIMP}$ in (\ref{eq:gammaWIMP}), together with the fact that $\Gamma_{\rm WIMP}(x_{\rm ann})=H(x_{\rm ann})$, gives the simple result
\ben
 Y_X(x_{\rm washout}) - Y_X(\infty) \approx \frac{2x_{\rm ann}}{x_{\rm washout}}\,Y_X(x_{\rm ann})\approx \frac{2x_{\rm ann}}{x_{\rm washout}}\,Y_X(\infty).
\een
Notice that for $x_{\rm washout}>x_{\rm ann}$, $Y_X$ is constant at leading order from $x_{\rm washout}$ to $\infty$. Also as mentioned earlier, by assuming both $X$ and $\psi$ have weak-scale masses and interactions, $x_{\rm ann}/x_{\rm washout}\sim m_\psi/m_X$. We can then obtain an estimate for the baryon asymmetry:
\ben
Y_{\Delta B} \approx \frac{\epsilon\,x_{\rm ann}}{x_{\rm washout}}\,Y_X(\infty) \approx \epsilon\left(\frac{m_\psi}{m_X}\right)\,Y_X(\infty).
\een
We see that the baryon asymmetry decreases linearly with $m_\psi$ when $m_\psi\ll m_X$.

The ratio of the baryon energy density to the dark matter energy density is
\ben
\frac{\Omega_B}{\Omega_X}\sim10^{-3}\times\epsilon\left(\frac{m_\psi}{m_X}\right)\left(\frac{\mathrm{TeV}}{m_X}\right).\label{eq:bdratio2}
\een
If there is no large hierarchy in $m_X$ and $m_\psi$ (i.e.~$m_\psi/m_X\gtrsim0.1$), and using our earlier estimate of $\epsilon\sim10^{-2}$ for $\mathcal O(1)$ couplings that give the correct WIMP relic density, we find that $\Omega_B/\Omega_X\gtrsim10^{-6}$. We emphasize, however, that even smaller asymmetries are possible if the imaginary parts of the couplings are tuned to be small or if there exist hierarchies in the masses of the new fields.

Considering equations (\ref{eq:bdratio1}) and (\ref{eq:bdratio2}), we find that the expected range for the baryon-to-dark matter ratio in WIMPy baryogenesis is
\ben
10^{-6}\lesssim\frac{\Omega_B}{\Omega_X}\lesssim10,
\een
 assuming $\mathcal O(1)$ couplings, and weak-scale masses for all new fields, i.e. $m_X, m_\psi\sim \mathcal O(0.1-1\rm TeV)$. The observed value of $\Omega_B/\Omega_X\approx0.2$ falls within this range, and thus WIMPy baryogenesis can account for the entire observed baryon asymmetry, but it does fall toward the upper end of the allowed region.  \\

\begin{figure}
\begin{center}
\includegraphics[width=10cm]{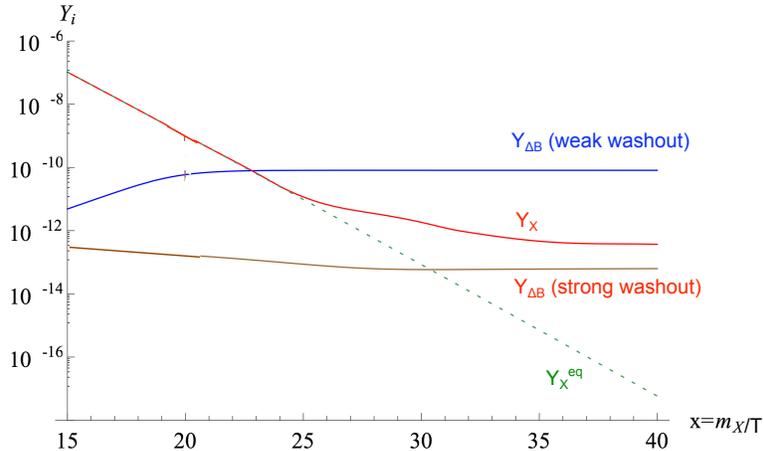}
\caption{The evolution of the number density per comoving volume for field $i$ ($Y_i$) as a function of $x=m_X/T$. The numerical solutions shown here are based on the WIMPy leptogenesis model discussed in Section \ref{sec:model}, where the dominant annihilation process is $XX\rightarrow L\psi$ and the dominant washout is $L\psi\rightarrow L^\dag\psi^\dag$. The input parameters are $y_X=2.7,~ \lambda_L=0.8,~ \epsilon=0.2,~ m_X=3\rm~TeV$, and $m_S=5\rm ~TeV$. $m_\psi=4$ TeV gives the behavior when washout freezes out well before WIMP annihilation freezes out (``weak washout'').  $m_\psi=2\rm ~TeV$ gives the behavior when washout becomes ineffective subsequent to WIMP freeze-out (``strong washout'').
}\label{fig:Yplots}
\end{center}
\end{figure}

To summarize, models of WIMPy baryogenesis predict a dark matter relic density inversely proportional to the WIMP annihilation cross section, as in conventional WIMP models, and a baryon asymmetry proportional to the dark matter density at the time when washout processes freeze out. In Figure \ref{fig:Yplots}, we illustrate the evolution of the dark matter abundance and the baryon asymmetry in one model of WIMPy baryogenesis for the two limiting washout cases.

\section{WIMP Annihilation to Leptons}\label{sec:model}
\subsection{Model Overview}
We have discussed baryogenesis in the generalized sense of either the direct production of a baryon asymmetry  through WIMP annihilation or leptogenesis, in which a lepton asymmetry is produced by WIMP annihilation and converted to a baryon asymmetry through sphalerons. In this section, we present a model of leptogenesis, where the lepton asymmetry is generated above the electroweak phase transition while sphalerons are still active. We discuss the field content and symmetries of the model Section \ref{sec:fields}, and we calculate the efficiency of generating a lepton number asymmetry in Section \ref{sec:asymmetry}. As we showed in Section \ref{sec:wimpybaryogenesis}, the final baryon asymmetry is determined by the time at which washout processes freeze out. We address washout in Section \ref{sec:washout}, discussing the implications for the WIMPy leptogenesis parameter space. Finally, we give the Boltzmann equations in Section \ref{sec:Boltzmann}.

\subsubsection{Field Content and Lagrangian}\label{sec:fields}
We  consider a simple model with the minimal ingredients for WIMPy leptogenesis.  Dark matter consists of a pair of gauge singlet Dirac fermions\footnote{A Majorana dark matter field $X$ does not work in this case because $X$ must carry a complex charge for the model to generate a non-zero lepton asymmetry, as we show later in this section.} $X$ and $\bar X$ that annihilate to the Standard Model lepton doublet $L_i$ and new weak-scale fields $\psi_i$. $X$ annihilates through weak-scale gauge singlet pseudoscalars\footnote{We consider pseudoscalars instead of scalars because they do not have a velocity-suppressed  $XX$ annihilation cross section.  A scalar $S$ with  couplings to $X$ that are $CP$-violating with a large imaginary part would work as well.}  $S_\alpha$. By gauge invariance, $\psi_i$ has charge $(2,1/2)$ under the $\mathrm{SU}(2)_{\rm L}\times\mathrm{U}(1)_Y$ gauge interactions. The Lagrangian is
\ben\label{eq:lagrangian}
\mathcal L = \mathcal L_{\rm kin} + \mathcal L_{\rm mass} -\frac{i}{2}\left(\lambda_{X\alpha}X^2+\lambda_{X\alpha}'\bar X^2\right)S_\alpha + i\,\lambda_{L\,\alpha i}\,S_{\alpha} L_i\psi_i +\mathrm{h.c.}
\een
To satisfy the Sakharov conditions, dark matter annihilation must also violate $CP$, and $\lambda_{L\,\alpha i}$ must be complex. To have physical $CP$ violation, there must be more than one scalar $S_\alpha$ so there is a relative phase in their amplitudes. $XX$ annihilation can then generate an asymmetry in $L_i$, and the lepton asymmetry is subsequently converted to a baryon asymmetry by sphalerons. Because the symmetry preserved by sphalerons is $B-L$, a negative lepton asymmetry must be generated to account for the observed positive baryon asymmetry.

A positive lepton number asymmetry also accumulates in $\psi_i$, and it is important that this positive  asymmetry does not erase the negative asymmetry in Standard Model leptons. In our model, $\psi_i$ decays into light gauge singlets $n_i$ that are decoupled from Standard Model fields at low temperatures. The  asymmetry produced in $\psi$ is therefore sequestered in a sterile sector and  the Standard Model asymmetry persists to the present time\footnote{If $m_\psi>m_S$, then $\psi$ can decay into $S+L^\dagger$ and wipe out the lepton asymmetry. However, $S$  then subsequently decays into either $L+H+n^\dagger$ or $L^\dagger+H^*+n$ (when $XX\rightarrow L\psi$ is kinematically allowed, $S\rightarrow XX$ is kinematically forbidden because $2m_X > m_\psi > m_S$), and the difference in the rates of these decays  generates another lepton asymmetry. If the efficiency of asymmetry generation from $S$ decays is comparable to that from $XX$ annihilations, the asymmetry is comparable to the original asymmetry created from $XX\rightarrow L\psi$.}. A $Z_4$ symmetry, with charges in Table \ref{tab:Z4lepton}, forbids other operators that allow $\psi$ to decay directly into Standard Model leptons, thus preventing the erasure of the Standard Model lepton asymmetry. The $Z_4$ symmetry also makes dark matter stable.

\begin{table}
\begin{center}

 \begin{tabular}{| c | c | c | c | c | c | c | c |}
 \hline
 & $X$ & $\bar X$ & $\psi$ & $\bar\psi$ & $S$ & $n$ & Standard Model \\ \hline
 $Z_4$ & $+i$ & $-i$ & $-1$ & $-1$ & $-1$ & $-1$ & $+1$ \\ \hline
 \end{tabular}
 \caption{$Z_4$ charges of  fields in the WIMPy leptogenesis model described by (\ref{eq:lagrangian}).}
  \label{tab:Z4lepton}
\end{center}
\end{table}

In the simplest model, $\psi_i$ decays to $n_i+H$ through the interaction
\ben
\Delta\mathcal L = \lambda_i\,H^\dagger n_i\psi_i+\mathrm{h.c.}
\een
We assume that $\psi_i$ is vectorlike with a partner $\bar\psi_i$ in order to more readily satisfy electroweak precision constraints. Its mass is restricted by the LEP bound $m_\psi\gtrsim100$ GeV (see Section \ref{sec:collider}).

After electroweak symmetry breaking, $\psi_i$ mixes with the sterile neutrino $n_i$, and we must ensure that the sterile neutrino satisfies overclosure constraints. Since $\psi$ is Dirac, we also include $\bar\psi$ when diagonalizing the mass matrix and find that there remains a massless eigenstate even after the Higgs condenses. This is good, because light, weakly interacting thermal relics (such as sterile neutrinos) with masses $\gtrsim\mathcal O(\mathrm{eV})$ would overclose the universe. $n$ could have a Majorana mass $m_n\lesssim$ eV and still satisfy observational constraints, since the light eigenstate would have a mass $\lesssim$ eV as well, but we take $n$ to be massless in our model.

The Lagrangian (\ref{eq:lagrangian}) is also invariant under a $\rm U(1)^3$ lepton flavor symmetry that prohibits flavor-changing neutral currents but allows flavor-dependent couplings. $L_i$, $n_i$, and $\psi_i$ have charges $+1$, $+1$, and $-1$, respectively, under the $\mathrm U(1)_i$ factor of the flavor symmetry. We assume that the only source of flavor-breaking in the low-energy theory is through the neutrino mass matrices, and this effect is very small.

\begin{figure}[t]
\begin{center}
\includegraphics[height=3cm]{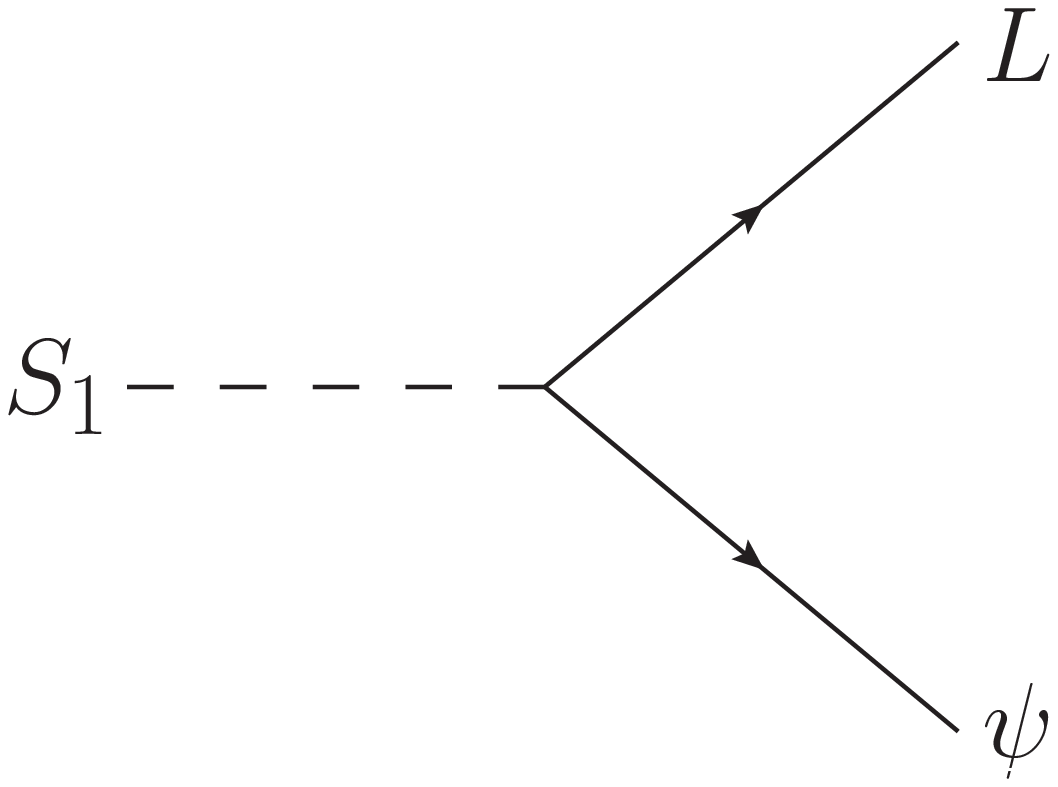}\hspace{1cm}
\includegraphics[height=3cm]{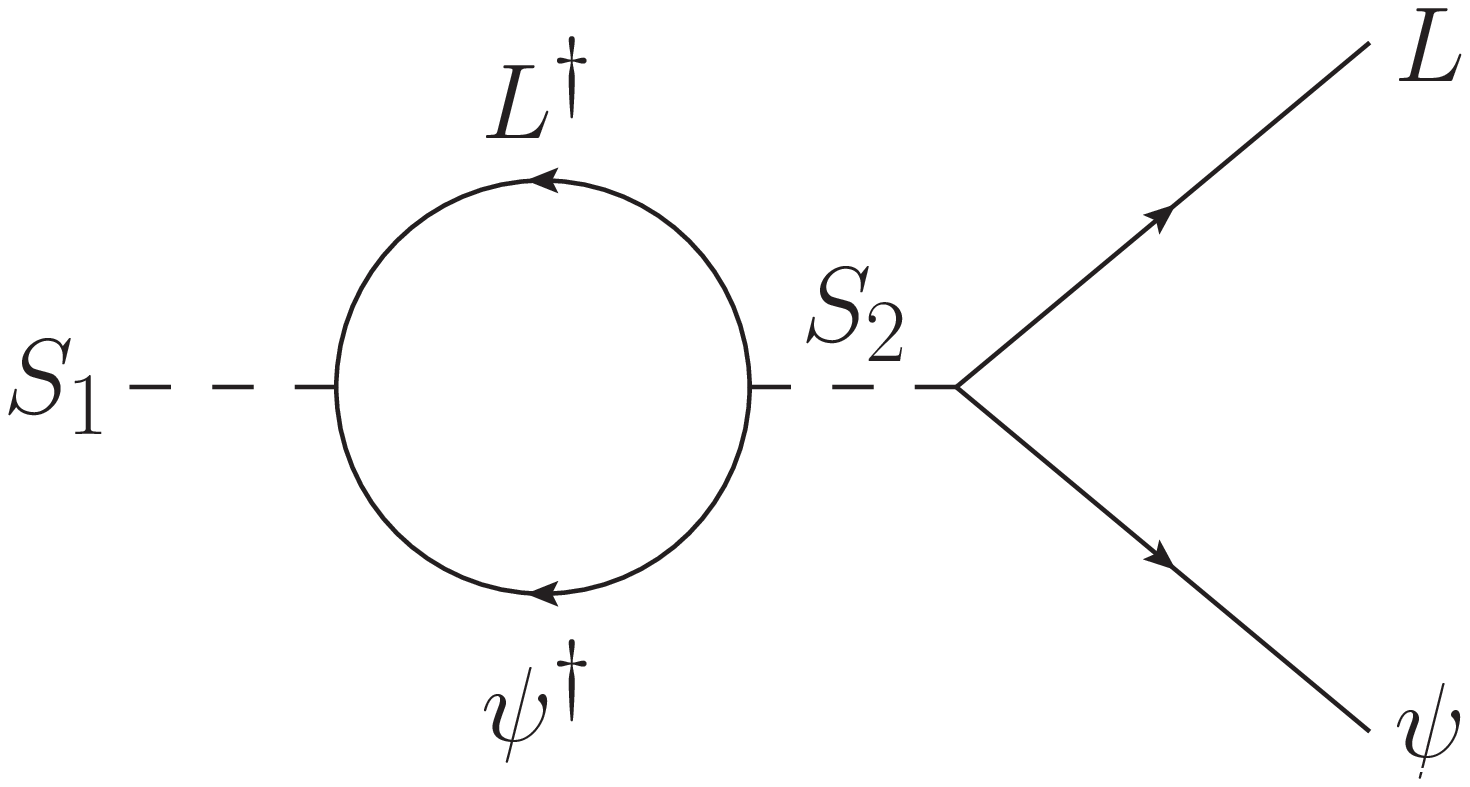}\hspace{1cm}
\includegraphics[height=3.25cm]{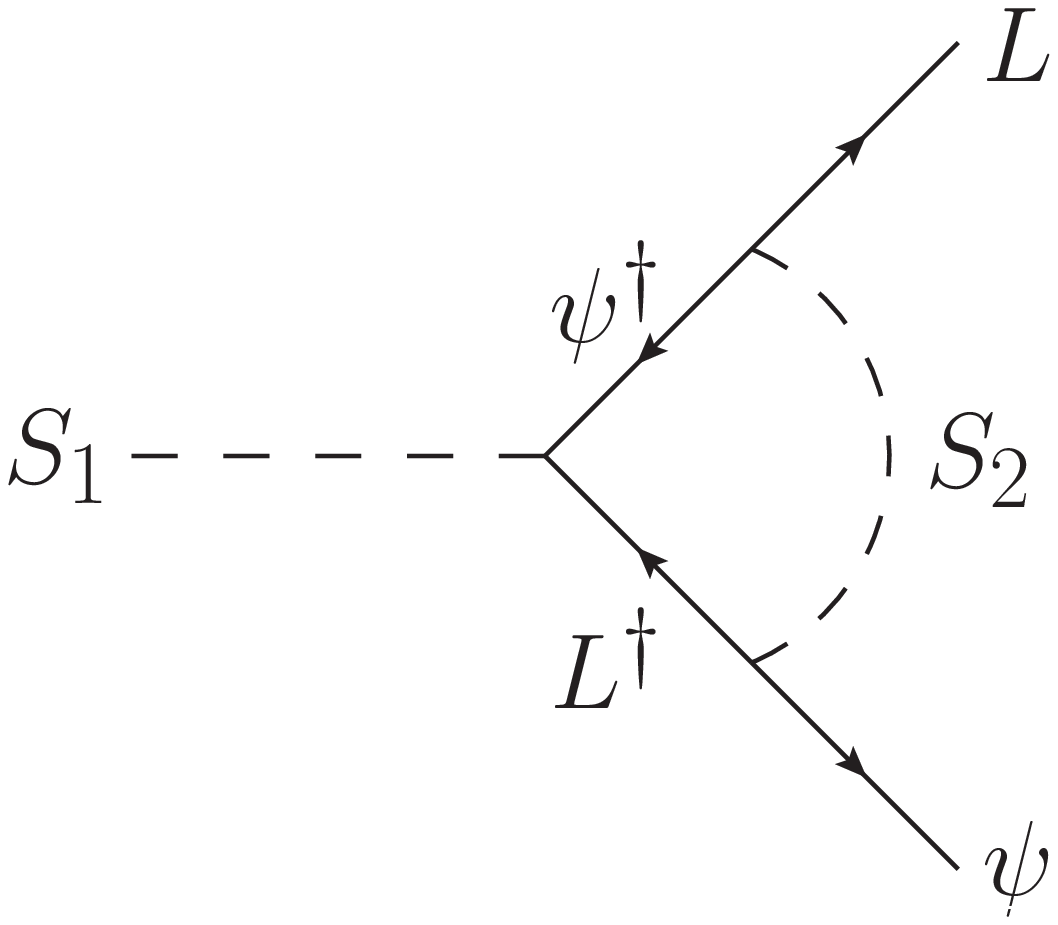}
\caption{Diagrams of tree and loop contributions to $S$ decay. The difference between these rates and their conjugates generates a lepton asymmetry.}
\label{fig:decayasymmetry}
\end{center}
\end{figure}
\begin{figure}[t]
\begin{center}
\includegraphics[height=2.3cm]{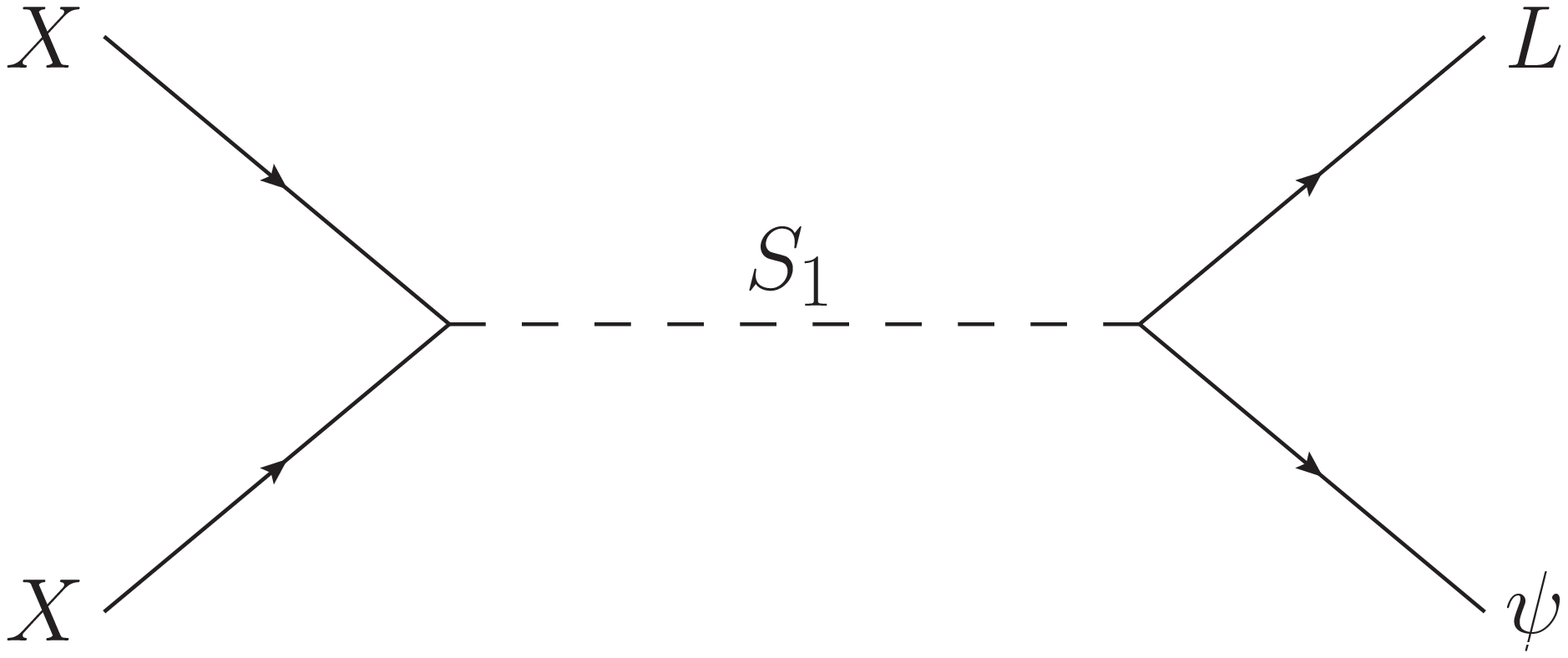}\hspace{0.7cm}
\includegraphics[height=2.3cm]{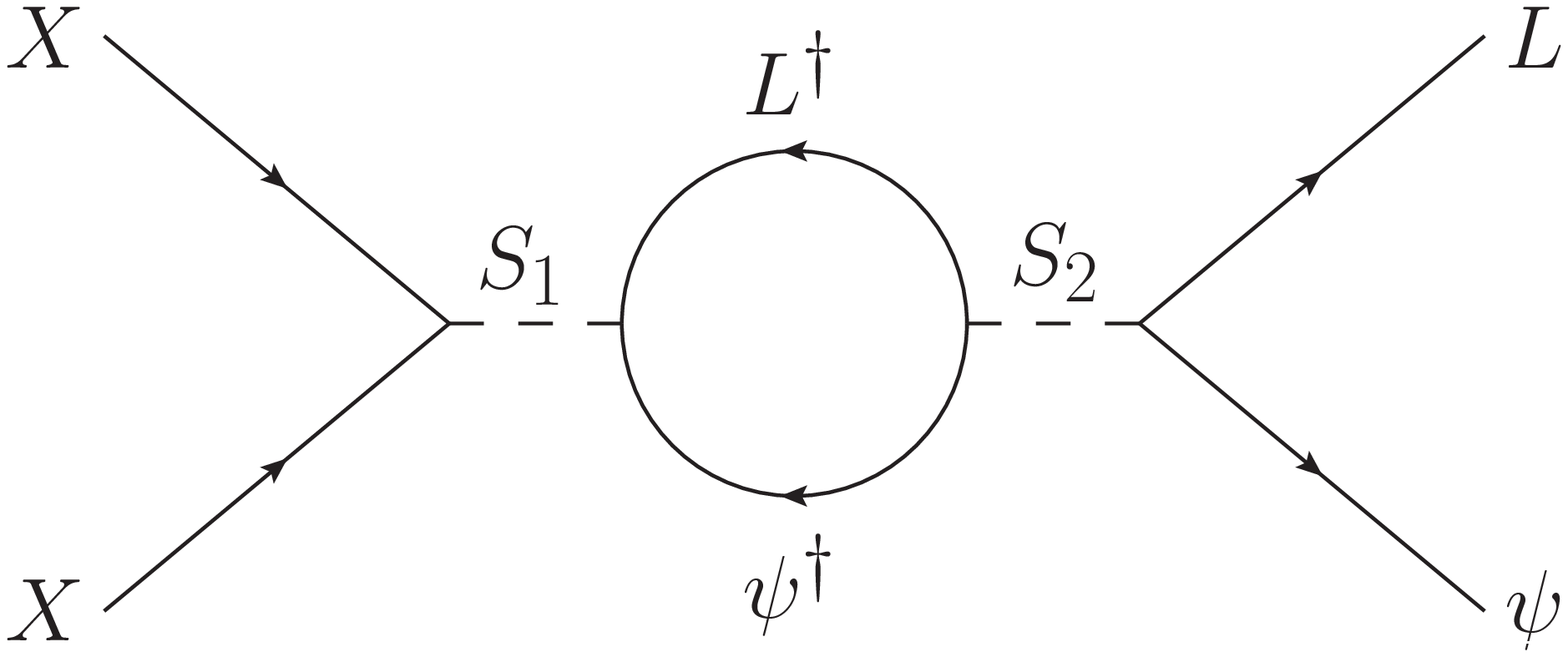}\hspace{0.7cm}
\includegraphics[height=2.36cm]{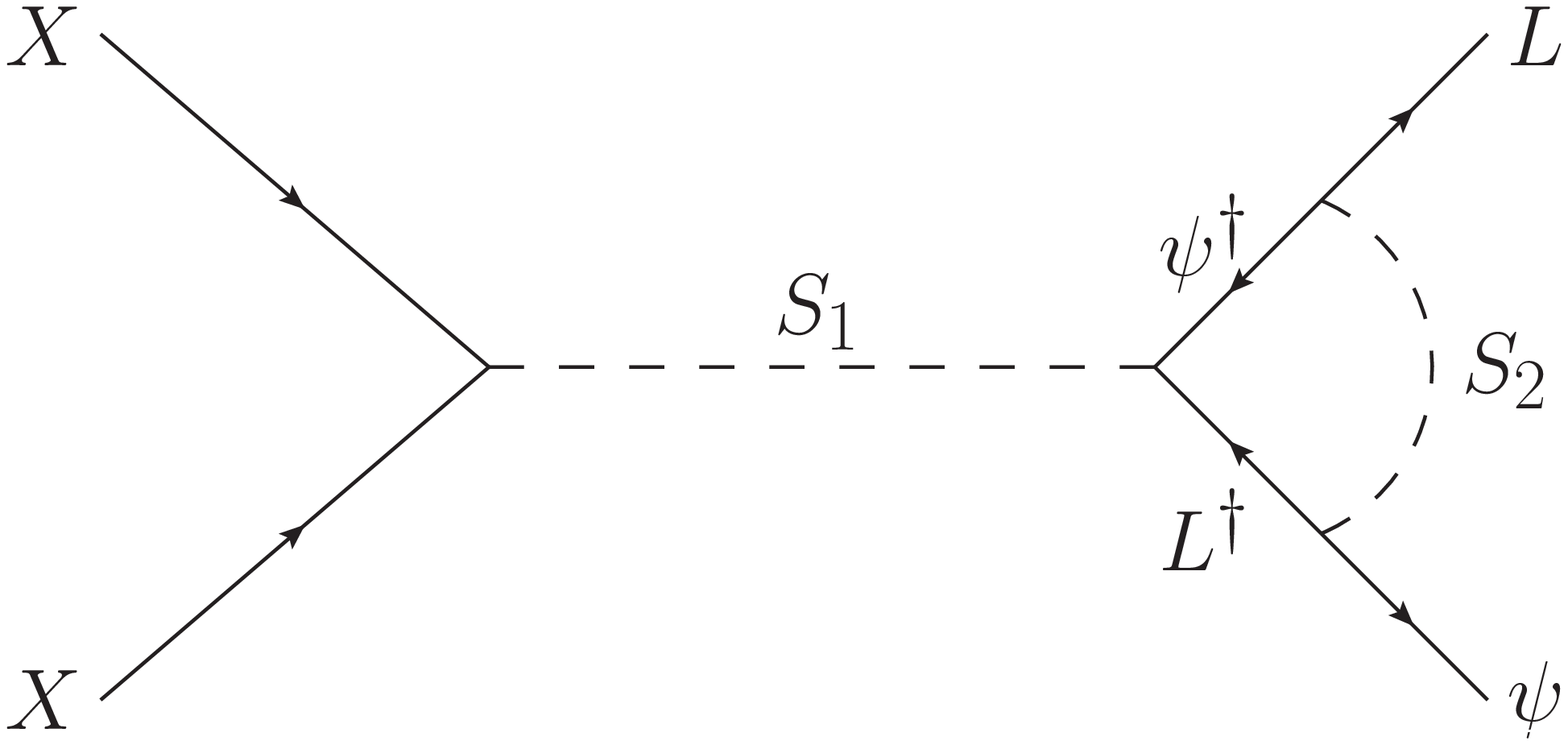}
\caption{Diagrams of tree and loop contributions to the $XX$ annihilation cross section. The difference between these rates and their conjugates generates a lepton asymmetry.}
\label{fig:asymmetry}
\end{center}
\end{figure}

\subsubsection{Asymmetry generation}\label{sec:asymmetry}

In this model, a lepton asymmetry can in principle be generated through two processes: the more conventional process of $S_\alpha$ decay into $L_i\psi_i$ and their conjugates (or directly into $L_i+n_i+H$ if $m_\psi>m_S$), and $XX$ annihilation into the same final states. We show these in Figures \ref{fig:decayasymmetry} and \ref{fig:asymmetry}, respectively, assuming that decay and annihilation occur predominantly through the lightest scalar $S_1$. Existing work discusses the relevant processes for generating a lepton asymmetry through $2\rightarrow2$ scattering \cite{Bento:2001rc}, although the authors consider only high-scale models ($T\gtrsim10^9$ GeV)  with qualitatively different features than WIMPy leptogenesis.   $CP$-violating phases in our model appear in the interference between tree-level and one-loop diagrams. We define asymmetry factors for the decay of the lightest scalar $S_1$ and for WIMP annihilations, respectively, in the manner of conventional leptogenesis:
\bean
\epsilon_1 &=& \frac{\Gamma(S_1\rightarrow\psi_i L_i)-\Gamma(S_1\rightarrow\psi_i^\dagger L_i^\dagger)}{\Gamma(S_1\rightarrow\psi_iL_i)+\Gamma(S_1\rightarrow\psi_i^\dagger L_i^\dagger)},\\
\epsilon_2 &=& \frac{\sigma(XX\rightarrow \psi_i L_i)+\sigma(\bar X\bar X\rightarrow \psi_i L_i)-\sigma(XX\rightarrow\psi_i^\dagger L_i^\dagger)-\sigma(\bar X\bar X\rightarrow\psi_i^\dagger L_i^\dagger)}{\sigma(XX\rightarrow \psi_i L_i)+\sigma(\bar X\bar X\rightarrow \psi_i L_i)+\sigma(XX\rightarrow\psi_i^\dagger L_i^\dagger)+\sigma(\bar X\bar X\rightarrow\psi_i^\dagger L_i^\dagger)}.
\eean
$\epsilon_1$ gives the fractional asymmetry generated per $S_1$ decay, while $\epsilon_2$ gives the fractional asymmetry generated per $XX$ annihilation. The precise values of $\epsilon_1$, $\epsilon_2$ in this case depend on the masses $m_{S\alpha}$ and the couplings $\lambda_{\alpha i}$.

To reduce the number of arbitrary parameters in our analysis, we make the following assumptions:
\begin{itemize}
\item Dark matter annihilation occurs dominantly to only one flavor of lepton, and the couplings of all other leptons to $S_\alpha$ are zero.  The non-zero couplings of the single lepton flavor are denoted $\lambda_{L\alpha}$.

\item Dark matter annihilation and washout  occur mostly through the lightest scalar, $S_1$, and we consider the rates of only these processes in our analysis. For concreteness, we require that the corresponding cross sections with intermediate $S_2$ to be less than 20\% of the corresponding cross sections with $S_1$, giving roughly
\bean
\frac{\lambda_{L2}^4}{m_{S2}^4} &\lesssim& \frac{\lambda_{L1}^4}{5m_{S1}^4},\label{eq:yls2constraint}\\
\frac{\lambda_{X2}^2\lambda_{L2}^2}{m_{S2}^4} &\lesssim& \frac{\lambda_{X1}^2\,\lambda_{L1}^2}{5m_{S1}^4}.
\eean
We also assume that $m_{S1}\ll m_{S2}$, so that  the loop integrals in $\epsilon_1$ and $\epsilon_2$ can be put in a simple analytic form (\ref{eq:epsilonexpr}).

\item The physical $CP$ phases are large i.e.~$\mathrm{Im}(a)\approx a$ where $a$ is some product of couplings appearing in scattering and decay amplitudes.

\end{itemize}
None of these assumptions  are required by WIMPy leptogenesis, and we make them only to simplify the analysis and its interpretation.  Relaxing these assumptions would introduce much complexity into the Boltzmann equations while giving qualitatively similar results. The phenomenology does, however, depend to some extent on the flavor of leptons to which dark matter predominantly annihilates (see Section \ref{sec:signals} for details). With the above assumptions\footnote{This expression is derived in the narrow-width approximation. For TeV WIMPs,  $\lambda_X\gtrsim1$ is often necessary to obtain the correct dark matter relic abundance, which may lead to $\Gamma_{S1},\,\Gamma_{S2}\sim m_S$. When $2m_X>m_{S1}$, $S_1$ is kinematically forbidden from decaying into $XX$ and the $S_1$ width is  narrow (because typically $\lambda_L\lesssim1$). $S_2$ may  be broad, but the imaginary part of its self-energy correction $\mathrm{Im}\Pi(p^2)$ as substituted into (\ref{eq:epsilonexpr}) must be evaluated at $p^2=4m_X^2\ll m_{S2}^2$ and satisfies $\mathrm{Im}\Pi(4m_X^2)\sim4m_X^2\ll m_{S2}^2$. Similarly, if $2m_X<m_{S1}$, then the partial width of $S_1$ to $X$ can be very large for $2m_X\ll m_{S1}$, but once again $\mathrm{Im}\Pi$ is evaluated in (\ref{eq:epsilonexpr}) as $\mathrm{Im}\Pi(4m_X^2)\sim 4m_X^2<m_{S1}^2$. Therefore, the narrow-width approximation holds true to a degree sufficient for our purposes.},
\ben\label{eq:epsilonexpr}
\epsilon_2\approx-\frac{1}{6\pi}\,\frac{\mathrm{Im}(\lambda_{L1}^2\lambda_{L2}^{*2})}{|\lambda_{L1}|^2}\,\frac{(2m_X)^2}{m_{S2}^2}\left[7-15\left(\frac{m_\psi}{2m_X}\right)^2+9\left(\frac{m_\psi}{2m_X}\right)^4-\left(\frac{m_\psi}{2m_X}\right)^6\right].
\een
The expression for $\epsilon_1$ is the same but with $2m_X\rightarrow m_{S1}$. Since we are most interested in the asymmetry from annihilation, $\epsilon_2$ is the relevant parameter for WIMPy baryogenesis and we denote its asymmetry factor by $\epsilon\equiv\epsilon_2$. $\epsilon$ is suppressed by $1/m_{S2}^2$ from the $S_2$ propagator, and is proportional to $(2m_X)^2$ because the momentum flowing through the $S_1$ propagator in $XX$ annihilation is  $\sqrt{\hat s} = 2m_X+\mathcal O(T)$, where $T\ll m_X,m_{S1}$ at freeze-out. Note  that (\ref{eq:epsilonexpr}) vanishes when $m_\psi=2m_X$, at which point the particles in the loop cannot go on shell and there is no imaginary part of the amplitude (and, hence, no $CP$ violation).

Using (\ref{eq:yls2constraint}), (\ref{eq:epsilonexpr}), and the assumption of large $CP$ phases, we can bound $\epsilon$ from above:
\ben\label{eq:epsilonapprox}
|\epsilon| \lesssim\frac{2\lambda_{L1}^2}{3\pi\sqrt 5}\,\frac{m_X^2}{m_{S1}^2}\left[7-15\left(\frac{m_\psi}{2m_X}\right)^2+9\left(\frac{m_\psi}{2m_X}\right)^4-\left(\frac{m_\psi}{2m_X}\right)^6\right].
\een
We  treat $\epsilon$ as a free parameter, subject to (\ref{eq:epsilonapprox}), and we can now express all rates and cross sections in terms of $\lambda_X\equiv\lambda_{X1}$, $\lambda_L\equiv\lambda_{L1}$, $\epsilon$, $m_X$, $m_\psi$, and $m_S\equiv m_{S1}$.

We have assumed that the lepton asymmetry from  $XX$ annihilations dominates over that from $S$ decays. We find that this assumption is true whenever $m_X<m_S$. Since the asymmetry is proportional to the number density of $X$ or $S$ at the time of washout freeze-out, the ratio of asymmetry from decay vs.~annihilation is the same as the ratio of the number of $S$ particles to the number of $X$ particles at the time of washout freeze-out. The assumption of annihilation-dominated asymmetry is therefore equivalent to $m_X< m_S$.

\subsubsection{Washout}\label{sec:washout}

\begin{figure}[t]
\begin{center}
\includegraphics[width=3cm]{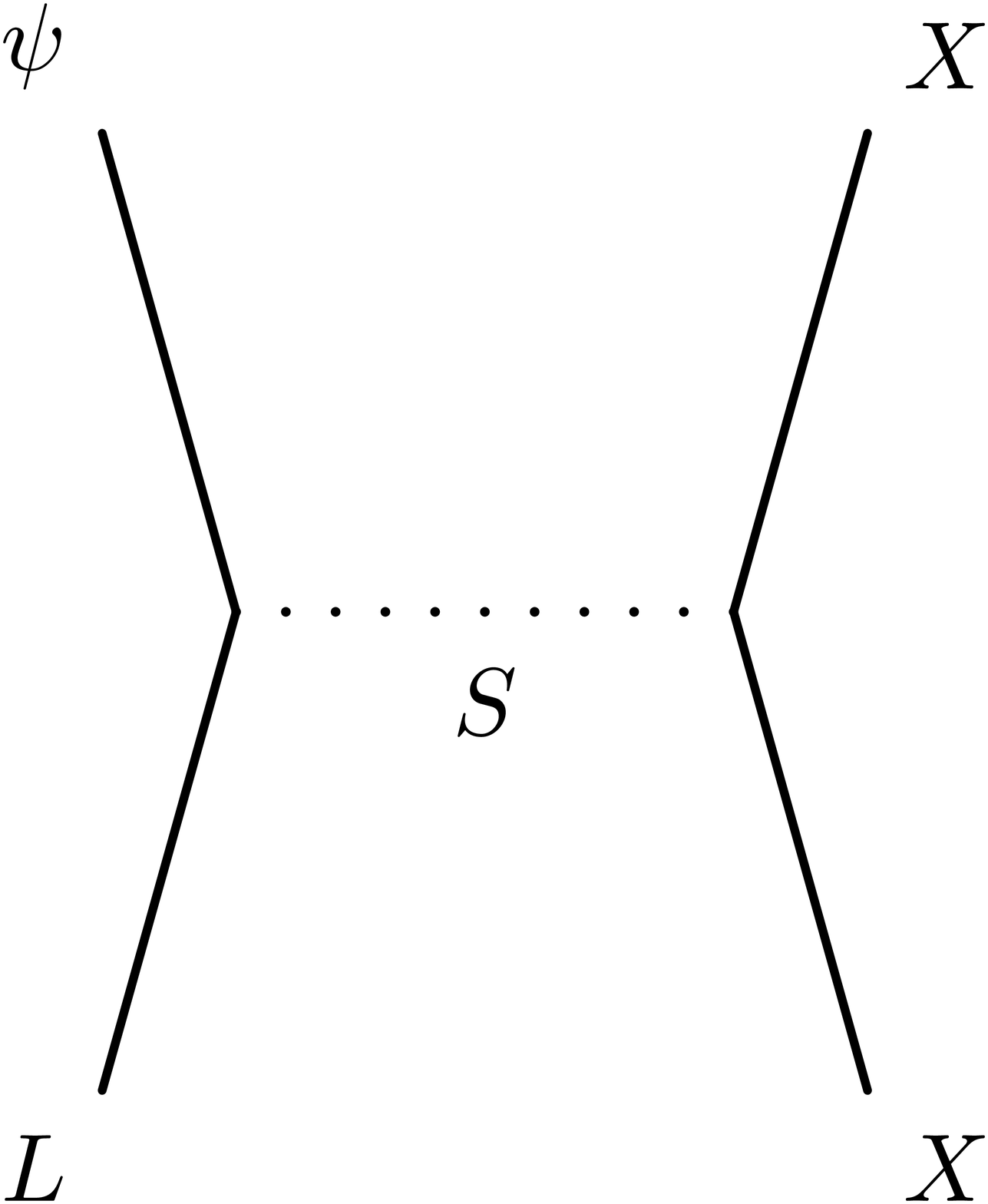}\hspace{1cm}
\includegraphics[width=3cm]{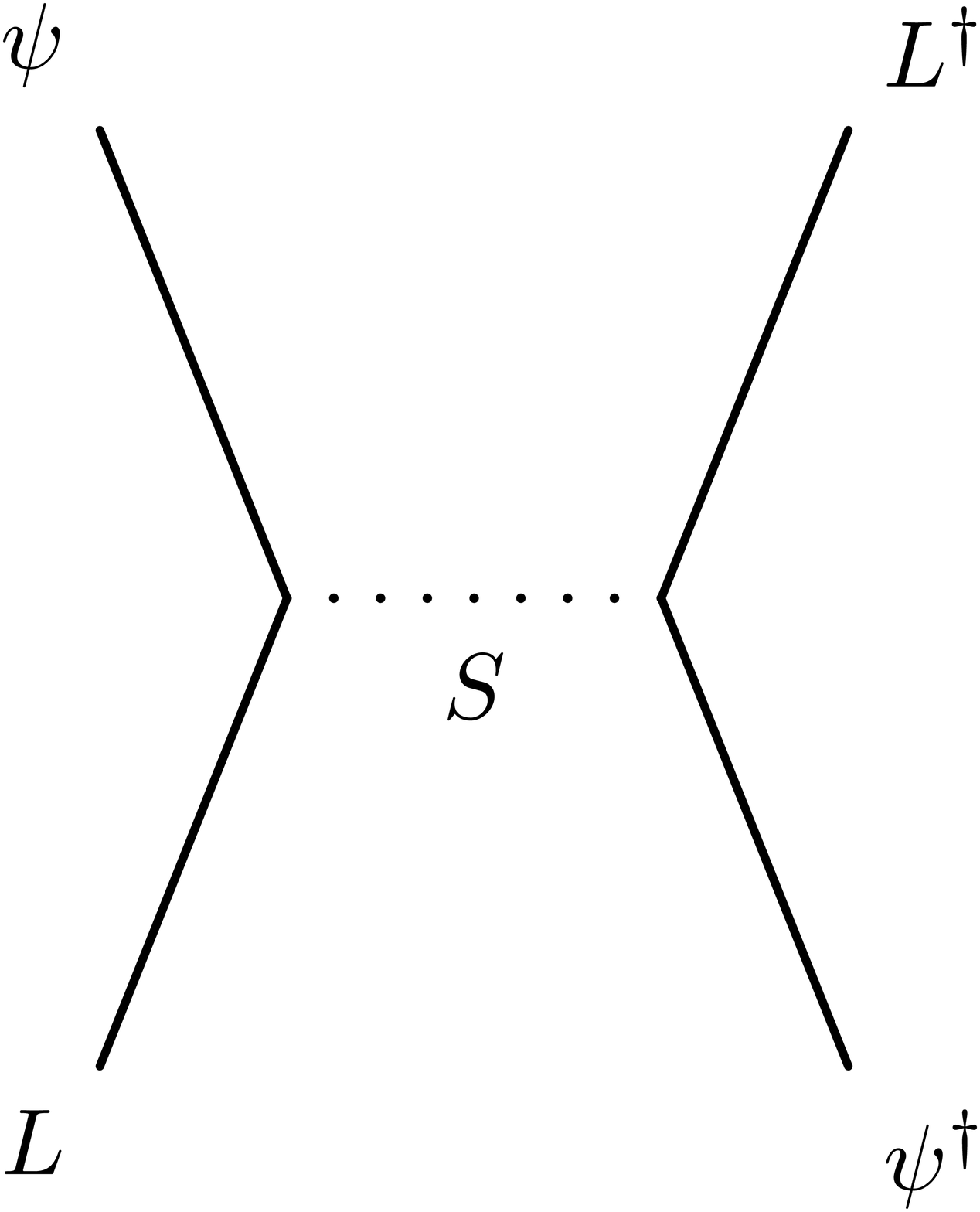}
\\
\vspace{0.5cm}
\includegraphics[width=3cm]{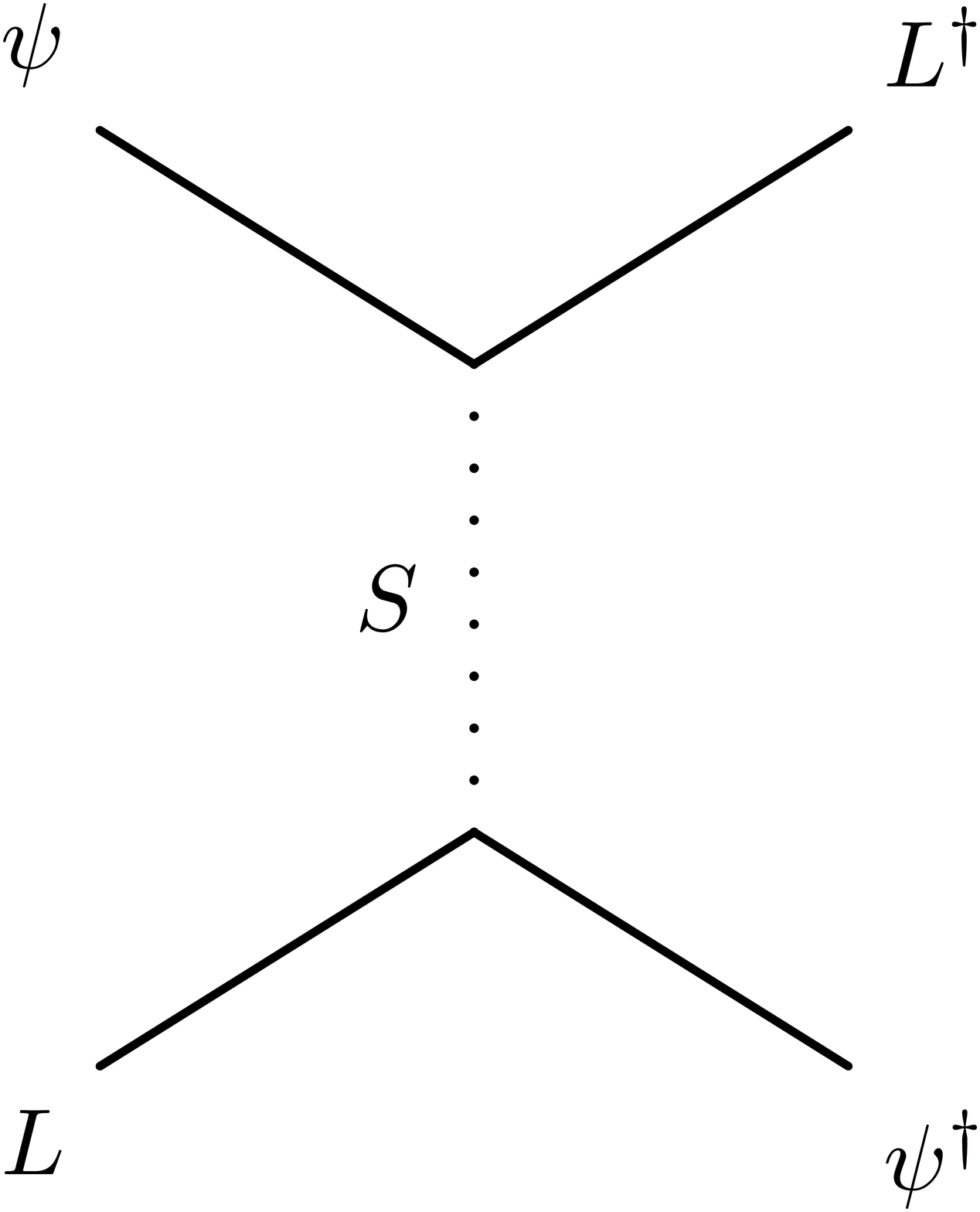}\hspace{1cm}
\includegraphics[width=3cm]{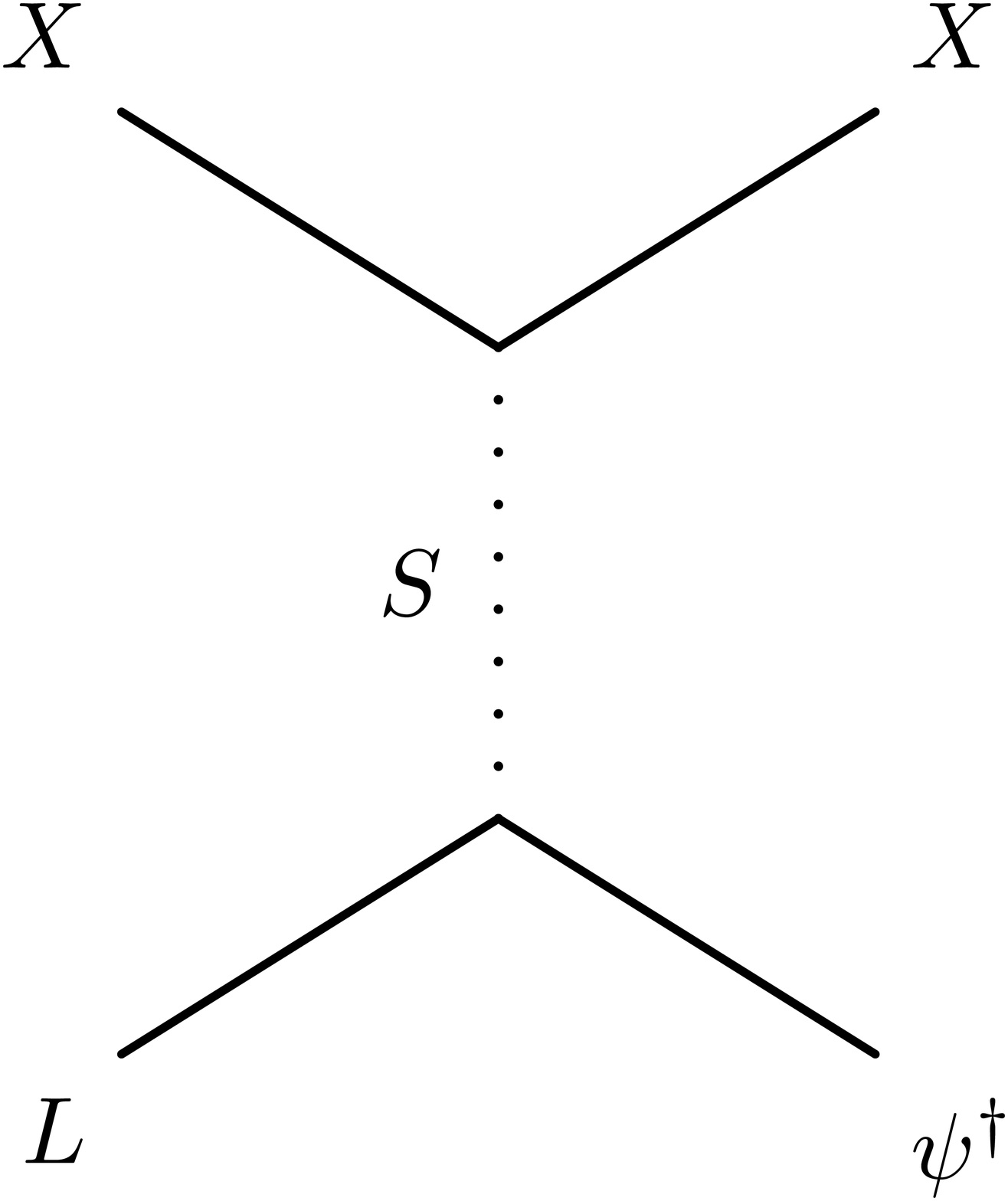}\hspace{1cm}
\includegraphics[width=3cm]{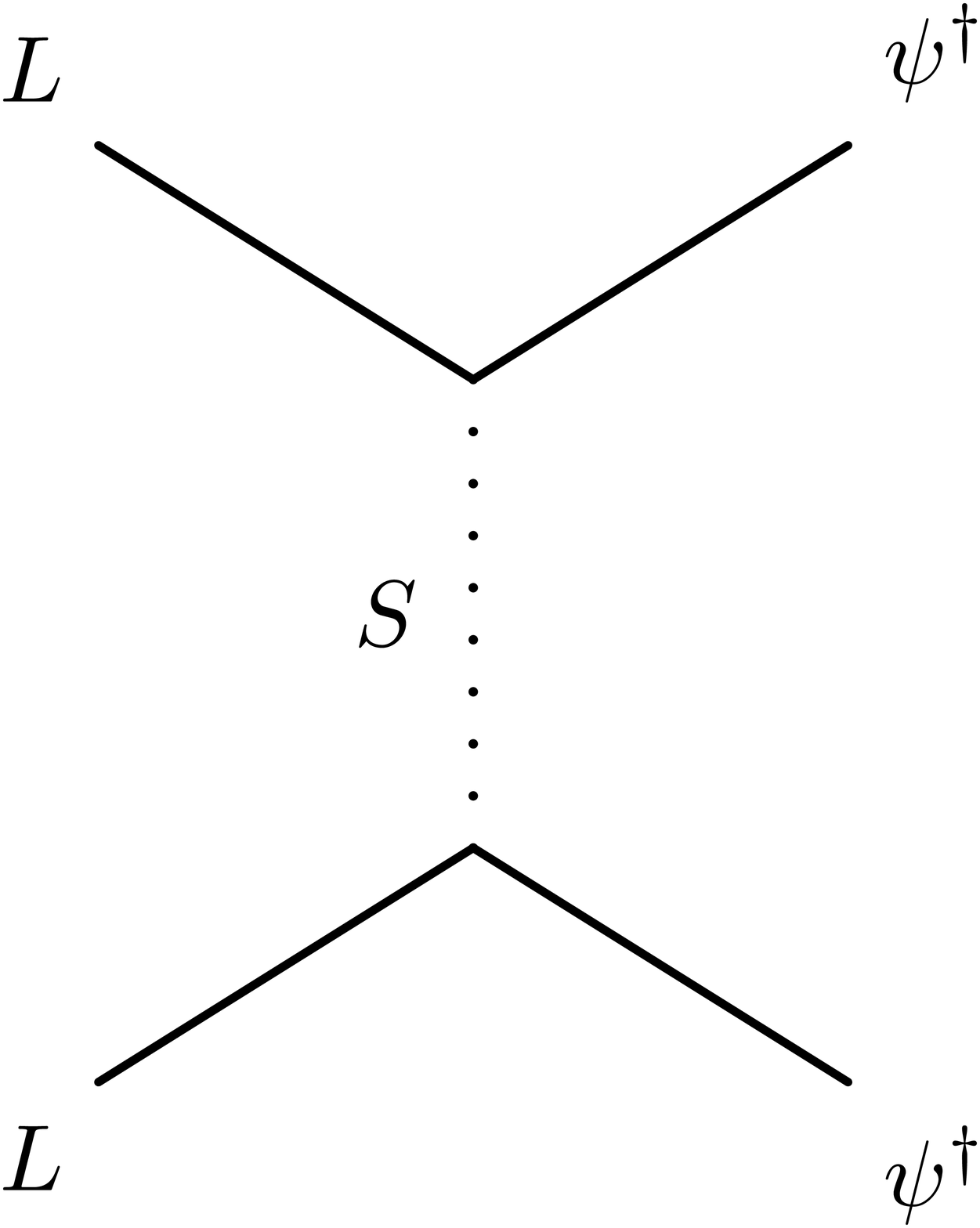}
\caption{Diagrams leading to washout of the lepton number from (top row) $s$-channel and (bottom row) $t$-channel scatterings.}
\label{fig:washout}
\end{center}
\end{figure}

As we demonstrated in Section \ref{sec:wimpybaryogenesis}, the final baryon asymmetry depends on the time of washout freeze-out. We now discuss  the implications for  WIMPy leptogenesis, finding that we need $m_\psi\gtrsim m_X$ for successful WIMPy leptogenesis. We show the lepton number washout processes in Figure \ref{fig:washout}. They include inverse annihilations, lepton $\rightarrow$ antilepton scatterings, and $\psi X\rightarrow L^\dagger X$ processes. The dominant washout is typically from $L\psi\rightarrow L^\dagger\psi^\dagger$ scatterings, because the inverse annihilation  $L\psi\rightarrow XX$ is kinematically suppressed  for $T<m_X$ and $\psi X\rightarrow L^\dagger X$ gets more Boltzmann suppression. Applying (\ref{eq:washoutrate}) for the specific model of WIMPy leptogenesis, the washout rate is proportional to
\ben
\Gamma_{\rm washout}(x) \approx \frac{s(x)}{\,Y_\gamma}\,\langle\sigma_{L\psi\rightarrow L^\dagger\psi^\dagger}\,v\rangle\,Y_L^{\rm eq}\,Y_\psi^{\rm eq}(x),
\een
where $s(x)$ is the entropy density at $x$. Washout freezes out when its rate is about equal to the Hubble scale, $\Gamma_{\rm washout}(x_{\rm washout})\thickapprox H(x_{\rm washout})$. $\Gamma_{\rm washout}(x_{\rm washout})$ can be small for one of two reasons:
\benum
\item $m_\psi\gtrsim m_X$ so that $Y_\psi^{\rm eq}(x_{\rm washout})$ is Boltzmann-suppressed while dark matter is annihilating.
\item $\langle\sigma_{L\psi\rightarrow L^\dagger\psi^\dagger}\,v\rangle$ is small relative to the annihilation cross section so that washout freezes out before annihilation. The washout cross section can be small if $\lambda_L\ll1$.
\eenum
One of these two conditions must hold for each washout process. We find that option \#1 leads to viable WIMPy leptogenesis. Option \#2, on the other hand, does not give a large asymmetry. According to (\ref{eq:epsilonapprox}), the asymmetry efficiency factor $\epsilon$ is also suppressed  when $\lambda_L\ll1$, and the potential gain in the baryon asymmetry from early washout freeze-out in option \#2 is offset  because leptogenesis occurs less efficiently. Therefore, $m_\psi\gtrsim m_X$ is generally required to generate the observed baryon asymmetry.

\begin{figure}[t]
\begin{center}
\includegraphics[width=7cm]{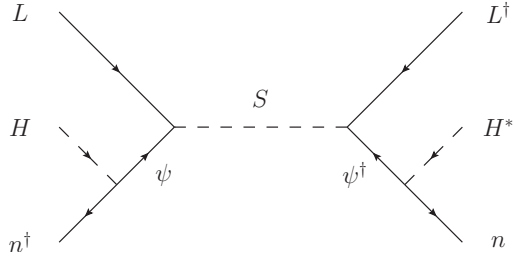}
\caption{$3\rightarrow3$ washout process that can dominate over $2\rightarrow2$ scattering when $T\ll m_\psi$.}
\label{fig:3body}
\end{center}
\end{figure}

When $m_\psi$ is much larger than $m_X$ (we find  this is typically true for $m_\psi \gtrsim 2m_X$), the exponential suppression of $Y_\psi$ is so large that $3\rightarrow3$ scatterings of $L\,n\,H\rightarrow L^\dagger\,n^\dagger\,H^*$ become important (see Figure \ref{fig:3body}). This region is, however, kinematically inaccessible in WIMPy leptogenesis since $2m_X>m_\psi$ for efficient annihilation to occur, and we neglect $3\rightarrow3$ processes.

\subsubsection{Boltzmann equations}\label{sec:Boltzmann}

We consider the evolution of a single component $L_a$, where $a$ is a gauge index (flavor indices are suppressed since we  consider only one flavor). We define $\mathrm{Br}_L$ ($\mathrm{Br}_X$) as the total branching fraction of $S$  into leptons ($X$), with $\mathrm{Br}_L+\mathrm{Br}_X=1$. Also, $\xi = 1+\mu_{\psi_a}/\mu_{L_a}$, where $\mu$ are chemical potentials, and $\eta$ is defined as the amount of $L_a$ asymmetry generated for each $L_a$ directly created/annihilated (this accounts for the fact that the asymmetry spreads among all baryons and leptons by rapid thermalization). We calculated  the cross sections and widths analytically and checked them with CalcHEP \cite{Pukhov:2004ca}.

The Boltzmann equations describing the evolution of the various particle species and the asymmetry in {\bf one} of the components of the $L_a$ doublet are
\small
\bean
\frac{H(m_X)}{x}\,\frac{dY_X}{dx}&=&-4s\langle\sigma_{XX\rightarrow L_a\psi_a}\,v\rangle[Y_X^2-(Y_X^{\rm eq})^2] - 2s\epsilon\frac{\xi\,Y_{\Delta L_a}}{Y_\gamma}\langle\sigma_{XX\rightarrow L_a\psi_a}\,v\rangle(Y_X^{\rm eq})^2\nonumber\\
&& -\mathrm{Br}^2_X\langle\Gamma_S\rangle Y_S^{\rm eq}\left(\frac{Y_X}{Y_X^{\rm eq}}\right)^2+\mathrm{Br}_X\langle\Gamma_S\rangle\left(Y_S-\mathrm{Br}_L\,Y_S^{\rm eq}\right)-\epsilon\frac{\xi\,Y_{\Delta L_a}}{2Y_\gamma}\mathrm{Br}_X\mathrm{Br}_L\langle\Gamma_S\rangle Y_S^{\rm eq};\label{eq:xabundance}\\
\frac{H(m_X)}{x}\,\frac{dY_S}{dx} &=& -\langle\Gamma_S\rangle Y_S+\langle\Gamma_S\rangle Y_S^{\rm eq}\left[\mathrm{Br}_L+\mathrm{Br}_X\left(\frac{Y_X}{Y_X^{\rm eq}}\right)^2\right];\\
\frac{H(m_X)}{x\,\eta}\frac{dY_{\Delta L_a}}{dx} &=& \frac{\epsilon}{2}\mathrm{Br}_L\langle\Gamma_S\rangle\left[Y_S+Y_S^{\rm eq}\left(1-2\mathrm{Br}_L-\mathrm{Br}_X\left[1+\frac{Y_X^2}{(Y_X^{\rm eq})^2}\right]\right)\right]+2s\,\epsilon\langle\sigma_{XX\leftrightarrow L_a\psi_a}\,v\rangle\left[Y_X^2-(Y_X^{\rm eq})^2\right]\nonumber\\
&& {} -\frac{\xi\,Y_{\Delta L_a}}{Y_\gamma}\left[s\,\langle\sigma_{XX\leftrightarrow L_a\psi_a}\,v\rangle(Y_X^{\rm eq})^2+2s[\langle\sigma_{L_a\psi_a\leftrightarrow L^\dagger_a\psi_a^\dagger}\,v\rangle+\langle\sigma^{(a\neq b)}_{L_a\psi_a\leftrightarrow L^\dagger_b\psi_b^\dagger}\,v\rangle] Y_L^{\rm eq}Y_\psi^{\rm eq}+2s\,\langle\sigma_{L_a\psi_b\leftrightarrow L^\dagger_b\psi_a^\dagger}\,v\rangle Y_L^{\rm eq}Y_\psi^{\rm eq}\right]\nonumber\\
&& {}-\frac{\xi\,Y_{\Delta L_a}}{Y_\gamma}\left[s\,\langle\sigma_{X\psi_a\leftrightarrow X L^\dagger_a}\,v\rangle Y_X Y_\psi^{\rm eq}+2s\,\langle\sigma_{\psi_a\psi_a\leftrightarrow L_a^\dagger L_a^\dagger}\,v\rangle(Y_\psi^{\rm eq})^2+2s\,\langle\sigma^{(a\neq b)}_{\psi_a\psi_b\leftrightarrow L_a^\dagger L_b^\dagger}\,v\rangle(Y_\psi^{\rm eq})^2\right]\nonumber\\
&& {} +\frac{\xi\,Y_{\Delta L_a}}{4Y_\gamma}\mathrm{Br}_L\langle\Gamma_S\rangle Y_S^{\rm eq}\left(\epsilon^2\mathrm{Br}_L+\mathrm{Br}_X\right).\label{eq:leptonasymmetry}
\eean
\normalsize
We assume that all abundances are in thermal equilibrium at $x=1$ and that all fields remain in equilibrium except for $S$ and $X$. In the evolution of the scalar $S$, we only include the decay terms, as they dominate over $SS$ annihilation for $T\ll m_S$.

To determine the relationship between $\mu_{\psi_a}$ and $\mu_{L_a}$, we assume that all abundances other than $S$, $X$ and $\Delta L_a$ are in thermal equilibrium and that all processes except those involving $S$ are in chemical equilibrium. We also take sphaleron processes to be in equilibrium. The non-$S$ couplings in (\ref{eq:lagrangian}) distribute the $L$ and $\psi$ asymmetries among the light fields. Solving the chemical potential relations gives
\bean
\xi &=& \frac{16+12\,n_{\psi_a}^{\rm eq}/n_{L_a}^{\rm eq}}{3+12\,n_{\psi_a}^{\rm eq}/n_{L_a}^{\rm eq}},\\
\eta &=&\frac{2(7+28\,n_{\psi_a}^{\rm eq}/n_{L_a}^{\rm eq})}{79+355\,n_{\psi_a}^{\rm eq}/n_{L_a}^{\rm eq}}.
\eean
The precise values of $\xi$ and $\eta$ are model-dependent.

The assumption of thermal equilibrium for $\psi$ is consistent provided the decay width $\Gamma_\psi > H(T_{\rm lep})$, which in our model constrains
\ben\label{eq:equil}
\lambda_i^2 \gtrsim \frac{16\pi H(T_{\rm lep})}{m_{\psi\,i}} \approx \frac{80\sqrt{g_*}\,T_{\rm lep}^2}{M_{\rm Pl}\,m_{\psi\,i}}.
\een
For $T_{\rm lep}=100$ GeV and $m_{\psi\,i}=2$ TeV, this gives $\lambda_i \gtrsim 3\times10^{-8}$. This is not a very stringent requirement, since this value is smaller than any of the Standard Model Yukawa couplings.

The final lepton asymmetry is also determined by the chemical potential relations. The relation between the total lepton asymmetry $\Delta L_{\rm tot}$ and the asymmetry in a single component of the doublet field $\Delta L_a$ as determined by equation (\ref{eq:leptonasymmetry}) is
\ben
Y_{\Delta L\,\rm{tot}} = \frac{51+243n_{\psi_a}^{\rm eq}/n_{L_a}^{\rm eq}}{7+28n_{\psi_a}^{\rm eq}/n_{L_a}^{\rm eq}}\,Y_{\Delta La}.
\een
The final baryon asymmetry $Y_{\Delta B}$ follows from the sphaleron chemical potential relations, and is
\ben
Y_{\Delta B}(x) = -4\left(\frac{7+28n_{\psi_a}^{\rm eq}/n_{L_a}^{\rm eq}}{51+243n_{\psi_a}^{\rm eq}/n_{L_a}^{\rm eq}}\right) Y_{\Delta L\,\rm{tot}} \\
=-4Y_{\Delta L_a}.
\een
In the limit $x\rightarrow\infty$, the ratio of total baryon to lepton number reduces to the same expression as conventional leptogenesis \cite{Harvey:1990qw}.

The total dark matter relic abundance is
\ben
Y_{\rm DM}(\infty) = Y_X(\infty) + Y_{\bar X}(\infty) = 2Y_X(\infty).
\een

\subsection{Numerical Results}\label{sec:leptonnum}
There are six free parameters in our model: three masses ($m_S$, $m_X$, and $m_\psi$) and three dimensionless parameters ($\lambda_X$, $\lambda_L$, and $\epsilon$). To determine over what range of parameters WIMPy leptogenesis can be successful, we perform scans over two parameters at a time while holding others fixed. In particular, we are interested to see what range of masses is allowed, and if any tuning of the mass and coupling constant relations is necessary to generate the correct baryon asymmetry and WIMP relic density.\\

\begin{figure}[t]
\begin{center}
\includegraphics[width=9cm]{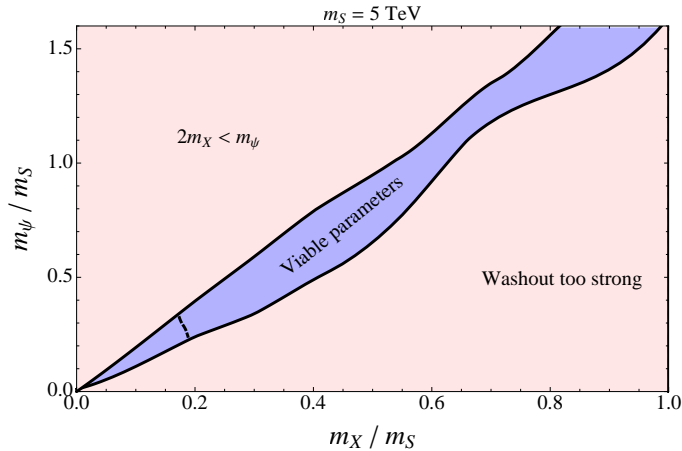}
\caption{Regions in the $m_X$-$m_\psi$ plane with the correct WIMP relic density and baryon asymmetry from WIMPy leptogenesis, with $m_S=5$ TeV and some choice of perturbative couplings. The masses giving both observed abundances are shown in blue (middle stripe).  We plot the ratios $m_X/m_S$, $m_\psi/m_S$  to show the relationship between the $X$ and $\psi$ masses and the  mediator scale $m_S$.  The excluded regions are shown in red: the upper region is not viable because $2m_X<m_\psi$ and the thermal annihilation cross section is  Boltzmann-suppressed, while the lower region has $Y_\psi$ too large to prevent rapid washout of the asymmetry. The dashed line indicates the lower boundary of allowed $m_X$ and $m_\psi$; below the line, the electroweak phase transition occurs before the baryon asymmetry is large enough to account for the observed value.  For $m_X/m_S>1$, the asymmetry is dominated by $S$ decay. }
\label{fig:mxmpsimain}
\end{center}
\end{figure}

\noindent\underline{Range of allowed masses}: We hold $m_S$ fixed and determine for which $m_X$ and $m_\psi$ masses there exists \emph{some} perturbative couplings that give the observed dark matter density and baryon asymmetry. We place no other restrictions on the couplings. If $m_\psi>m_S$, we assume that the $S$ width is dominated by the three-body decay $S\rightarrow L\,H\,n$. We show in Figure \ref{fig:mxmpsimain} the masses that give rise to successful WIMPy leptogenesis. The viable $\psi$ masses satisfy $m_\psi\approx (1-2)m_X$, while there is no correlation between $m_X$ and $m_S$ as long as $m_X<m_S$. For smaller values of $m_\psi/m_X$, the Boltzmann suppression of the washout rate is insufficient to generate the observed baryon asymmetry, while $m_\psi\gtrsim2m_X$ is not allowed because dark matter annihilation is not kinematically allowed and because the asymmetry efficiency $\epsilon$ is zero ($CP$ violation is zero if $L$ and $\psi$ cannot go on-shell in the dark matter annihilation loop diagrams).

The lower boundary of the allowed region has a meandering shape around $m_\psi\approx m_S$. The reason is that $s$-channel washout processes have a resonant enhancement in this region, leading to a smaller baryon asymmetry and a restricted parameter space. Above resonance, $t$-channel washout processes are also important, explaining why the bend in the curve is  centered at $m_\psi$ slightly larger than $m_S$.

The principal reason that it is difficult to generate a large baryon asymmetry is because the efficiency of asymmetry generation $\epsilon$ is tied to the washout cross section through its dependence on $\lambda_L^2$ in (\ref{eq:epsilonapprox}). A large asymmetry can only be generated when washout effects are also large, limiting how much of an asymmetry can be generated. The viable parameter space is larger if (\ref{eq:epsilonapprox}) can be relaxed, as is the case when $S_1$ and $S_2$ are nearly degenerate and the asymmetry is resonantly enhanced, but this is not a required feature of WIMPy leptogenesis.

In leptogenesis, the asymmetry must be generated prior to the electroweak phase transition, at which point sphalerons decouple and the conversion of a lepton asymmetry into a baryon asymmetry ceases. Since WIMPy leptogenesis is a weak-scale model, the timing of the asymmetry generation relative to the phase transition is important. To illustrate this, we computed the critical temperature $T_{\rm c}$ of the phase transition assuming a Standard Model Higgs with mass $m_h=120$ GeV, and we required that the baryon asymmetry at $T_{\rm c}$ be equal to the observed asymmetry\footnote{Since the Standard Model phase transition is of second order, sphalerons do not suddenly shut off, and a more proper treatment would account for the gradual departure from equilibrium of the sphaleron effects. Since the asymmetry is typically generated over a very short time period (we find numerically that it is on the order of $\Delta x\sim2-3$ or $\Delta T\sim5-10$ GeV), the dynamics of sphaleron shut-off are largely irrelevant and the most important factor is the rate of $L\rightarrow B$ transfer at the washout freeze-out time $x_{\rm washout}$. }. This typically yields a much smaller baryon asymmetry than lepton asymmetry at late times because the baryon asymmetry stops accumulating at $T_{\rm c}$. Accounting for the effects of the phase transition, the allowed region is above the dashed line in Figure \ref{fig:mxmpsimain}. If the phase transition is modified by additional Higgs fields or other new physics, then this boundary line changes.\\

\begin{figure}[t]
\begin{center}
\includegraphics[width=7.5cm]{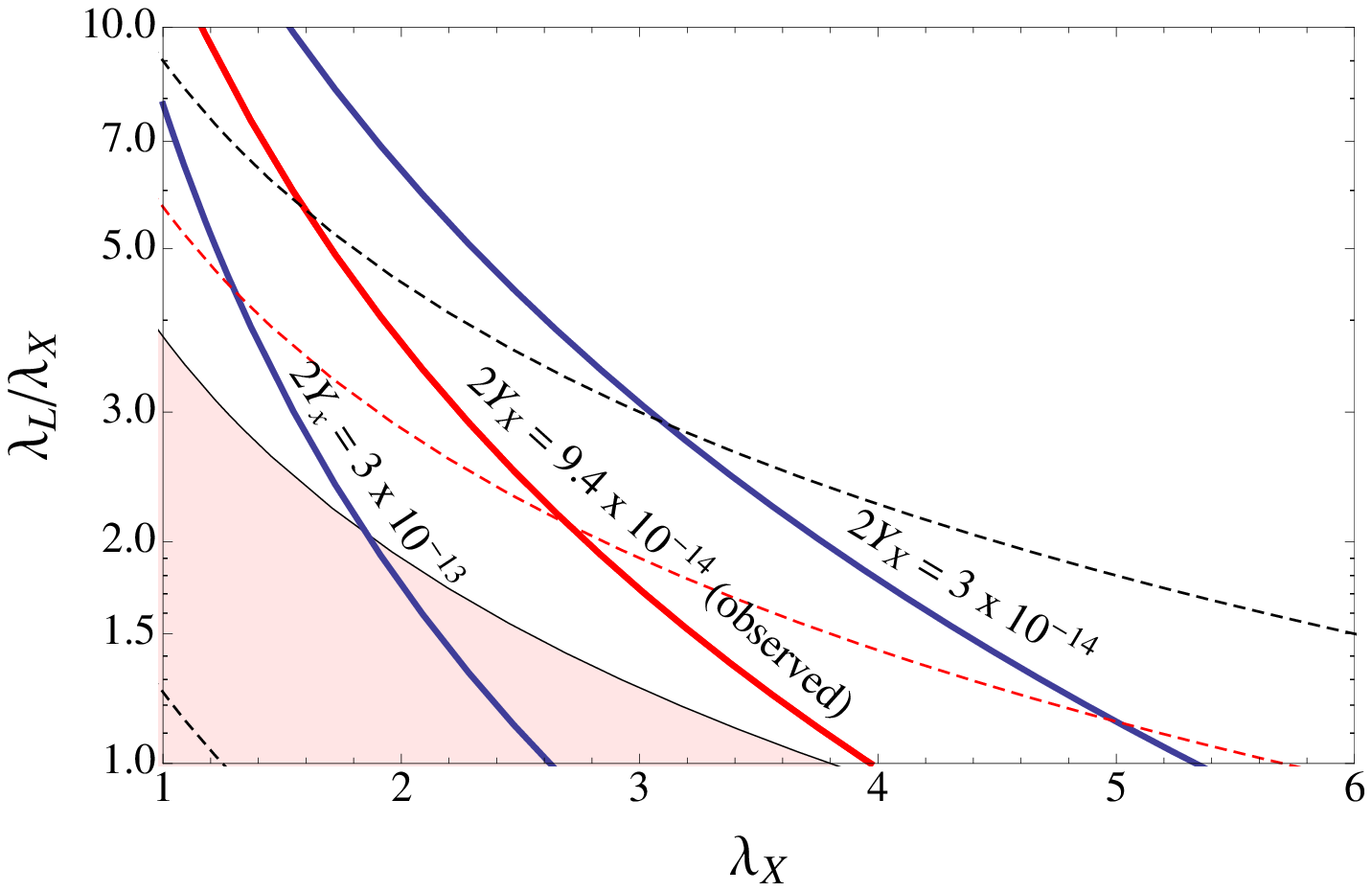}\hspace{1cm} \includegraphics[width=7.5cm]{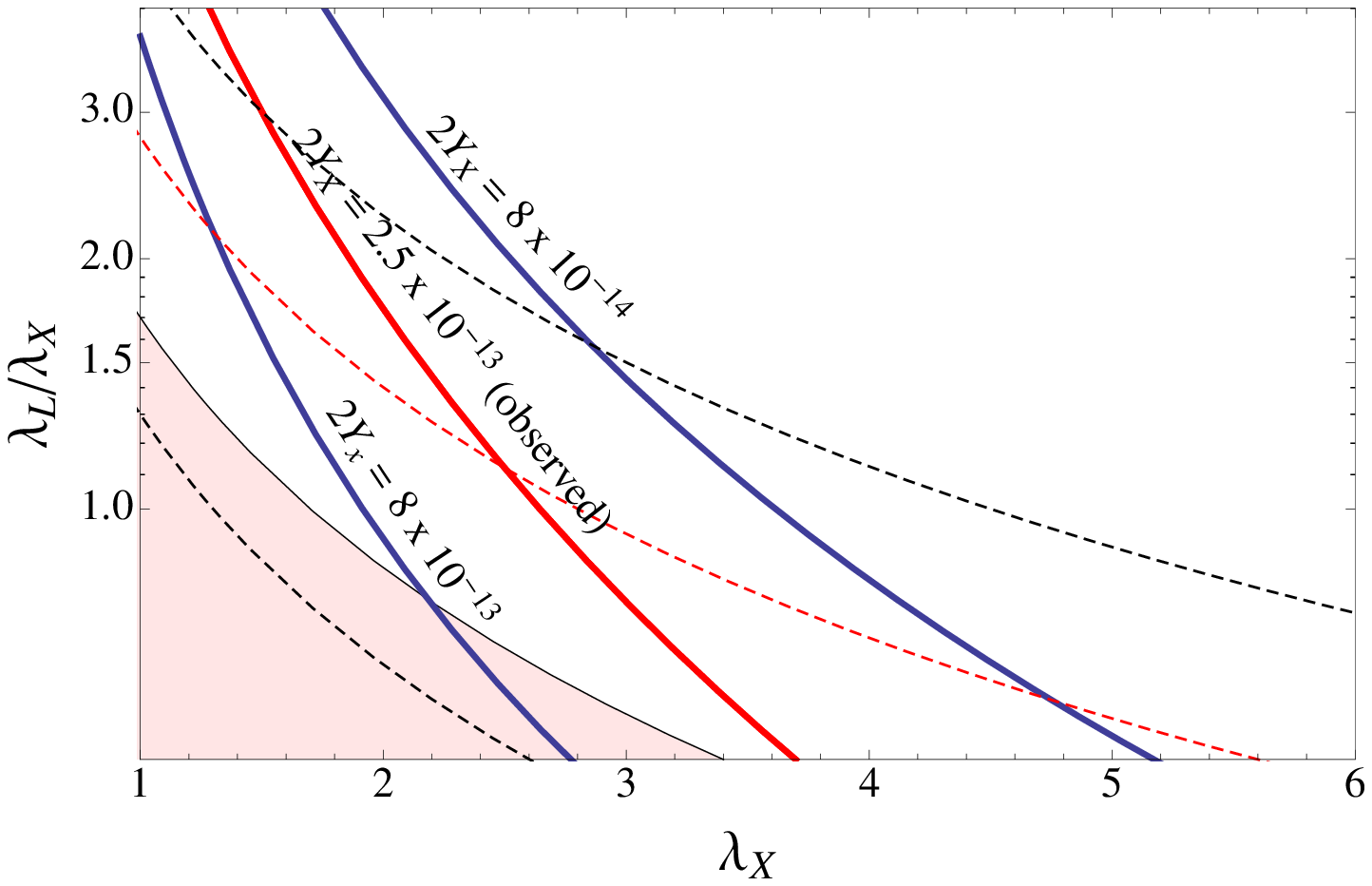}
\caption{Dark matter relic density \emph{(solid lines)} and baryon asymmetry \emph{(dotted lines)} as functions of couplings $\lambda_X$ and $\lambda_L/\lambda_X$. We consider two sets of massses, with $m_S=5$ TeV for both: {\bf (left)} $m_X=4.25$ TeV, $m_\psi=7.5$ TeV, and $|\epsilon|=0.075$; {\bf (right)} $m_X=1.5$ TeV, $m_\psi=2.25$ TeV, and $|\epsilon|=0.0075$. The asymmetry contours are, from top to bottom: {\bf (left)} $Y_{\Delta B} = 5\times10^{-11}$, $8.5\times10^{-11}$ (observed asymmetry), and $3\times10^{-10}$; {\bf (right)} $Y_{\Delta B} = 3\times10^{-11}$, $8.5\times10^{-11}$ (observed asymmetry), and $3\times10^{-10}$. The dark matter abundances are printed on the plots. In the shaded regions, the numerical value of $\epsilon$ is not consistent with our assumptions according to the bound (\ref{eq:epsilonapprox}).}
\label{fig:ylyxmain}
\end{center}
\end{figure}

\noindent\underline{Range of allowed couplings}: We choose representative values of the masses, with $m_S=5$ TeV for all cases, and $m_X$ and $m_\psi$ chosen in the middle of the allowed bands in Figure \ref{fig:mxmpsimain}. For one set of parameters, dark matter annihilates above the $S$ resonance, with parameters $m_X=4.25$ TeV, $m_\psi=7.5$ TeV, and $|\epsilon|=0.075$, and we determine the dark matter relic abundance and baryon asymmetry as  functions of the two couplings. We also study $XX$ annihilation below resonance, with $m_X=1.5$ TeV, $m_\psi=2.25$ TeV, and $|\epsilon|=0.0075$. We plot the results in Figure \ref{fig:ylyxmain} as contours of constant relic density and baryon asymmetry. We focus on the ratio $\lambda_L/\lambda_X$ because we are interested in seeing if any relationship between these two theoretically unrelated quantities is required to obtain a particular relic abundance and asymmetry. In both cases shown, WIMPy leptogenesis gives the correct dark matter relic abundance and asymmetry when both couplings are $\mathcal O(1)$. Thus, a perfectly natural choice of couplings, and the very same couplings that satisfy the WIMP miracle, can also generate the correct baryon asymmetry if $CP$ phases are large! Specifically, with $m_X=4.25$ TeV, $m_\psi = 7.5$ TeV, and $|\epsilon|=0.075$, the observational constraints are satisfied with $\lambda_X=2.7$ and $\lambda_L=5.7$; with $m_X=1.5$ TeV, $m_\psi=2.25$ TeV, and $|\epsilon|=0.0075$, the couplings are $\lambda_X=2.8$, $\lambda_L=2.5$.\\

In deriving our results, we assumed that dark matter only annihilates through lepton-number-violating interactions. In a more general model, dark matter may also have lepton-number-preserving interactions that contribute to the total annihilation cross section. We parameterize this possibility with the quantity
\ben
\alpha \equiv \frac{\langle\sigma_{XX\rightarrow\mathrm{anything}}\,v\rangle}{\langle\sigma_{XX\rightarrow L\psi}\,v\rangle} \geq1.
\een
When $\alpha>1$, the asymmetry generated by WIMPy leptogenesis is smaller, because only $1/\alpha$ of dark matter annihilations proceed through lepton-number-violating couplings and can create an asymmetry\footnote{When $\alpha>1$, the WIMP annihilation cross section is also larger, and $\lambda_X$, $\lambda_L$ are smaller to give the same WIMP relic density. As discussed in Section \ref{sec:washout}, however, decreasing the couplings results in \emph{both} a smaller washout rate and a smaller efficiency of generating an asymmetry. These two effects counteract one another, and the change in couplings for $\alpha>1$ does not substantially affect the asymmetry.}.
%
\begin{figure}[t]
\begin{center}
\includegraphics[width=7cm]{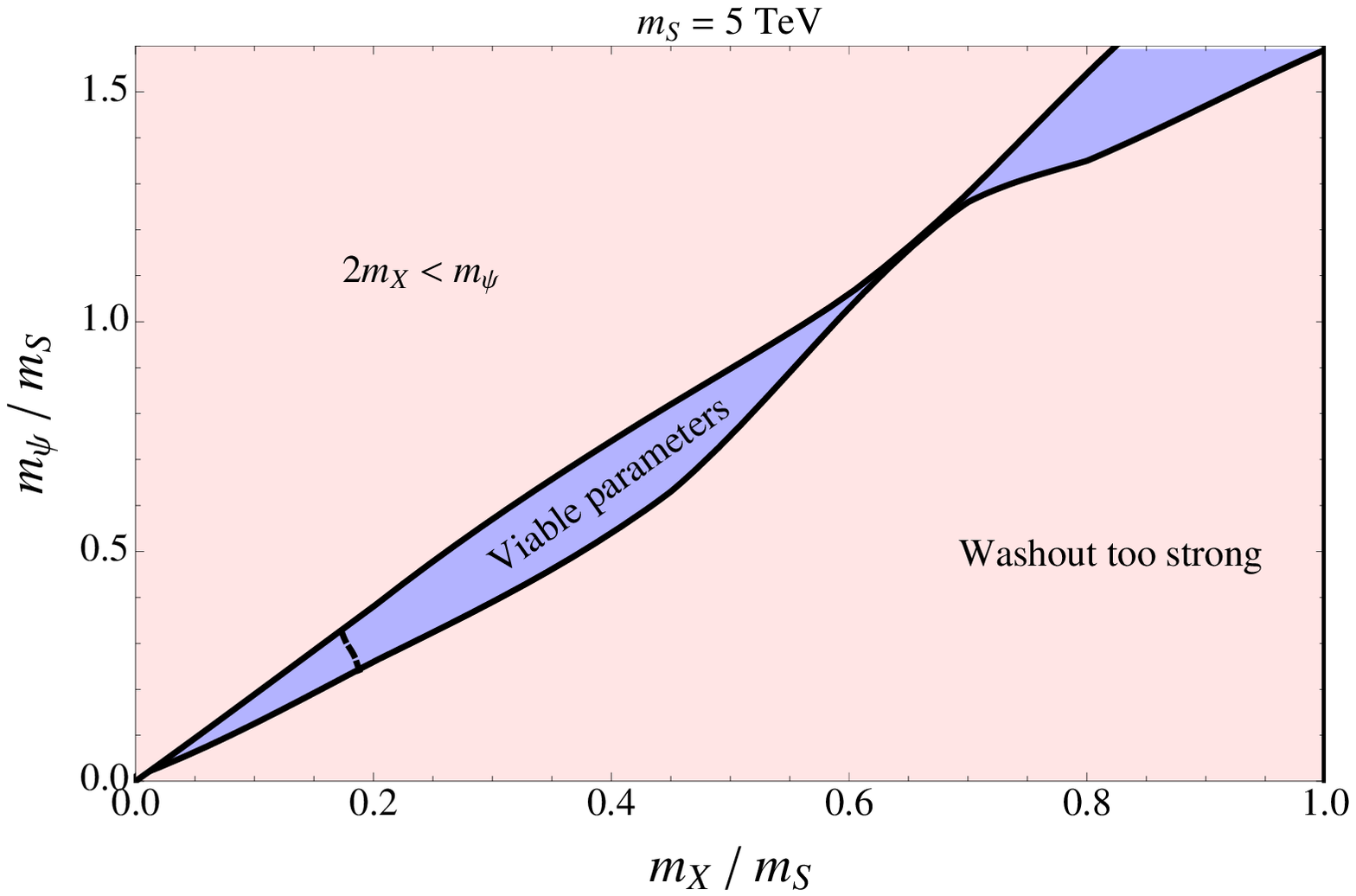}\hspace{1cm}
\includegraphics[width=7cm]{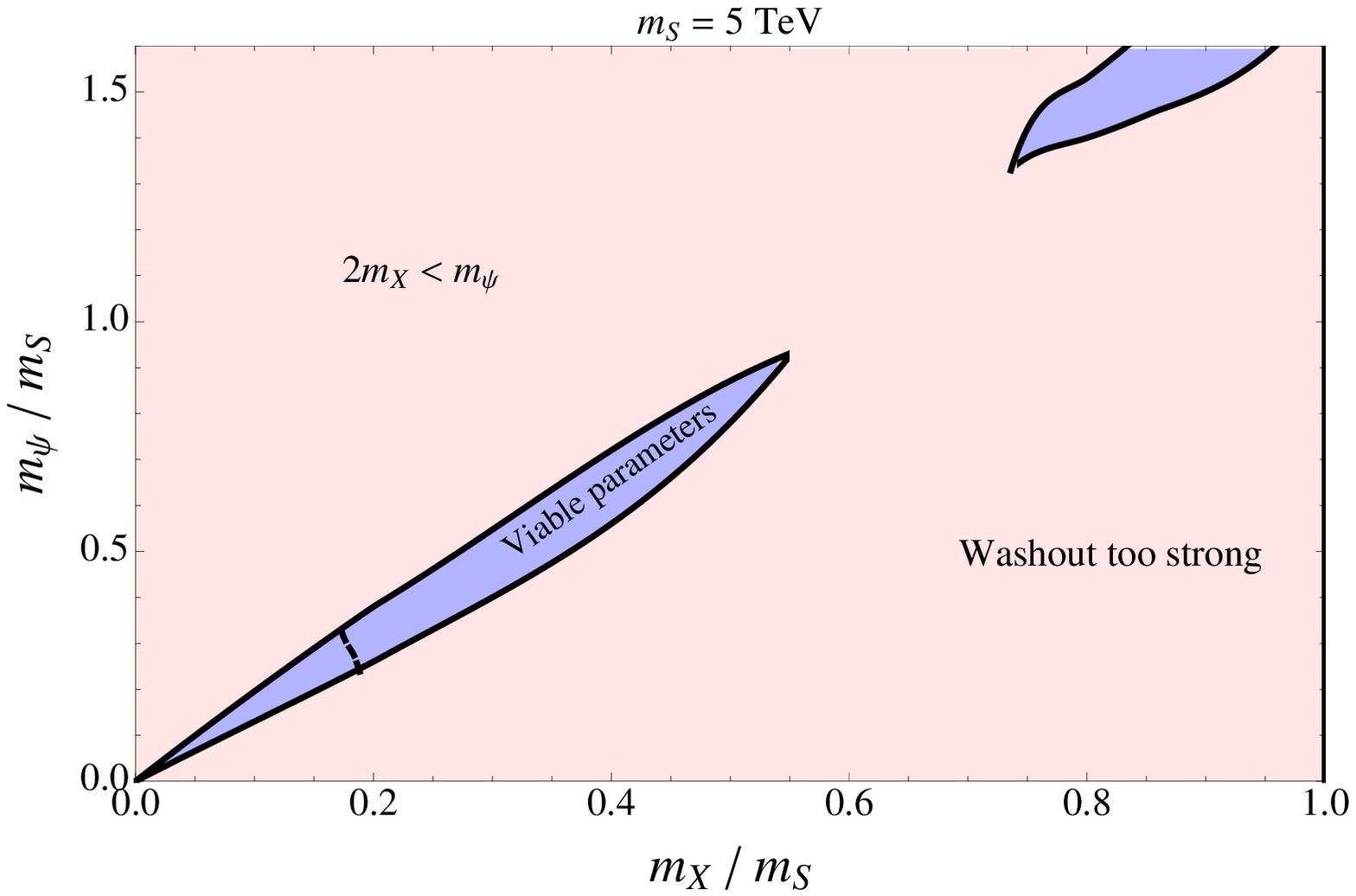}
\caption{Regions in the $m_X$-$m_\psi$ plane for viable WIMPy leptogenesis with additional lepton-number-preserving dark matter annihilation modes: {\bf (left)} $\alpha=2$, {\bf (right)} $\alpha=3$. The masses giving the correct dark matter density and baryon asymmetry for some choice of perturbative coupling are shown in blue (middle stripe). As the lepton-number-preserving annihilation cross section increases, the efficiency of asymmetry generation drops, and the marginal regions of parameter space become inaccessible, particularly the enhanced washout region $m_\psi\sim m_S$.  The descriptions of the regions on the plot are the same as those in Figure \ref{fig:mxmpsimain}. }
\label{fig:leptonalphamass}
\end{center}
\end{figure}
%
As a result, the viable parameter space for WIMPy leptogenesis is reduced. In Figure \ref{fig:leptonalphamass}, we show the masses $m_X$, $m_\psi$ giving successful WIMPy leptogenesis with $\alpha>1$. In particular, we find that the lepton asymmetry from WIMPy leptogenesis is too small in regions with large washout ($m_\psi\sim m_S$, where wash-out scattering is on-shell). While WIMPy leptogenesis is possible with  lepton-preserving annihilation channels, $m_\psi$ lies in a more restricted region when $\alpha>1$.

To summarize, we have presented a model of WIMPy leptogenesis where the WIMP miracle has been extended to the WIMPy baryogenesis miracle: the correct baryon asymmetry and WIMP relic density can be generated simultaneously with TeV-scale masses and $\mathcal O(1)$ couplings. We find that, depending on the ratio $m_\psi/m_X$ and the Yukawa couplings, larger and smaller asymmetries are also possible over a range of about seven orders of magnitude.  Generating the observed baryon asymmetry does require some correlation between $m_X$ and $m_\psi$, which may be explained if the masses have some common dynamical origin. For $m_X$ lighter than about 1 TeV, sphalerons decouple in the middle of asymmetry generation and the resulting baryon asymmetry is typically smaller than the observed $Y_{\Delta B}$.

\section{WIMP Annihilation to Quarks}\label{sec:alternatives}
\subsection{Model Overview}

If the final products of dark matter annihilation are quarks, WIMP annihilation can directly generate a baryon number asymmetry. The lower bound of $\sim$ TeV on $m_X$ in WIMPy  leptogenesis (the dashed line in Figure \ref{fig:mxmpsimain}) does not apply when WIMPs annihilate to quarks,  since the production of baryon number  no longer depends on efficient sphaleron interactions. Just as the leptogenesis model included new weakly charged vectorlike doublets, this model requires new vectorlike colored states to couple to quarks. Such states can be pair-produced at the LHC, leading to much stronger constraints and better detection prospects, which we discuss in Section \ref{sec:collider}.

The model content is similar to the leptogenesis model discussed in Section \ref{sec:model}: vectorlike gauge singlet dark matter $X$ and $\bar X$, singlet pseudoscalars $S_\alpha$, and vectorlike exotic quark color triplets $\psi_i$ and $\bar\psi_i$. The Lagrangian is
\ben\label{eq:lagrangianquark}
\mathcal L = \mathcal L_{\rm kin} + \mathcal L_{\rm mass} -\frac{i}{2}\left(\lambda_{X\alpha}X^2+\lambda_{X\alpha}'\bar X^2\right)S_\alpha + i \lambda_{B\,\alpha}\,S_\alpha\bar u\psi.
\een
A baryon asymmetry can be generated in $\bar u$ along with an equal negative baryon asymmetry in $\psi$. $\psi$ must decay, because it would otherwise overclose the universe and violate bounds on stable colored particles.

The negative baryon asymmetry in $\psi$ must not destroy the positive baryon asymmetry in $\bar u$ when it decays. This can happen in two ways:
\benum
\item $\psi$ decays into a sector decoupled from Standard Model quarks at low energies. Total baryon number is preserved, but the negative baryon number carried by $\psi$ is sequestered from quarks at late times and does not eliminate the Standard Model asymmetry.
\item $\psi$ decays into Standard Model quarks through baryon-number-violating couplings. The final baryon asymmetry is different from the asymmetry  created initially in $\bar u$ from WIMP annihilations because $\psi$ decays give an additional contribution to the baryon asymmetry.
\eenum
We now implement each of the above scenarios.
\begin{enumerate}
\item $\psi_i$ decays to light, baryon-number-carrying singlets $n_i$ plus Standard Model antiquarks. It can do so through a colored scalar $\phi$ with Standard Model gauge representation $(3,1,-1/3)$. The additional terms in the Lagrangian are
\ben\label{eq:lagrangianquark1}
\Delta\mathcal L = \lambda_i\,\bar\psi_i\,\bar d_i\,\phi^*+\lambda_i'\,\phi\,\bar d_i\,n_i+\mathrm{h.c.}
\een
A $Z_4$ symmetry prevents $\psi$ from decaying directly into Standard Model quarks through a $QH\bar\psi$ term and eliminating the baryon asymmetry. We show the charges in Table \ref{tab:Z4quark}. This Lagrangian has a $\rm U(1)^3$ flavor symmetry and satisfies all quark flavor constraints. In particular, $\bar u_i$, $\psi_i$, and $n_i$ have charges $-1$, $+1$, and $-3$, respectively, under the $\mathrm U(1)_i$ factor of the flavor symmetry. $\phi$ has charge $-2$ under all $\rm U(1)$ flavor symmetries. We assume that the only sources of flavor violation are the Standard Model Yukawa matrices.

\item In this scenario, $\psi_i$ also decays to two antiquarks plus a singlet $n$, but $n$ is  a Majorana fermion that does not carry any charge. Baryon number is now explicitly violated, and dark matter annihilations  generate $-1$ unit of baryon number for each $\psi+\bar u$ produced from dark matter annihilations (because $\psi\rightarrow \bar d\bar dn$). There is a new colored scalar $\tilde d_i$ in the $(3,1,-1/3)$ representation of the Standard Model gauge group that mediates $\psi$ decays. The additional terms in the Lagrangian are
\ben\label{eq:lagrangianquark2}
\Delta\mathcal L =\lambda\,\epsilon^{ijk}\,\bar\psi_i\,\bar d_j\,\tilde d^*_k+\lambda_i'\,\bar d_i\,\tilde d_i\,n+\mathrm{h.c.}
\een
A $Z_4$ symmetry, with charges given in Table \ref{tab:Z4quark}, prevents $\psi$ from decaying to Standard Model quarks through other interactions that would destroy the baryon asymmetry. This Lagrangian can be naturally realized in supersymmetric models, where $\tilde d$ is the down squark and $n$ is the neutralino, although this is not the only possible realization of this scenario.
(\ref{eq:lagrangianquark2}) has a $\rm U(3)$ flavor symmetry, which is the diagonal subgroup of the full $\mathrm U(3)_u\times\mathrm U(3)_d$ flavor group. The quark, $\psi_i$ and $\tilde d_i$ fields transform in the fundamental of $\rm U(3)$.  The Yukawa couplings between $S\bar u\psi$ in (\ref{eq:lagrangianquark}) have a flavor-independent piece and a flavor-dependent piece proportional to the up Yukawa matrix $Y_u$, consistent with minimal flavor violation.

\end{enumerate}

\begin{table}
\begin{center}

 \begin{tabular}{| c | c | c | c | c | c | c | c | c | c | c | c | c |}
 \hline
 & $X$ & $\bar X$ & $\psi$ & $\bar\psi$ & $S$ & $\bar u$ & $\bar d$ & $\phi/\tilde{d}$ &  $Q$ & $H$ & $n$ & leptons\\ \hline
 $Z_4$ & $+i$ & $-i$ & $+1$ & $+1$ & $-1$ & $-1$ & $-1$ & $-1$ & $-1$& $+1$ & $+1$ & $+1$ \\ \hline
 \end{tabular}
 \caption{$Z_4$ charges of  fields in  models with WIMP annihilation to quarks.}
  \label{tab:Z4quark}
\end{center}
\end{table}

In both scenarios, $\psi$ decays to a singlet $n$ plus quarks. Operators allowing $\psi$ to decay entirely to quarks (such as $\phi^*\bar d\,\bar u$ or $\tilde d^*\bar d\,\bar u$) are forbidden by the $Z_4$ symmetry. The $Z_4$ symmetry also ensures the stability of dark matter and of the proton. The proton is  stable provided $m_{\rm p} < 2m_X, m_S$ because baryons have charge $(-1)^3=-1$ and can never decay into the lighter meson and lepton fields, which are uncharged under the $Z_4$.

The $Z_4$ symmetry in principle allows neutral baryons to oscillate into one another. For scenario \#1, the generalized baryon number symmetry prohibits neutron-antineutron oscillation. In scenario \#2, the baryon-number-violating term is antisymmetric in flavor indices, and  the dominant contribution to neutron-antineutron mixing involves loops of $W$ bosons and off-diagonal CKM matrix element insertions $V_{bd}$ and $V_{sd}$. Since the bound on the neutron-antineutron oscillation operator $c/\Lambda^5(\bar u\bar d\bar d)^2$ is $\Lambda\gtrsim10-100$ TeV for $c=1$ \cite{Mohapatra:2009wp}, the loop- and CKM-suppression is sufficient to lower the oscillation rate well below current constraints for $m_S,\,m_\psi,\,m_{\tilde d}\sim$ TeV and $\mathcal O(1)$ couplings.

\begin{figure}[t]
\begin{center}
\includegraphics[width=7.5cm]{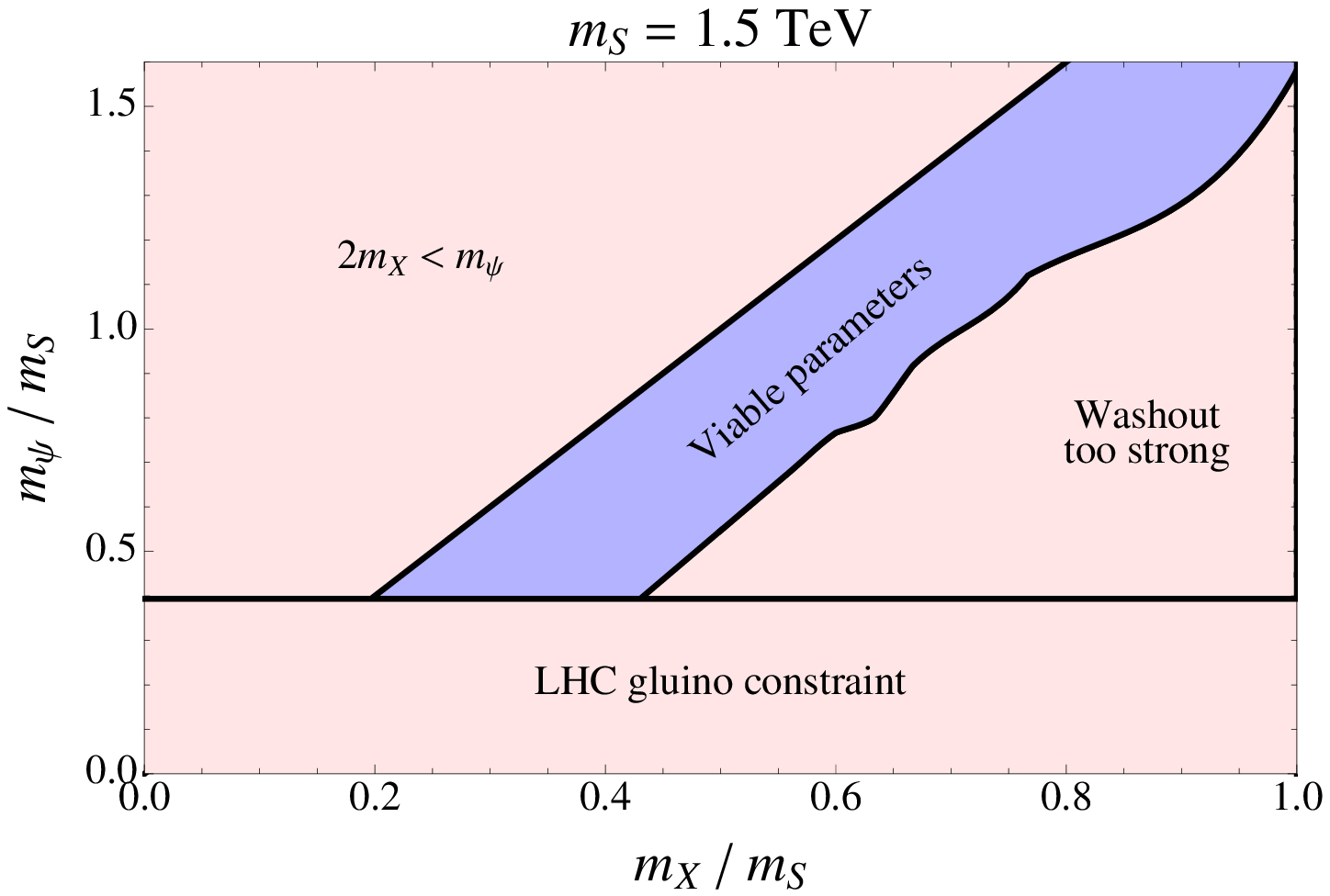}\hspace{1cm}
\includegraphics[width=7.5cm]{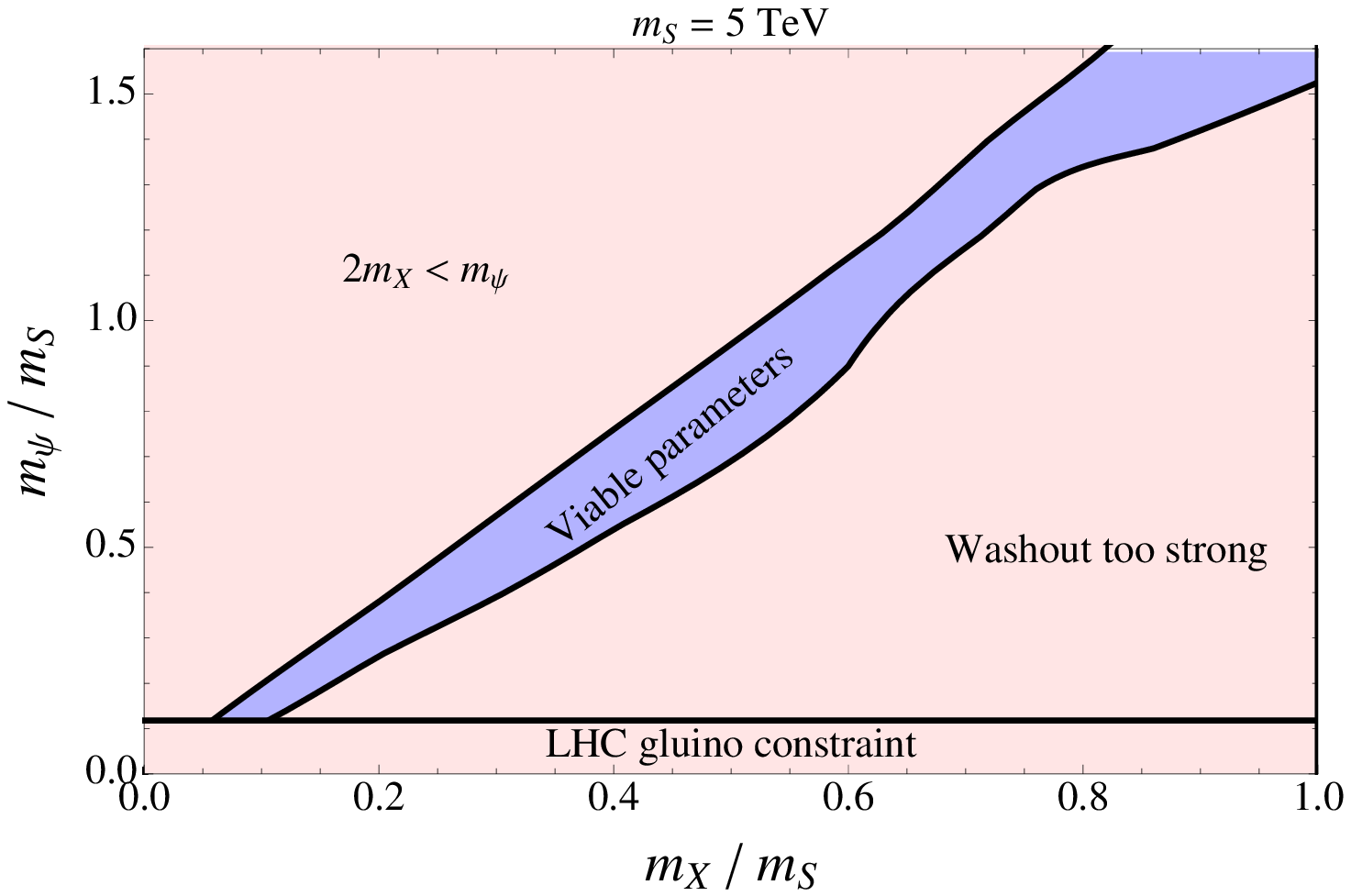}
\caption{Regions in the $m_X$-$m_\psi$ plane with the correct WIMP relic density and baryon asymmetry from WIMPy baryogenesis, with {\bf (left)} $m_S=1.5$ TeV  and {\bf (right)} $m_S=5$ TeV, and any choice of  perturbative couplings. The masses giving both observed abundances are shown in blue (middle stripe).  The descriptions of the regions on the plot are the same as those in Figure \ref{fig:mxmpsimain}. }
\label{fig:quarkheavymass}
\end{center}
\end{figure}

The Boltzmann equations for WIMP annihilation to quarks are changed only by group theory factors from the corresponding equations for leptons. Similarly, the chemical potentials relations are modified to reflect the new interactions (\ref{eq:lagrangianquark1}) or (\ref{eq:lagrangianquark2}).

As with WIMPy leptogenesis, we assume that $\psi$ is in equilibrium, and this places constraints on the couplings through which it decays. In scenario \#1, we considered the decay of $\psi$ according to interactions given in equation (\ref{eq:lagrangianquark1}). If $m_\psi>m_\phi$, the $\psi$ decay is two-body and the  constraint (\ref{eq:equil}) applies, giving $\lambda_i\gtrsim6\times10^{-8}$. If $m_\psi < m_\phi$, $\psi$ undergoes a three-body decay to $\bar d\,\bar d\, n$, and the constraint is
\ben
(\lambda_i\lambda_i')^2 \gtrsim \frac{80\sqrt{g_*}\,T_{\rm lep}^2\,m_\phi^2}{M_{\rm Pl}m_\psi^3}.
\een
For $T_{\rm lep}=50$ GeV and $m_\psi=1$ TeV, we find that the constraint on the geometric mean of the couplings is
\ben
\sqrt{\lambda_i\lambda_i'} \gtrsim 6\times10^{-5}\,\sqrt{\frac{m_\phi}{2\,\,\rm TeV}}.
\een
The constraints on scenario \#2 are comparable.

\subsection{Numerical Results}\label{sec:quarknum}
Because of its similarity to  leptogenesis, we use the quark flavor structure of scenario \#1 in our analysis, since it allows for a more direct comparison of numerical results in both cases. We consider two scenarios: $m_S=5$ TeV, to compare the results for quarks with that for leptons, and $m_S=1.5$ TeV, because dark matter can be much lighter than in WIMPy leptogenesis since there are no constraints from sphaleron decoupling. For simplicity, we consider sphalerons to be out of equilibrium for the $1.5$ TeV case and in equilibrium for the 5 TeV case to avoid considering sphaleron decoupling effects, although the calculation can be easily extended to include them. \\

\noindent\underline{Range of allowed masses}: We show the range of allowed masses in Figure \ref{fig:quarkheavymass}. Gluino searches at the LHC constrain this scenario (see Section \ref{sec:collider}), in contrast with the leptogenesis model, for which the entire parameter space is unconstrained by collider searches. This is particularly true for  $m_S=1.5$ TeV, where LHC searches will cover almost the entire parameter space for dark matter annihilation to quarks during the 14 TeV run. The WIMP mass is already constrained to be $m_X\gtrsim295$ GeV by the gluino bound discussed in Section \ref{sec:collider} along with the kinematic requirement that $2m_X>m_\psi$. This is true independent of all other parameters.

In both cases, the results are qualitatively similar to leptogenesis. With WIMP annihilation to quarks, the annihilation and washout cross sections are enhanced because the final states are charged under $\mathrm{SU}(3)_{\rm C}$. As a result, the baryon asymmetry is suppressed by the increased washout rate. This is partially offset by the fact that the self-energy contribution to $\epsilon$ is enhanced by a group theory factor as well. With $m_X=0.9$ TeV, the parameter space is actually larger than for $m_X=4.25$ TeV or WIMPy leptogenesis. Since in this case, sphalerons no longer inter-convert baryon and lepton number, the asymmetry created in quarks is distributed among fewer fields, enhancing the asymmetry.\\

\begin{figure}[t]
\begin{center}
\includegraphics[width=7.5cm]{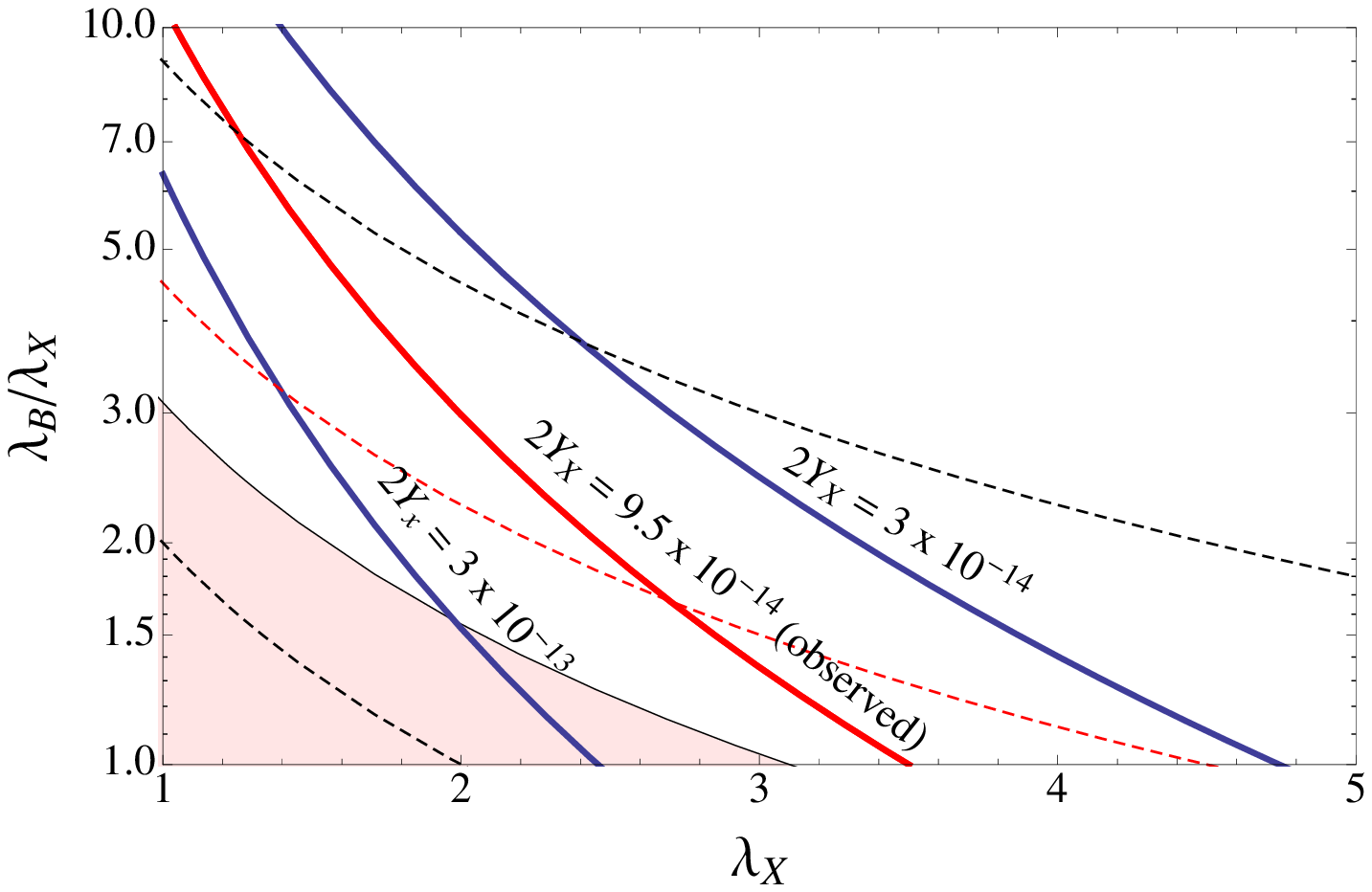}\hspace{1cm}
\includegraphics[width=7.5cm]{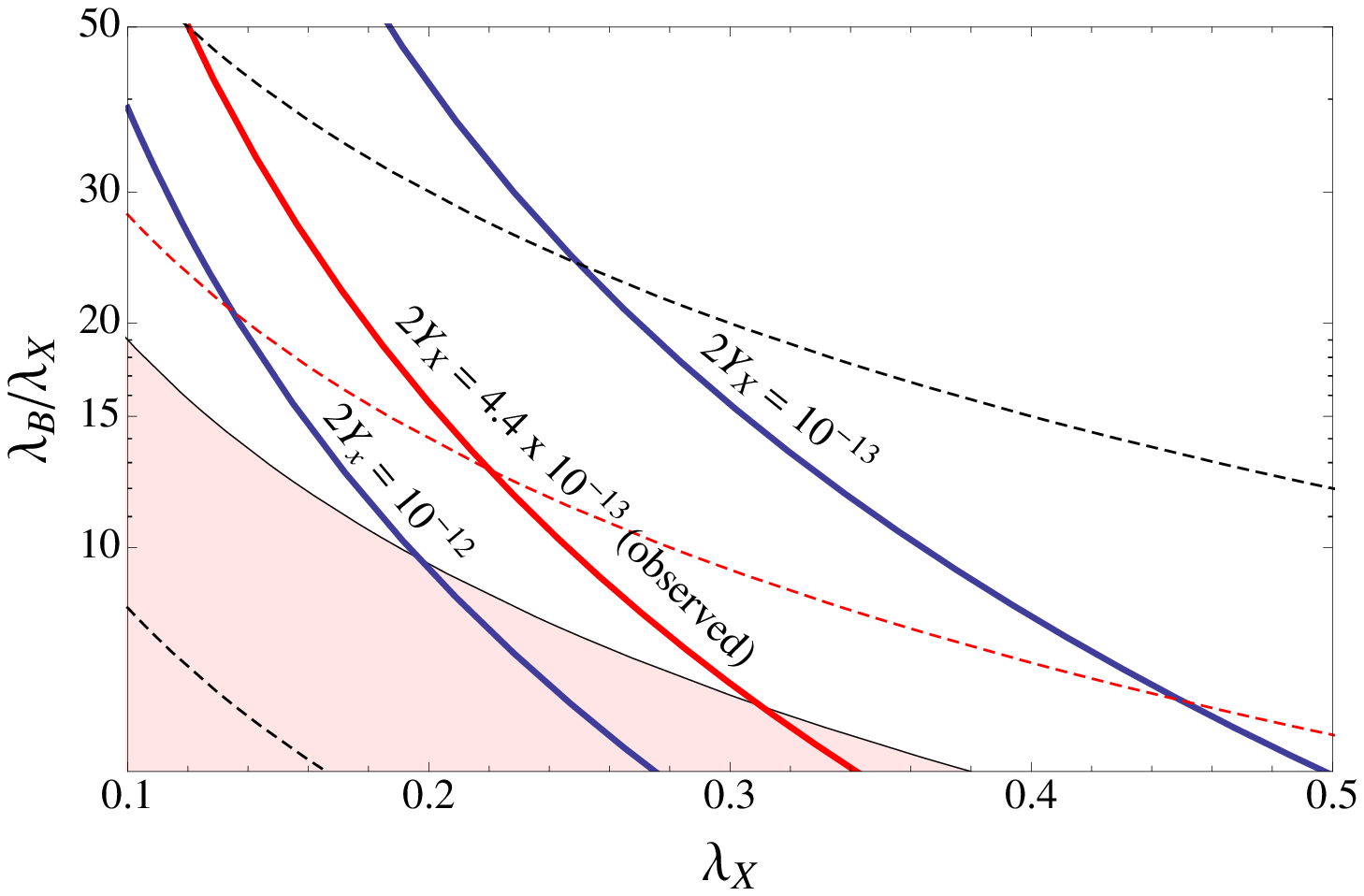}
\caption{Dark matter relic density (solid lines) and baryon asymmetry (dotted lines) as functions of couplings $\lambda_X$ and $\lambda_B/\lambda_X$. We consider two sets of massses: {\bf (left)}, $m_X=4.25$ TeV, $m_\psi=7.25$ TeV, $m_S=5$ TeV, and $|\epsilon|=0.075$. The asymmetry contours are, from top to bottom: $Y_{\Delta B} = 4\times10^{-11}$, $8.5\times10^{-11}$ (observed asymmetry), and $1.5\times10^{-10}$.  {\bf (right)} $m_X=0.9$ TeV, $m_\psi=1.2$ TeV, $m_S=1.5$ TeV, and $|\epsilon|=0.075$; the asymmetry contours are, from top to bottom: $Y_{\Delta B} = 5\times10^{-11}$, $8.5\times10^{-11}$ (observed asymmetry), and $3\times10^{-10}$. In the shaded regions, $\epsilon$ is not consistent with our assumptions according to the bound (\ref{eq:epsilonapprox}).  }
\label{fig:quarkcouplings}
\end{center}
\end{figure}

\noindent\underline{Range of allowed couplings}: We do the same analysis as in Section \ref{sec:leptonnum}. To compare the results of direct baryon asymmetry production with those for WIMPy leptogenesis, we choose a set of parameters used in the leptogenesis analysis: $m_X=4.25$ TeV, $m_\psi=7.25$ TeV, $m_S=5$ TeV, and $|\epsilon|=0.075$. We also consider a corresponding point with light $S$  and broken phase chemical potential relations: $m_X=0.9$ TeV, $m_\psi=1.2$ TeV, $m_S=1.5$ TeV, and $|\epsilon|=0.075$.  We show the results in Figure \ref{fig:quarkcouplings}.   The parameter points giving the correct dark matter density and baryon asymmetry are $\lambda_X=2.7$, $\lambda_B=4.5$ for $m_X=4.25$ TeV, and $\lambda_X=0.22$, $\lambda_B=2.8$ for $m_X=0.9$ TeV.

\section{Experimental Constraints and Detection Prospects}\label{sec:signals}
In this section, we survey the possible experimental constraints and signals for models of WIMPy baryogenesis, considering both annihilation to leptons and annihilation to quarks.  For WIMP annihilation to leptons, the experimental bounds on $m_X$ and $m_\psi$ are too weak to constrain leptogenesis because $m_X,m_\psi\gtrsim$ TeV are required to generate a sufficiently large baryon asymmetry. The prospects are better for WIMP annihilation to quarks, which predicts signals at indirect and direct detection experiments, as well as at the LHC. We first give a preview of our results in Table \ref{tab:result}.

\begin{table}[h]
\begin{center}

 \begin{tabular}{| c | c | c |}
 \hline
 & Annihilation to leptons & Annihilation to quarks\\ \hline
Direct detection & $--$ &  $m_X\lesssim5$ TeV for some parameters\newfootnotemark{1}\\
& &($\sigma_{X-\rm nucleon}\sim 10^{-46}-10^{-44}\,\,\mathrm{cm}^2$)\\\hline
Indirect detection & $m_X\lesssim200$ GeV & $m_X\lesssim1$ TeV\\
(antideuterons) & & \\\hline
Colliders &  $m_\psi\lesssim$ few hundred GeV, & $m_\psi\lesssim1.44$ TeV with \\
& possible improvements & 100 $\,\,\mathrm{fb}^{-1}$ LHC (14 TeV)\\
&  with targeted searches & \\\hline
EDM &  $--$ & $--$ \\ \hline
 \end{tabular}
 \footnotetext[1]{Precise reach depends on $m_\psi$, $\lambda_B$, $\lambda_X$, and $\epsilon$.}
 \caption{Search reach for minimal models of WIMPy baryogenesis/leptogenesis in current and near-future experiments. `$--$' indicates no signal in that search channel.}
 \label{tab:result}
\end{center}
\end{table}

\newfootnotetext{0}{Precise reach depends on $m_\psi$, $m_S$, $\lambda_X$, $\lambda_B$, and $\epsilon$.}

We provide details for each class of experiment in the following sections.

\subsection{Dark Matter Detection}

    \subsubsection{Direct Detection}\label{sec:didt}
Dark matter direct detection experiments are typically important probes of weak-scale dark matter models. As we show in this section, however, only WIMPy baryogenesis with dark matter annihilation to quarks is expected to give a  signal in conventional direct detection experiments. This is because the dark matter scattering cross section is suppressed by loops of heavy fields, and it is only when dark matter couples directly to quarks that the WIMP-nucleon cross section is large enough to give a signal in upcoming experiments. We assume in this section that dark matter annihilates predominantly to first generation quarks/leptons. The baryon-number-violating interactions in WIMPy baryogenesis can also induce proton decay due to WIMP scattering, but we find that our models are consistent with all current and projected experimental bounds.

\begin{figure}[t]
\begin{center}
\includegraphics[width=6cm]{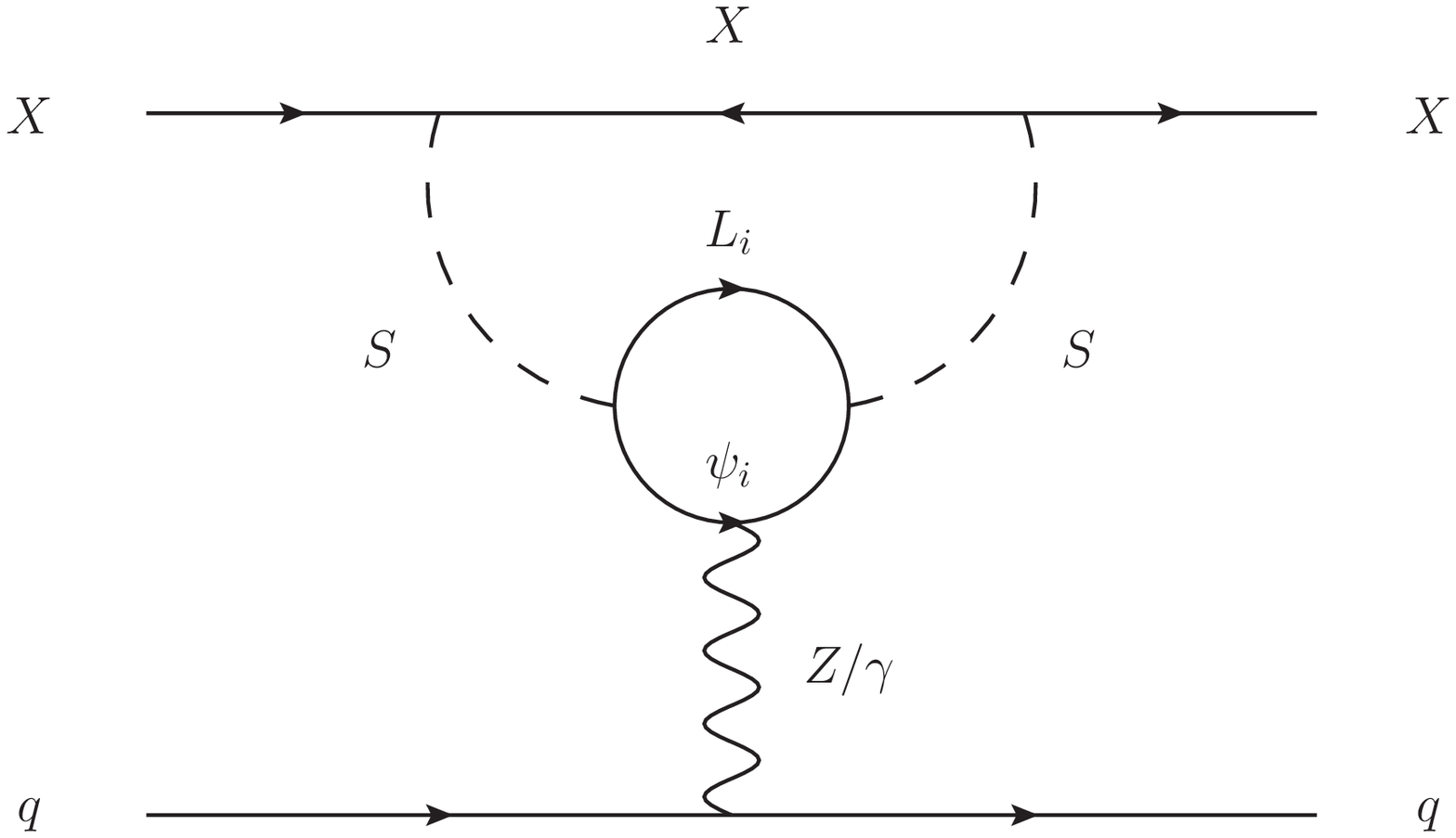}\hspace{2cm}
\includegraphics[width=6cm]{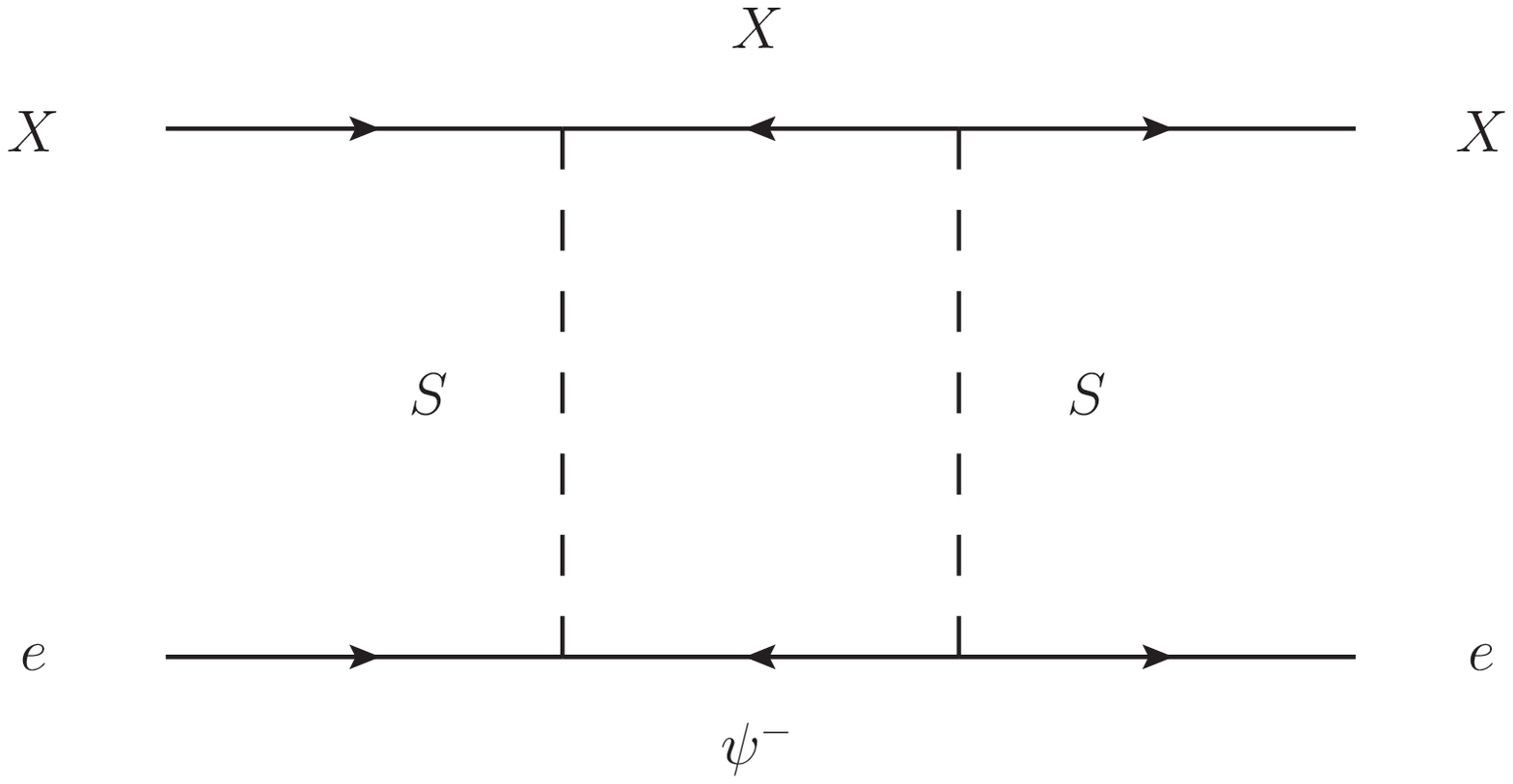}
\caption{Feynman diagrams for dark matter scattering off  {\bf (left)} nucleons and {\bf (right)} electrons in direct detection experiments for WIMPy leptogenesis.}
\label{fig:directlepton}
\end{center}
\end{figure}
\begin{figure}[t]
\begin{center}
\includegraphics[width=6cm]{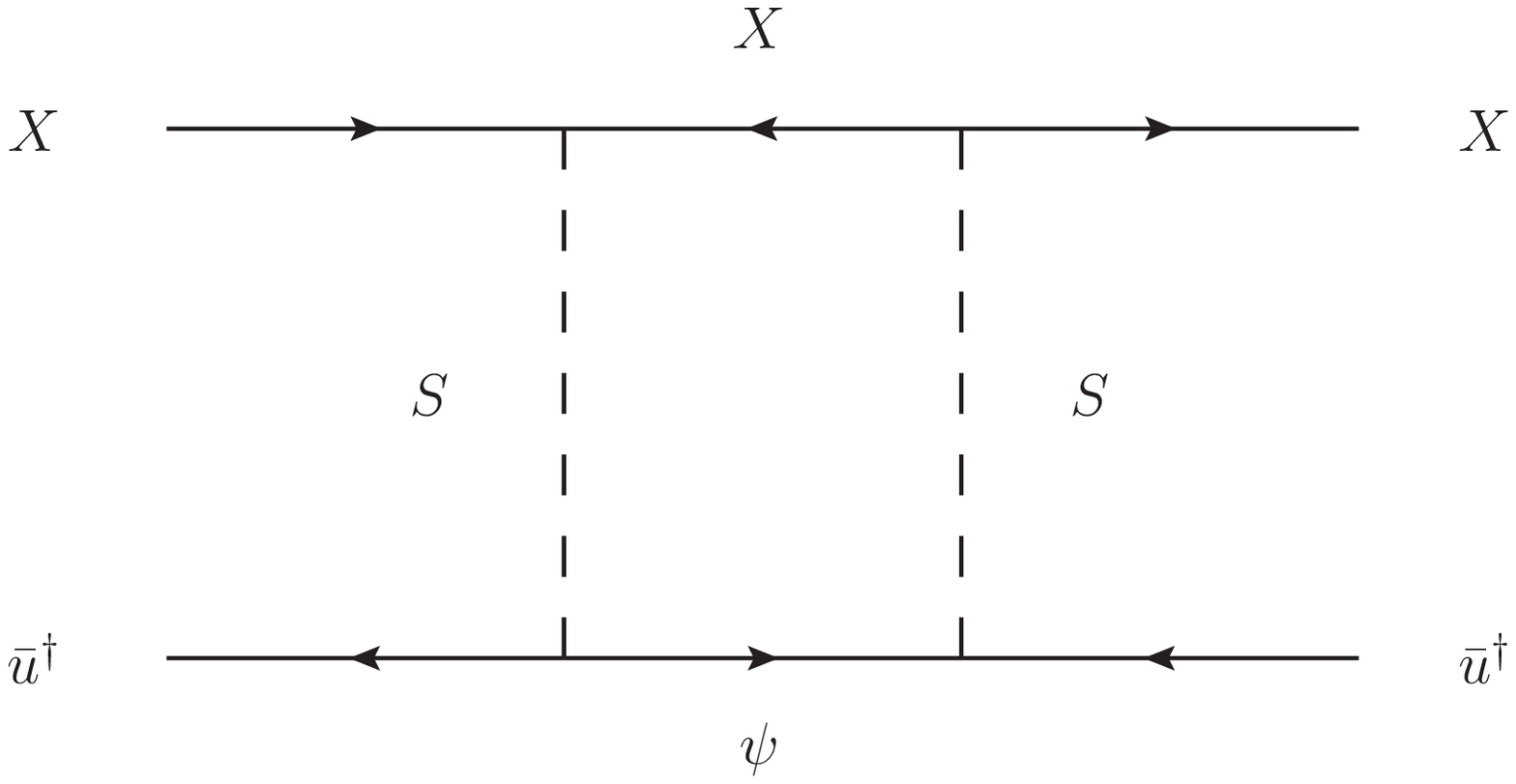}\hspace{2cm}
\includegraphics[width=6cm]{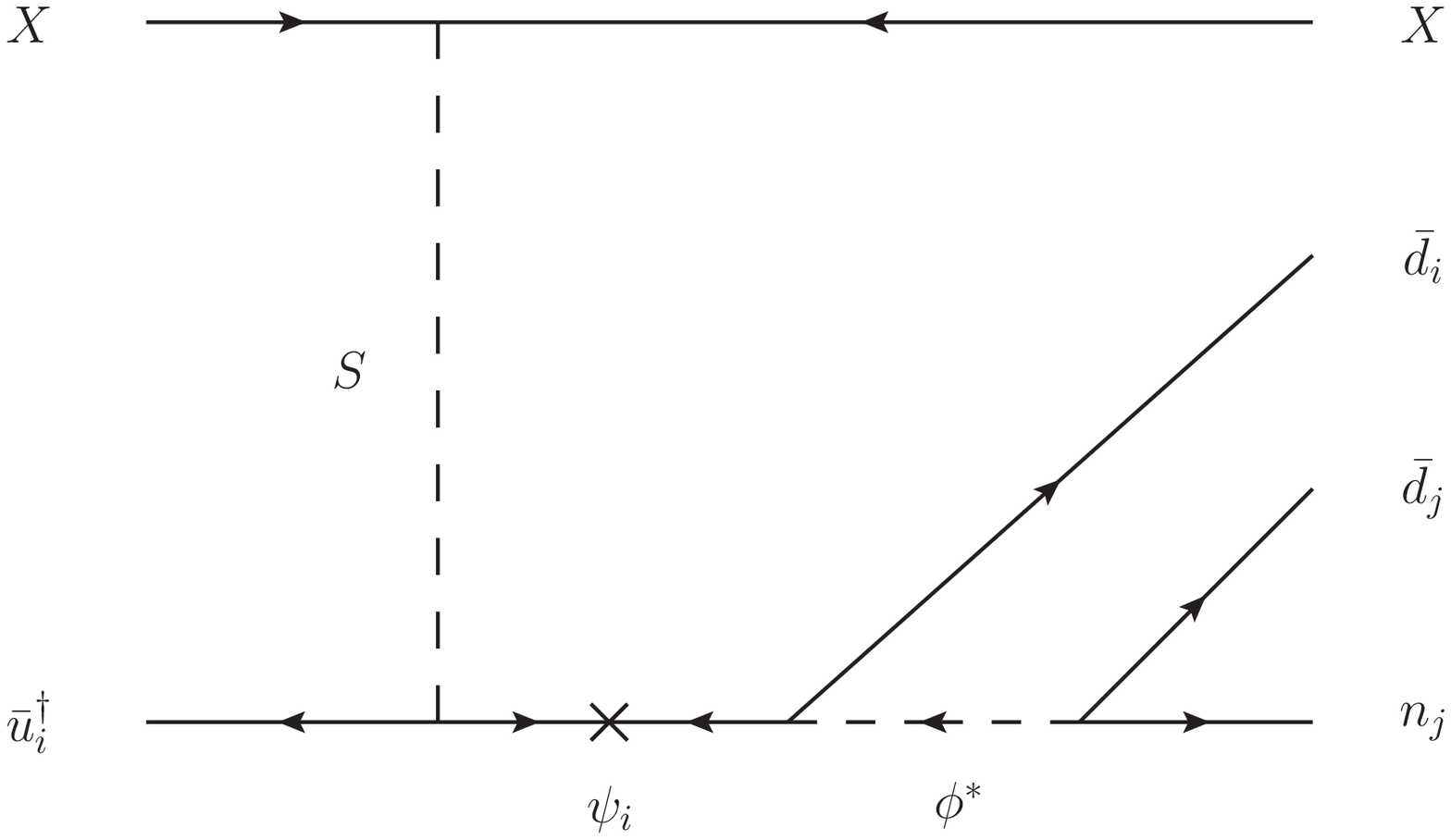}
\caption{Feynman diagrams for dark matter scattering off  nucleons in direct detection experiments when WIMPs annihilate to quarks: {\bf (left)} standard signal and {\bf (right)} inelastic induced nucleon decay.}
\label{fig:directquark}
\end{center}
\end{figure}
We first present the Feynman diagrams for the leading processes relevant to direct detection. With WIMP annihilation to leptons, we show the diagrams for scattering in direct detection experiments in Figure \ref{fig:directlepton}. We show the corresponding diagrams with WIMP  annihilation to quarks in Figure \ref{fig:directquark}.
\\

\noindent\underline{WIMP annihilation to leptons}: $X$ can elastically scatter off electrons at one loop and nucleons at two loops. However, direct detection experiments are on the verge of testing dark matter models with nucleon scattering at one loop \cite{Raby:1987nb,Kaplan:2009ag,Cohen:2009fz,Cui:2011qe} and electron scattering at tree level \cite{Kopp:2009et}. Therefore, the elastic scattering signals from our models are too small to be detected at near-future experiments.

As we discussed in Section \ref{sec:leptonnum}, there can be lepton-number-preserving dark matter annihilation channels in addition to those responsible for baryogenesis, although the parameter space is  more restricted in this case. The lepton-number-preserving WIMP interactions can lead to conventional WIMP signals in direct detection experiments, but do not probe the WIMP's lepton-number-violating couplings, which are the crucial ingredients of WIMPy leptogenesis.\\

\noindent\underline{WIMP annihilation to quarks}: The dominant contribution to the direct detection cross section is the one-loop scattering of dark matter off the right-handed up quark. We estimate the dark matter-nucleon cross section:
\ben\label{eq:ddxsec}
\sigma_{X-N} \sim \frac{1}{16\pi}\left(\frac{\lambda_B^2\lambda_X^2}{16\pi^2}\right)^2\,\frac{\mu^2}{m_X^4},
\een
where $\mu$ is the reduced mass of the dark matter-nucleon system\footnote{Because the masses of all fields running in the loop are similar in mass, there is no significant mass suppression to the cross section from evaluating the loop integral.}. In our models, $m_X\gg m_N$, where $m_N$ is the nucleon mass, so $\mu\approx m_N$.

We determine the direct detection cross section for the benchmark points in Section \ref{sec:quarknum}. For the point $m_X=4.25$ TeV, $m_\psi=7.25$ TeV, $m_S=5$ TeV, $\lambda_X=2.7$ and $\lambda_B=4.5$, we find $\sigma_{X-N}\approx 1\times10^{-44}\,\,\mathrm{cm}^2$. For the point $m_X=0.9$ TeV, $m_\psi=1.2$ TeV, $m_S=1.5$ TeV, $\lambda_X=0.22$ and $\lambda_B=2.8$, we find $\sigma_{X-N}\approx4\times10^{-46}\,\,\mathrm{cm}^2$.  The current limits from the XENON100, CDMS experiments \cite{xenon,{cdms}} on dark matter direct detection
have a minimum bound of $\sim10^{-44}\,\,\mathrm{cm}^2$ for WIMPs with masses of $\sim 50\,\,\rm GeV$. The upper limit on the cross section for a TeV WIMP is $\sim10^{-43}\,\,\mathrm{cm}^2$. We therefore see that the cross sections for our benchmark points are below current bounds but are large enough that they can give a signal in upcoming direct detection experiments such as XENON1T \cite{xenon1t}. We leave a detailed study of the direct detection reach for future work.

There also exists an inelastic scattering process that converts an up-type quark to two down-type antiquarks, as we show in the right-hand graph in Figure \ref{fig:directquark}. Such an inelastic process can lead to  nucleon decays induced by WIMP scattering. The dominant process is $X\,p\rightarrow X\,n\,\pi^+$, along with the corresponding processes with strange quark production ($b$ quark production is kinematically suppressed). To avoid conflict with proton decay experiments, the induced proton decay rate should satisfy bounds outlined in \cite{Davoudiasl:2011fj}. Comparing our model to the Hylogenesis model in \cite{Davoudiasl:2011fj}, we find that the operator giving rise to induced proton decay in our model is dimension-9 ($X^2\bar u\bar d\bar dn/\Lambda^5$), whereas the corresponding Hylogenesis operator is dimension-7. At the hadronic level, the Hylogenesis process is $2\rightarrow2$, in contrast with our $2\rightarrow3$ process,  which gives our model a relative phase space suppression $\sim1/(2\pi)^3$. Furthermore, the Hylogenesis model has a dark matter mass $\sim\mathcal O(\rm GeV)$, while the dark matter mass in WIMPy baryogenesis is $m_X\sim \mathcal O(\rm TeV)$. As a result, the dark matter number density is smaller by a factor of (GeV/TeV) in WIMPy baryogenesis, and the incident flux of dark matter particles is suppressed. Taking into account all factors, the induced proton decay rate for WIMPy baryogenesis has a $\sim(\frac{1}{2\pi})^3\left(\frac{\rm GeV}{\rm TeV}\right)^5\sim10^{-17}$ suppression compared to that of Hylogenesis. Since induced proton decay is on the verge of current bounds for Hylogenesis models with a heavy scale $\Lambda\sim$ TeV, the proton lifetime in our model is safely above the current bound, and not within the reach of near-future proton decay experiments.\\

To summarize, models with WIMP annihilation to leptons typically predict the absence of a signal in conventional dark matter direct detection experiments, while models with WIMP annihilation to quarks have WIMP-nucleon cross sections below the current bounds but accessible in upcoming experiments. Baryon/lepton-number-preserving WIMP interactions can also give a signal in direct detection experiments, but such models have a smaller viable parameter space. WIMP scattering can induce nucleon decay in WIMPy baryogenesis models, but the proton decay rate is far lower than current experimental constraints.

\subsubsection{Indirect Detection}

Models of WIMPy baryogenesis have indirect detection prospects similar to those in conventional WIMP scenarios because the dark matter relic abundance is symmetric and is established by thermal freeze-out. This is in contrast with many asymmetric dark matter models, which typically have suppressed indirect detection signals due to the fact that dark matter is largely asymmetric  today. The only asymmetric dark matter models with indirect detection signals are those in which the symmetric component is regenerated after WIMP freeze-out \cite{Buckley:2011ye}. In the following summary, we assume that WIMPs annihilate predominantly through the interactions that generate the baryon asymmetry.

We find that indirect detection is most promising with WIMPy baryogenesis with dark matter annihilating to quarks. In this scenario, the final states are color-connected quarks and sterile fields $n_i$, and the quarks hadronize in the dark matter rest frame. This populates the low-energy anti-deuteron spectrum, leading to a clean, low-background signal at GAPS and AMS-02. The mass reach in this scenario is $m_X\lesssim1$ TeV. Annihilation of dark matter in WIMPy leptogenesis also leads to $q\bar q$ production via Higgs decay, but the quarks hadronize in the Higgs rest frame. Fewer low-energy antideuterons are produced, and the mass reach is only $m_X\lesssim200$ GeV, which is too low for viable models of WIMPy leptogenesis. We give more details below.\\

\noindent\underline{WIMP annihilation to leptons}: The indirect detection signals are energetic neutrinos, positrons, and secondary photons from the leptons produced in WIMP annihilations, along with antiprotons and antideuterons ($\bar D$) from $\psi^0\rightarrow h+n\rightarrow b\bar{b}+n$. Unfortunately, the dark matter mass of $\mathcal O(\rm TeV)$ in  leptogenesis gives a flux lower than the sensitivities of most upcoming indirect direction experiments. With the standard cross section for thermal WIMP annihilation ($\langle\sigma_{\rm ann}\rangle\approx3\times10^{-26}\,\,\rm cm^3/s$), the reach of most  current experiments like Fermi-LAT \cite{collaboration:2011wa} is in the mass range $\lesssim\mathcal O(100\rm\,\, GeV)$. One exception is in the scenario with a very steep dark matter profile in the galactic center, which occurs in halo models favored by hydrodynamical simulations. In this case, HESS measurements of gamma rays from the galactic center are within a factor of two of constraining a 3 TeV WIMP with standard annihilation cross section to leptonic final states \cite{Abazajian:2011ak}. Based on the HESS analysis, it is likely that with more data Fermi-LAT could rule out WIMPy leptogenesis models with masses $\lesssim$ few TeV if the similar assumptions on dark matter distribution are applied. Such constraints suffer from large uncertainties in the dark matter profile, however, and we caution that such strong limits on WIMP masses may not be possible.

 According to the general analysis performed in \cite{Cui:2010ud}, the mass reach of  low energy antideuteron detection experiments at AMS-02 and GAPS could be up to $\sim 1\rm TeV$ if  hadronization happens mostly in the rest frame of dark matter annihilation, as occurs in the $gg$ channel in  \cite{Cui:2010ud}. However, hadronic decay products in the leptogenesis scenario are secondary or tertiary, and hadronization typically happens in the boosted frame, similar to the $WW$ channel in the same reference. The resulting mass reach could  be only $\sim200\,\,\rm GeV$, which is too low for WIMPy leptogenesis because sphalerons are decoupled during the era of asymmetry generation for dark matter masses in this range.\\

\noindent\underline{WIMP annihilation to quarks}: The possible signals are $\bar{p}$, $\bar{D}$, and $\gamma$.
    In contrast  with  leptogenesis, the baryon asymmetry can be generated after the electroweak phase transition and the dark matter mass can be as low as $\sim 290\,\,\rm GeV$ according to the bound in Section \ref{sec:collider}. This is  promising for detection at upcoming experiments, particularly low energy anti-deuteron searches.
      Because the primary products of WIMP annihilation now involve color-connected $u$ and $\psi$, a large proportion of  hadronization proceeds in the rest frame, resulting in a larger rate of $\bar{D}$ production \cite{Cui:2010ud}. This extends the mass reach at GAPS and AMS-02 \cite{Ahlen:1994ct} to $\sim1\,\,\rm TeV$ and covers a large part of the WIMPy baryogenesis parameter space. Higher WIMP mass regions ($\sim$ TeV) may be constrained by Fermi-LAT gamma ray observations of the galactic center, but as discussed above with WIMPy leptogenesis, these constraints are highly dependent on the dark matter profile \cite{Abazajian:2011ak}. \\

    In general, models of WIMPy baryogenesis and leptogenesis satisfy all current constraints from indirect detection experiments, and future searches for antideuterons are promising discovery channels for models with WIMP annihilation to quarks.

\subsection{Collider Detection}\label{sec:collider}
We consider the LHC constraints and detection prospects for  new charged particles predicted in WIMPy baryogenesis.  We find that the LHC can strongly constrain the scenario with WIMP annihilation to quarks but may not constrain WIMP annihilation to leptons. Searches for supersymmetry (SUSY) with missing energy are  relevant to our models, since WIMPy baryogenesis predicts new charged fields decaying to Standard Model particles and   neutral fermions. We focus on existing LHC searches  for SUSY and leave for later work the optimization of collider searches for the particular charged fields found in our models.

The strongest LHC constraints are bounds on new colored fields, such as gluinos and squarks. Current searches at the LHC therefore constrain the scenario with WIMP annihilation to quarks, which has new colored fields $\psi$. The luminosity at the LHC is not yet large enough to bound  electroweak production of new particles, due to the smaller production cross section, and softer jets and missing energy. As a result, current collider constraints on $m_\psi$ in the scenario with dark matter annihilation to leptons are well below the range needed for viable WIMPy leptogenesis. Higher luminosity and new targeted searches can improve the LHC reach for $m_\psi$ depending on its decay modes.

We now consider each scenario in more detail.\\

\underline{WIMP annihilation to leptons}: A characteristic feature of the WIMPy leptogenesis model in section \ref{sec:model} is the presence of an exotic vectorlike $\mathrm{SU}(2)_{\rm L}$ doublet $\psi$. The neutral and charged components of $\psi$ can be pair-produced via  electroweak gauge bosons. According to our arguments in Section \ref{sec:fields},  $\psi$ decays promptly.

\begin{figure}[t]
\begin{center}
\includegraphics[width=4cm]{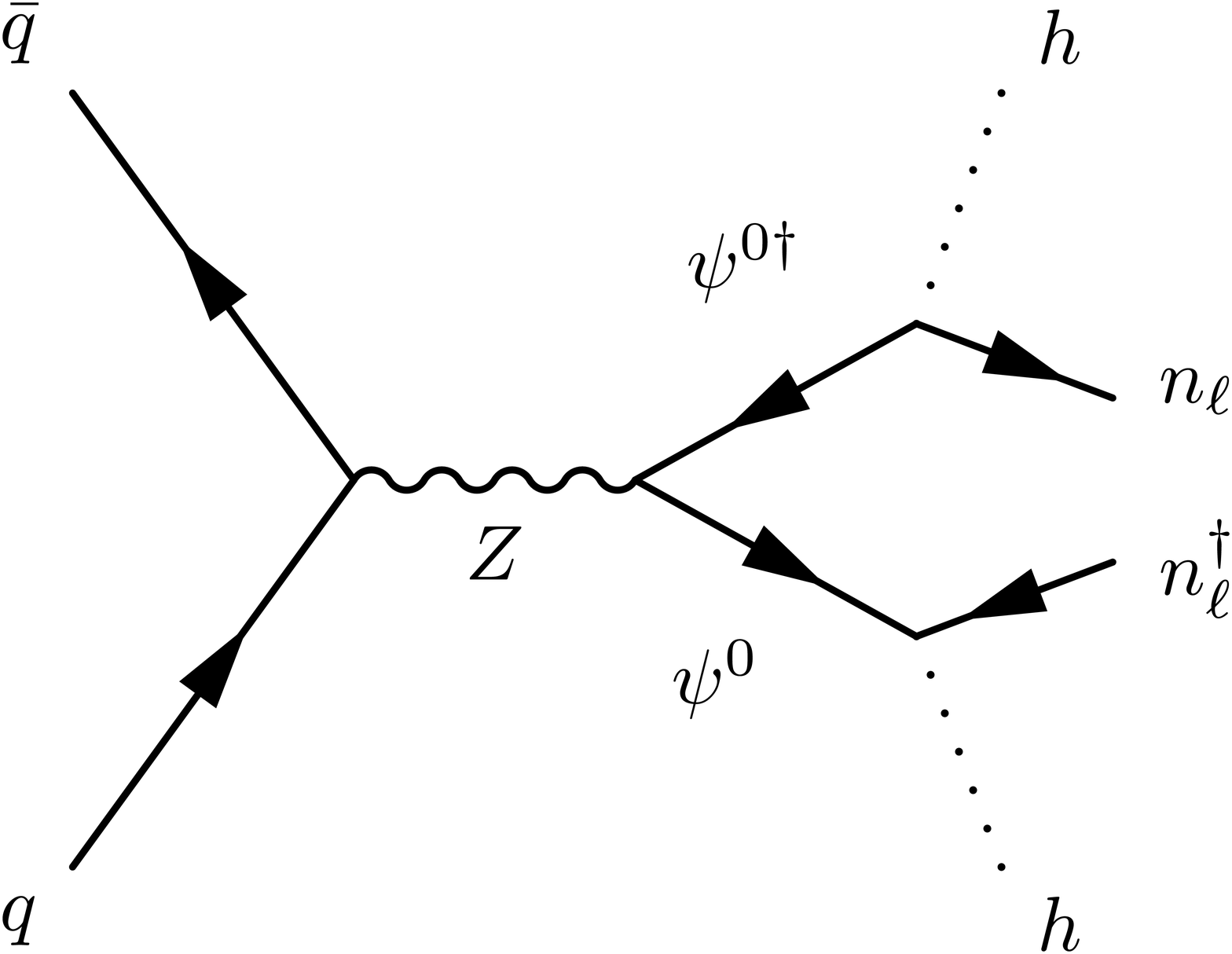}\hspace{2cm}
\includegraphics[width=4cm]{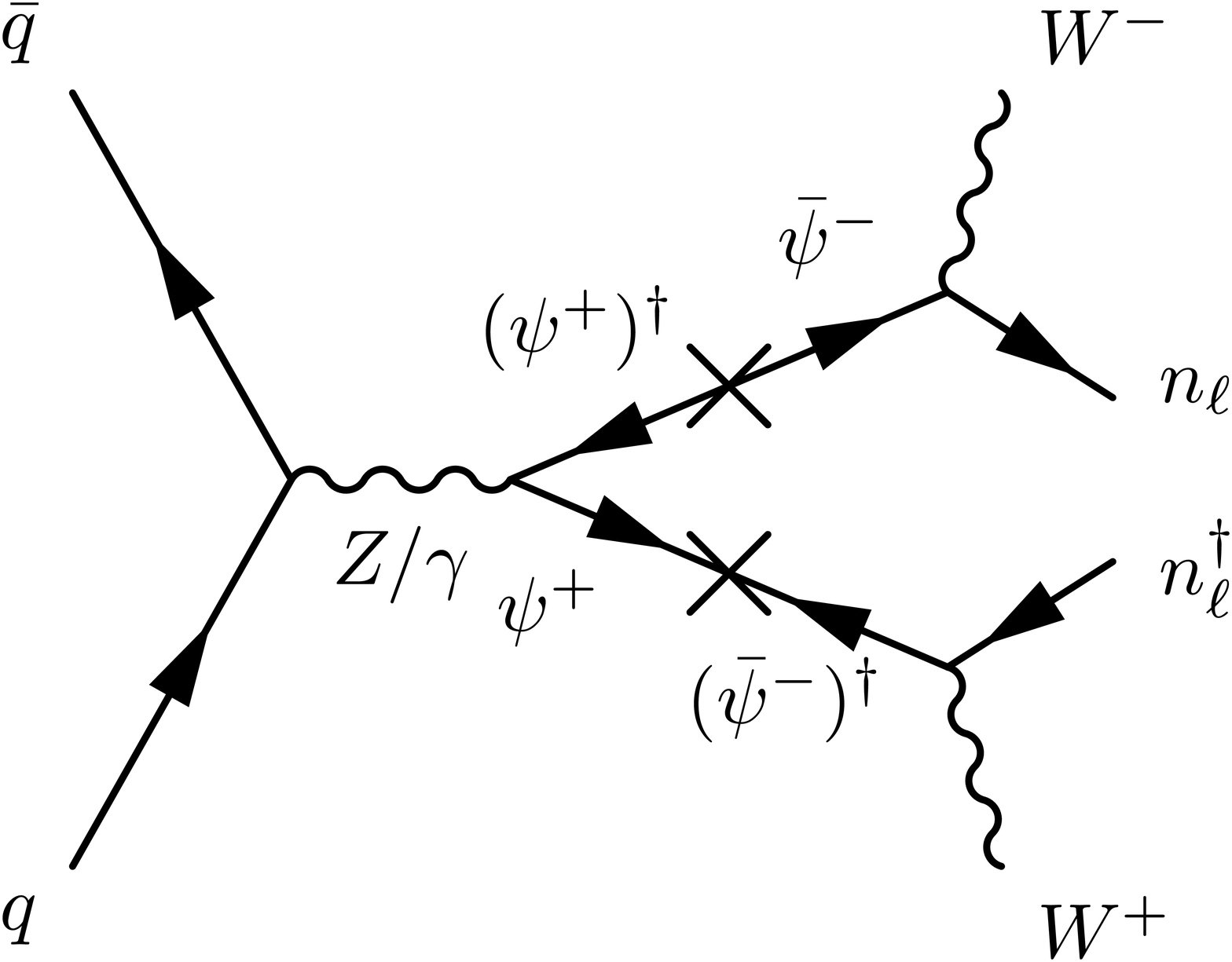}\hspace{2cm}
\includegraphics[width=4cm]{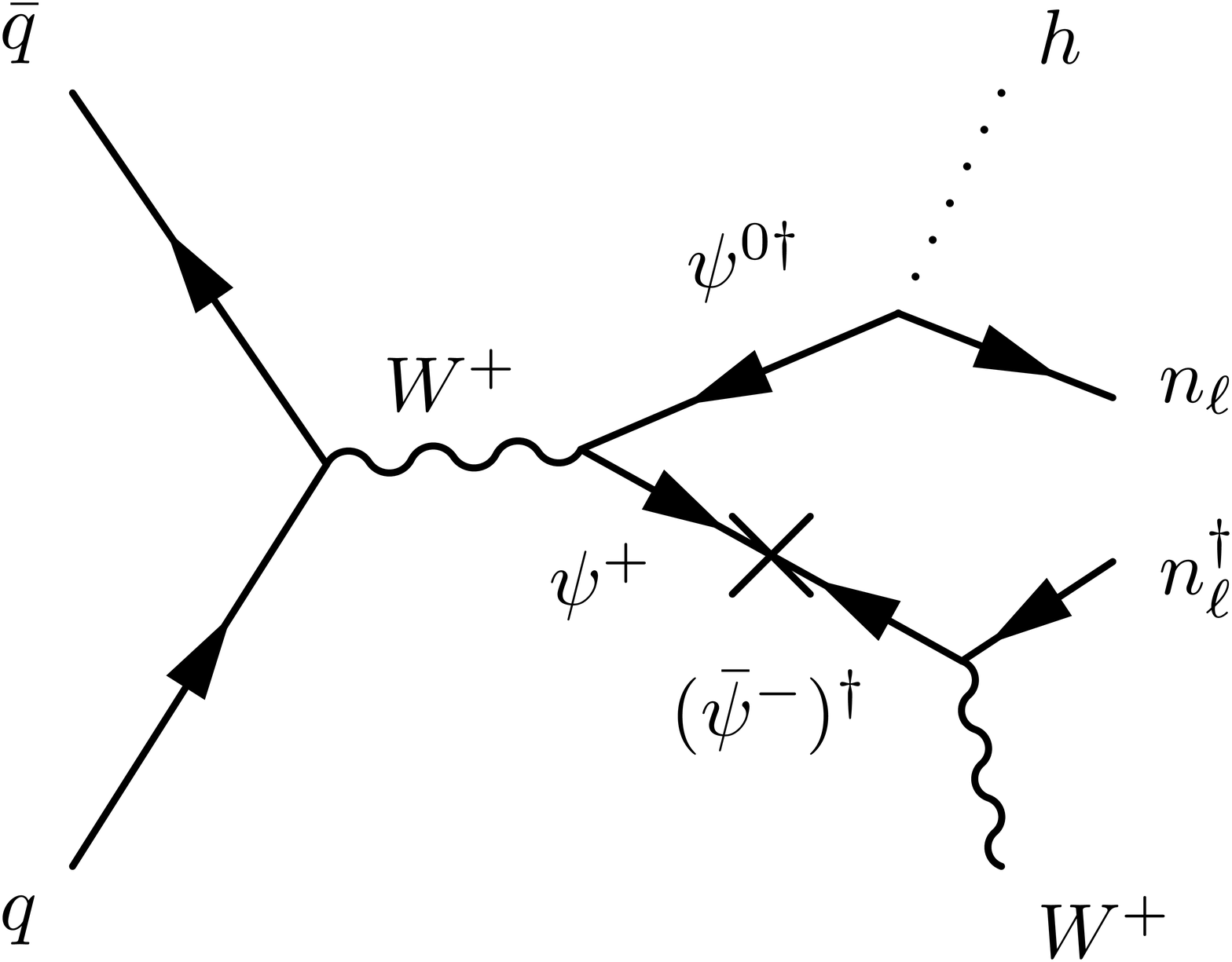}
\caption{LHC electroweak pair production of $\psi$ and its subsequent decay in the model with WIMP annihilation to leptons. $\psi^0$ decays to a Higgs boson and the light neutral fermion $n_\ell$, while $\psi^\pm$ decays to $W^\pm$ and $n_\ell$.}
\label{fig:leptonLHC}
\end{center}
\end{figure}

The dominant decay of $\psi^0$ is to Higgs + $n_\ell$ through the interactions in (\ref{eq:lagrangian}), where $n_\ell$ is the light neutral mass eigenstate after Higgs-induced mixing between $\psi$ and $n$. The resulting collider signature for pair production  is $\psi^0\psi^0\rightarrow4b(4j)+ \cancel E_{\rm T}$. The charged component, $\psi^\pm$, decays to $W^\pm+n_\ell$, with a collider signature for pair production of $\psi^+\psi^-\rightarrow W^+ W^-+\cancel E_{\rm T}$. The relevant diagrams are shown in Figure \ref{fig:leptonLHC}.

Searches at LEP constrain the masses of the charged and neutral components of $\psi$ with bounds on pair production of charginos ($\tilde\chi^\pm\rightarrow W^\pm\tilde\chi^0$), $m_{\chi\pm}\gtrsim100$ GeV \cite{Barate:1999fs}. $\psi^\pm$ decays look identical to chargino decays, so $m_{\psi\pm}\gtrsim100$ GeV as well. Hadronic chargino decays, which have a $4j+\cancel E_{\rm T}$ final state,  constrain the $\psi^0$ mass. The LEP  bound  groups hadronic chargino decays  with  other decay modes, so the bound is not directly applicable to $\psi^0$. A more careful analysis (that we leave for future work) is needed to determine the precise bound on $\psi^0$, but we expect it to be on the order of 100 GeV as well. The bounds on both $\psi^\pm$ and $\psi^0$ are  well below the typical $m_\psi$ required for WIMPy leptogenesis.

With the current luminosity of $5\,\,\mathrm{fb}^{-1}$ at $\sqrt s = 7$ TeV, the LHC  bounds the masses of weakly charged particles appearing in cascade decays of colored particles, but does not constrain particles such as $\psi$ that are only  produced directly from electroweak gauge bosons. Therefore, the LHC does not bound $m_\psi$ at present, and the LEP constraint remains the most important. Searches for direct chargino and slepton production with future LHC data will improve the bounds on $m_\psi$ to masses on the order of a few hundred GeV, but this is still smaller than $m_\psi$ needed in WIMPy leptogenesis.

New LHC searches at 14 TeV could possibly yield stronger constraints. For example, If the Higgs mass is known, we could require a reconstruction of the Higgs mass among final state jet pairs,  greatly reducing backgrounds. If $m_\psi\gg m_h$, the final state Higgses are boosted and can be studied with jet substructure techniques, as suggested in \cite{Kribs:2009yh}.

In summary, the collider constraints on $\psi$ are currently too weak to place any bounds on WIMPy leptogenesis models. Future LHC running will improve the bounds on $m_\psi$, and we have outlined some of the possible signals here. A more detailed collider analysis is deferred to later work.
\\

\begin{figure}[t]
\begin{center}
\includegraphics[width=5cm]{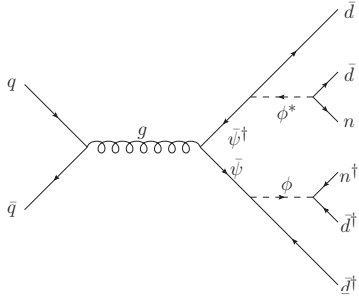}
\caption{LHC pair production of $\bar\psi$ and its subsequent decay in the model with WIMP annihilation to quarks. }
\label{fig:quarkLHC}
\end{center}
\end{figure}

\noindent\underline{WIMP annihilation to quarks}: As with WIMP annihilation to leptons, there are new charged states at the weak scale. In this model, $\psi$  carries color charge, and the bounds are consequently much stronger than for WIMPy leptogenesis: $m_\psi\gtrsim590$ GeV as inferred from the current LHC gluino search at 7 TeV. The phenomenology depends on how $\psi$ decays, and we outlined two possible models in (\ref{eq:lagrangianquark1}) and (\ref{eq:lagrangianquark2}). In both, $\psi$ decays to two jets plus a singlet, and so the collider phenomenology is identical. For the purposes of notation in this section, we assume that $\psi$ decays through an intermediate colored scalar $\phi$ to 2 Standard Model quarks and a singlet fermion, $n$.

$\psi$ can be pair-produced at the LHC with the signature $pp\rightarrow\psi\bar\psi\rightarrow 4j+ \cancel{E}_{\mathrm T}$. We show the relevant diagram in Figure \ref{fig:quarkLHC}.

Gluino searches at LHC7 bound the $\psi$ mass. Both gluinos and $\psi$ decay to $jj+\cancel E_{\mathrm T}$, and the bounds on both gluinos and $\psi$ are comparable, as their cross sections differ only by a group theory factor. We correct the gluino bounds for this factor. We apply simplified model searches from ATLAS, which place bounds on gluino and squark masses in the presence of a massless neutralino \cite{arXiv:1109.6572}.  The corresponding fields in WIMPy baryogenesis are $\psi$, the colored scalar $\phi$, and the massless singlet fermion $n$. The lower bound on the $\psi$ mass is $m_\psi\gtrsim590$ GeV, which comes from the gluino bound when the squark is much heavier than the gluino. In our scenario, this means that $m_\psi\gtrsim590$ GeV when $m_\phi\gg m_\psi$ (numerically, $m_\phi \gtrsim1.2$ TeV). The bounds on $m_\psi$ cut significantly into the allowed parameter space for  dark matter because $2m_X>m_\psi$, and so $m_X\gtrsim295$ GeV for heavy $\phi$. The bounds are stronger for lighter $\phi$ because $\phi$ and $\psi$ can be jointly produced. For example, the bound on $m_\psi$ is about 30\% higher for $m_\phi=1$ TeV.

The LHC search reach for gluinos is expected to be  $m_{\tilde g}\approx1.44$ TeV at $100\,\,\mathrm{fb}^{-1}$ and $\sqrt s =14$ TeV \cite{Baer:2003wx} (with the assumptions of mSUGRA and heavy squarks), so  models of WIMPy baryogenesis will be strongly constrained by future running of the LHC. The LHC will not reach the highest-mass regions of WIMPy baryogenesis, but will exclude models with masses $m_\psi\lesssim2$ TeV, $m_X\lesssim1$ TeV, and $\mathcal O(1)$ couplings.

LHC searches also constrain the mass of the colored scalar $\phi$. Since $m_{\phi}$ is not directly relevant to the outcome of WIMPy baryogenesis (apart from the requirement that it be light enough for $\psi$ decays to be in thermal equilibrium), bounds on $m_{\phi}$ do not directly constrain WIMPy baryogenesis. Nevertheless, the production rate of $\phi$ is comparable to that of squarks and is very high at the LHC. With the interaction (\ref{eq:lagrangianquark1}), $\phi$ decays to $d_i+\cancel E_{\rm T}$ and has an event topology identical to squark pair production in the MSSM: two jets (possibly $b$-tagged) plus missing energy. The current model-independent constraint is $m_{\phi}\gtrsim875$ GeV for degenerate squarks of the first two generations  \cite{arXiv:1109.6572}. In WIMPy baryogenesis, however, only a single field $\phi$ is necessary, so the bound can be relaxed. Since $\phi$ can decay into $b$, the bound is approximately that of a sbottom squark from D\O, $m_{\tilde b}>250$ GeV \cite{Abazov:2010wq}. Future LHC running at 14 TeV will improve the bound to $\sim 2$ TeV at $100\,\,\mathrm{fb}^{-1}$ \cite{Baer:2003wx}, and has the potential to discover colored scalars in the mass range of WIMPy baryogenesis.

 \subsection{Electric Dipole Moment Constraints}
\begin{figure}[t]
\begin{center}
\includegraphics[width=3.5cm]{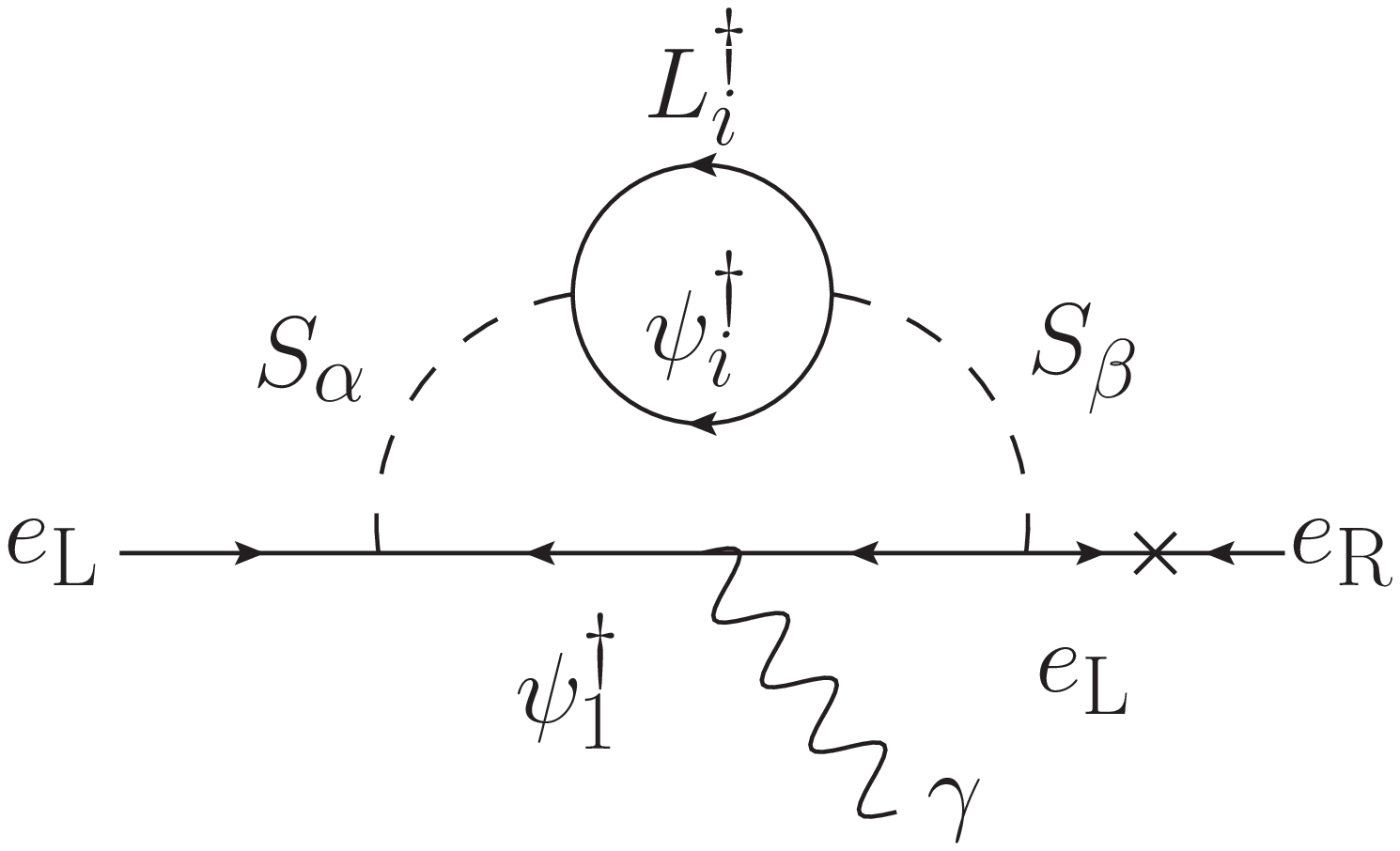}\hspace{0.4cm}
\includegraphics[width=3.5cm]{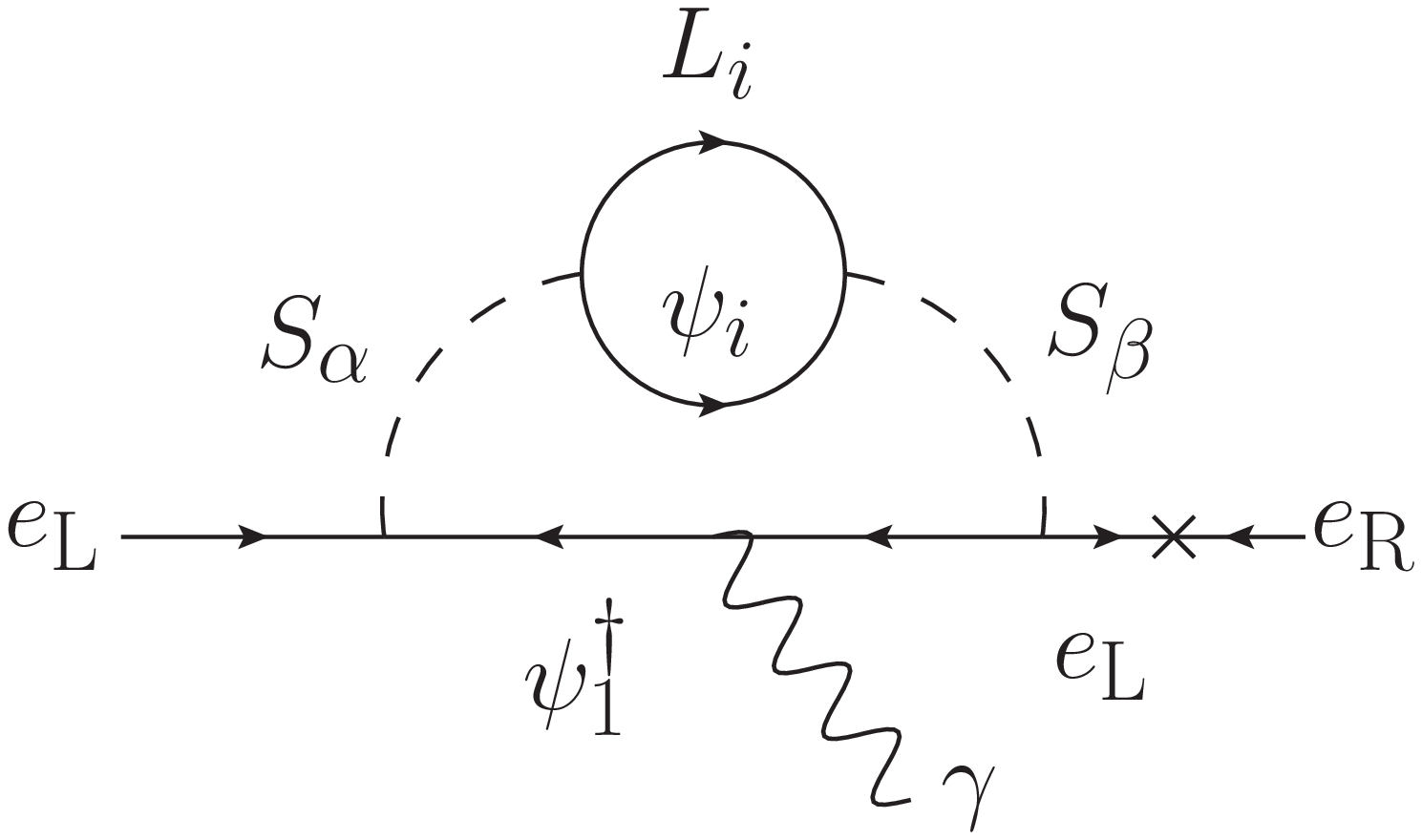}\hspace{0.4cm}
\includegraphics[width=3.5cm]{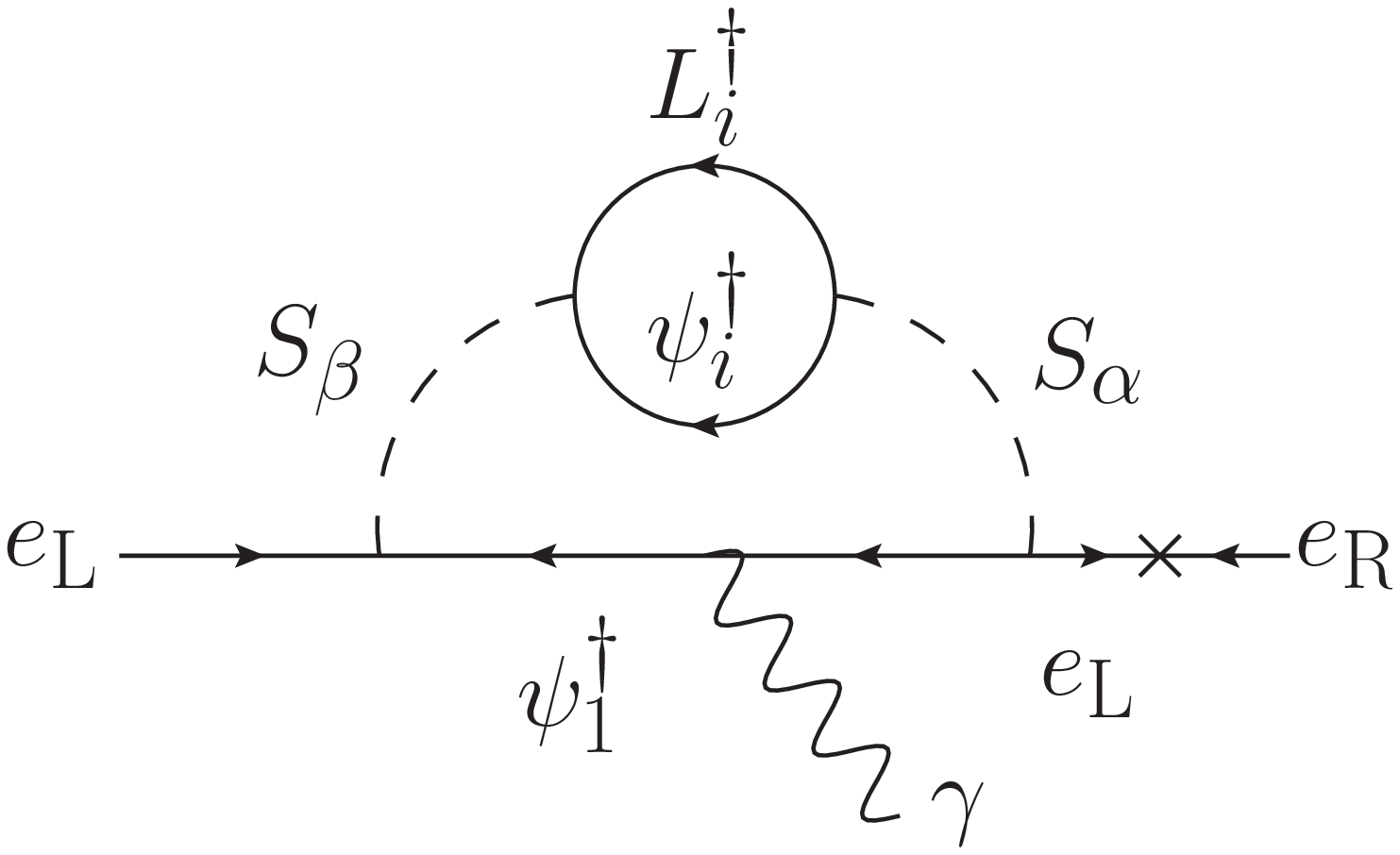}\hspace{0.4cm}
\includegraphics[width=3.5cm]{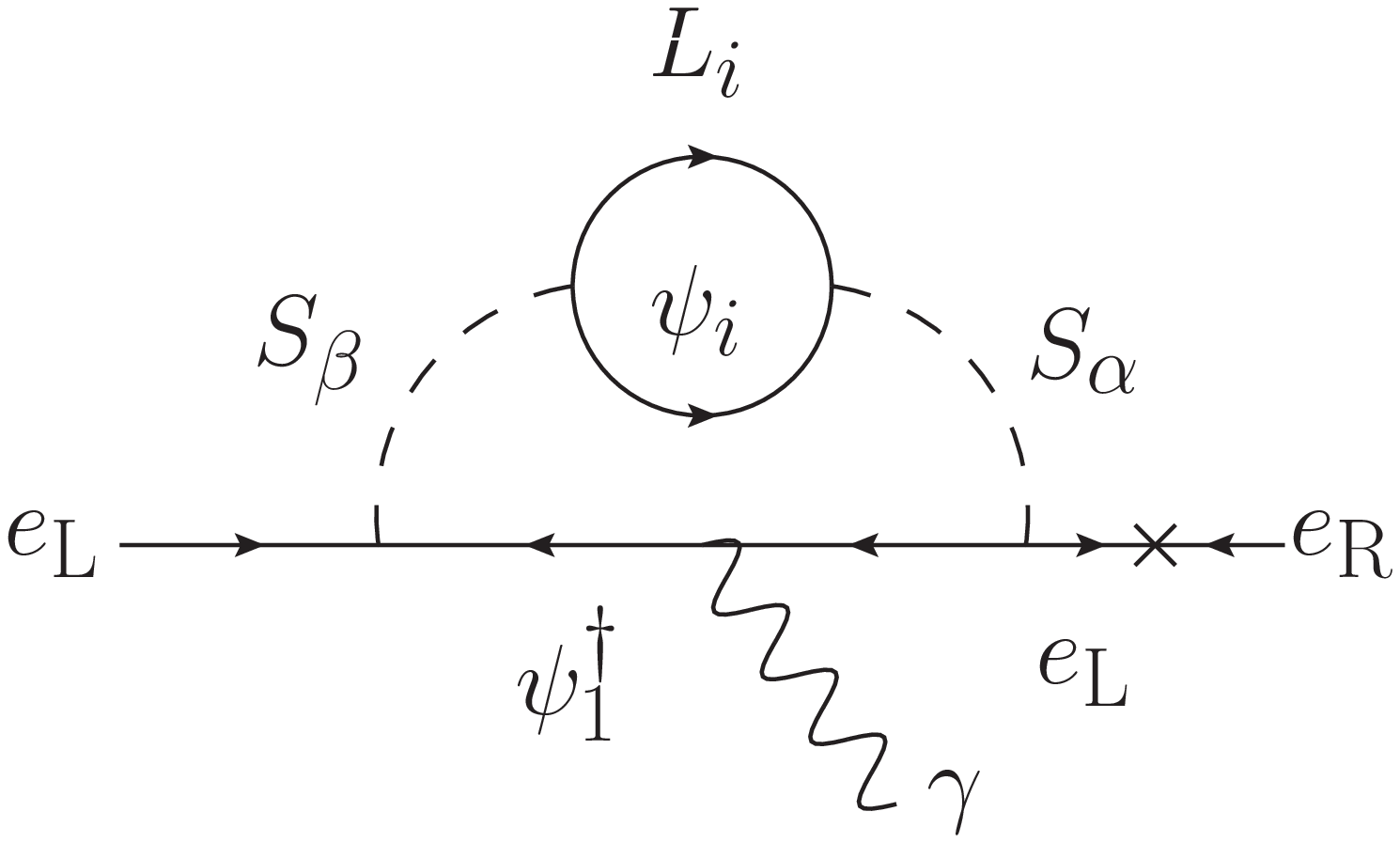}
\caption{A set of two-loop contributions to the electron EDM that vanishes when summed together.}
\label{fig:EDM}
\end{center}
\end{figure}
A viable mechanism for baryogenesis necessitates the existence of new $CP$ phases. Bounds on the electron and neutron electric dipole moments (EDMs) strongly constrain many new sources of $CP$-violation, but we find $CP$ phases in WIMPy baryogenesis are not constrained by EDM experiments. The minimal models of WIMPy baryogenesis presented in this paper couple new fields to \emph{either} left-handed or right-handed light fermions (but not both), resulting in suppressed EDMs that are consistent with current observations. As a result, minimal models of WIMPy baryogenesis do not have the $CP$ problem often associated with models of weak-scale physics.

In the models presented in Sections \ref{sec:model} and \ref{sec:alternatives}, the fields $S$ and $\psi$ couple exclusively to either left-handed or right-handed quarks and leptons. As a result, loops contributing to light fermion EDMs are helicity-preserving with an even number of Yukawa couplings, half of which are of the form $\lambda_{\alpha\,i}$ and the other half $\lambda_{\beta\,j}^*$. By summing over all permutations of different flavors of $S$, $L$ and $\psi$ on the internal lines, it can be shown that all one- and two-loop diagrams appear in pairs that are complex conjugates of one another. Summing over each set of pairs leads to a result that is real, and hence a vanishing EDM. As an example, we show in Figure \ref{fig:EDM} a set of four diagrams contributing to the electron EDM: the sum of the first two is proportional to $\lambda_{L\,\alpha1}\lambda^*_{L\,\beta1}$, while the sum of the second two is proportional to  $\lambda_{L\,\alpha1}^*\lambda_{L\,\beta1}$. Therefore, the sum of all four is real and does not contribute to the electron EDM.

\begin{figure}[t]
\begin{center}
\includegraphics[width=7cm]{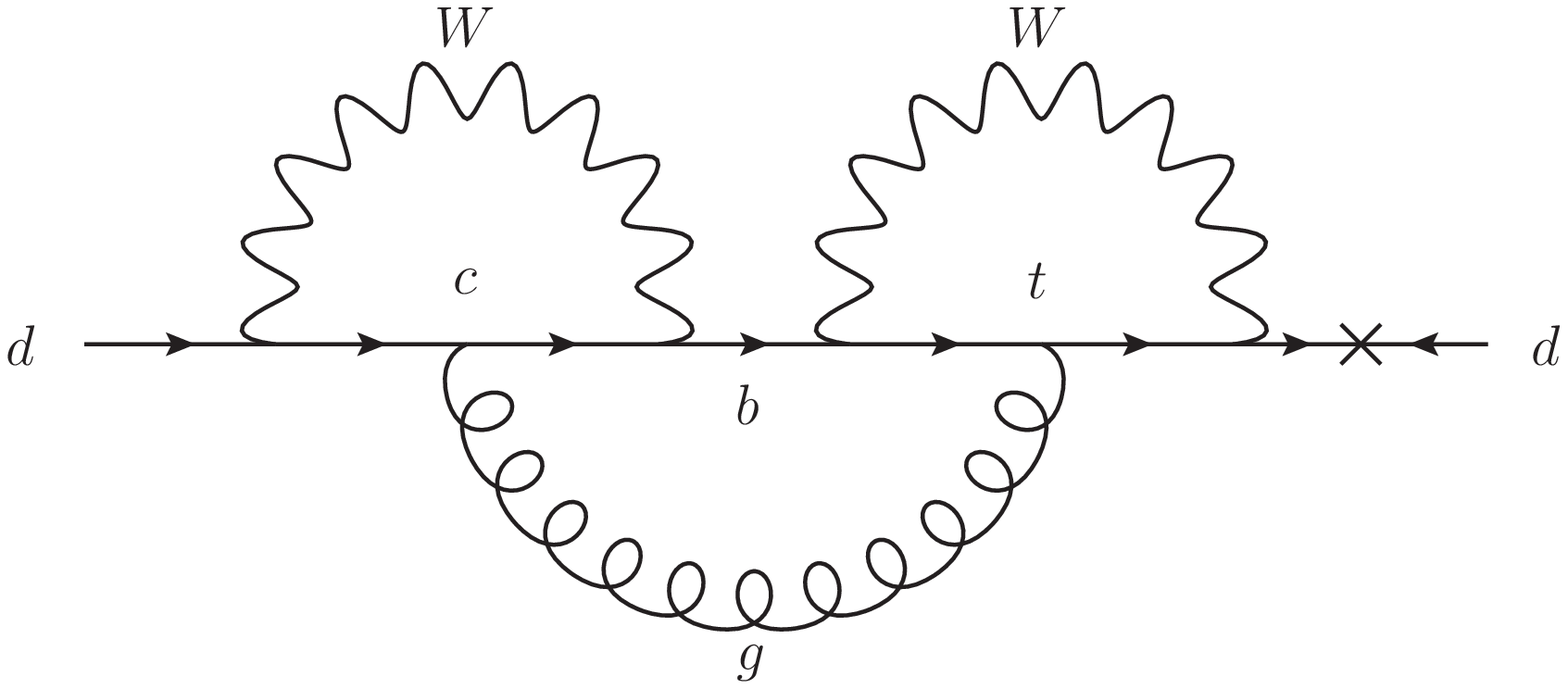}\hspace{0.7cm}
\includegraphics[width=7cm]{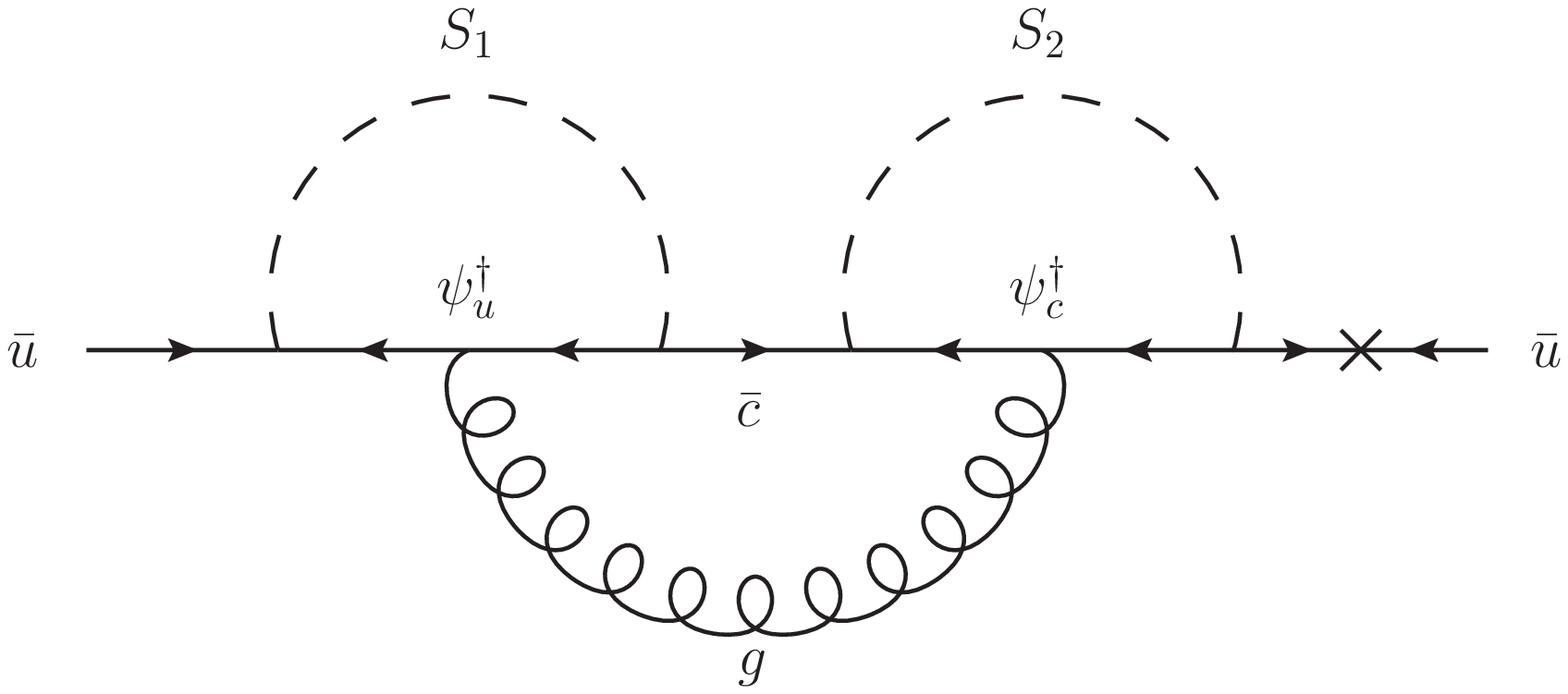}
\caption{{\bf (left)} Three loop EDM in the Standard Model. {\bf (right)} The analogous diagram for quark EDMs in WIMPy baryogenesis. The photon line attaches to any charged internal fields.}
\label{fig:SMEDM}
\end{center}
\end{figure}

The two-loop EDM in the Standard Model vanishes for the same reason as in WIMPy baryogenesis: $CP$-violation arises only in  couplings to one chirality of fermion. In both the Standard Model and WIMPy baryogenesis, the neutron EDM is non-zero at three loops, and we show the relevant diagrams in Figure \ref{fig:SMEDM}. The principal difference between the two  is that $CP$ violation vanishes in the Standard Model with fewer than three generations, so the Standard Model EDM is suppressed by mixings involving all three generations. By contrast, WIMPy baryogenesis has a contribution to the EDM with only two generations of quarks that couple to more than one flavor of $S$, and if the model is minimally flavor violating, the EDM will  be suppressed  by $\sin^2\theta_{\rm c}\approx0.05$, the square of the Cabibbo angle.  The na\"ive estimate for the neutron EDM in WIMPy baryogenesis with $\mathcal O(1)$ couplings is
\ben
\frac{d_n}{e} \sim \frac{\sin^2\theta_{\rm c}}{(16\pi^2)^3}\,\frac{m_u}{m_S^2}.
\een
Substituting $m_S\sim5$ TeV and $m_f\sim$ MeV gives $d_n/e\lesssim5\times10^{-32}\,\,\mathrm{cm}$, which is well below the current experimental bound of $d_n/e<2.9\times10^{-26}\,\,\mathrm{cm}$  \cite{Baker:2006ts}. The electron EDM from WIMPy leptogenesis  is even smaller than this, as flavor-changing effects in the charged lepton sector are suppressed by neutrino masses, and the EDM is also well below the experimental bound of $d_e/e<1.05\times10^{-27}\,\,\mathrm{cm}$ \cite{Hudson:2011zz}.

 Phases from WIMPy leptogenesis can also contribute to EDMs via other new, weak-scale fields not included in our minimal models. Since these contributions are model-dependent, we do not consider them further.

 Minimal models of WIMPy baryogenesis do not suffer from a $CP$ problem and are consistent with low-energy experiments, but it is possible that other, model-dependent contributions to the EDM could be constrained by and give rise to signals in electron and neutron EDM experiments.

    \section{Conclusions}\label{sec:concl}

In this paper, we explored a novel scenario called WIMPy baryogenesis that extends the WIMP miracle by generating the observed baryon asymmetry through annihilations of weak-scale dark matter and provides a dynamical connection between the dark matter and baryon abundances. We found that natural couplings and weak-scale masses for the new fields can lead to the correct baryon asymmetry. As a by-product of linking baryogenesis with dark matter annihilation, we introduced a new mechanism for weak-scale baryogenesis, avoiding any conflicts with reheat bounds in supersymmetric theories.

 The key observation is that, if dark matter annihilation proceeds via $CP-$ and (Standard Model) $B-$ or $L-$violating operators, all Sakharov conditions for baryogenesis are satisfied. Successful models  also suppress washout prior to dark matter freeze-out. In our models, such suppression results from the heaviness of the field $\psi$ carrying Standard Model gauge charges and $B-$ or $L-$number that is one of the final states in dark matter annihilation. Additional discrete symmetries forbid such  exotic fields from decaying back to Standard Model fields. We presented models where dark matter annihilates to either quarks or leptons, and  found viable parameter spaces with natural couplings and TeV-scale masses in both scenarios.

  In models where dark matter annihilates to leptons, the lepton asymmetry must be generated before the electroweak phase transition so that the asymmetry can be transferred to baryons via sphalerons. As a result, dark matter and $\psi$ masses must be $\mathcal O(\mathrm{TeV})$. Because the new states are heavy and dark matter does not couple directly to quarks, dark matter in this set-up is not in reach of near-future direct and indirect detection experiments. In this scenario, $\psi$ is  charged only under  weak interactions, making LHC searches challenging, although targeted searches at high integrated luminosity may allow discovery.

If dark matter annihilates to quarks, baryogenesis can occur after the electroweak phase transition, allowing smaller dark matter and $\psi$ masses. With lighter new states and colored objects, dark matter in these models can be within reach of direct detection experiments and indirect detection searches for antideuterons. LHC searches for $\psi$  are similar to gluino searches and can exclude $m_\psi\lesssim1.44$ TeV at $100\,\,\mathrm{fb}^{-1}$ and $\sqrt s=14$ TeV.

In both scenarios, WIMPy baryogenesis models can generate both the correct dark matter relic density and the baryon asymmetry at the weak scale. Such models predict new weak-scale particles that can lead to signals in dark matter direct and indirect detection experiments, and that may be accessible at the LHC.

\section*{Acknowledgements}
We wish to thank Zackaria Chacko, Tongyan Lin, Raman Sundrum, David Simmons-Duffin, and Neal Weiner for helpful conversations. LR would like to thank the Aspen Center for Physics and the KITP for their hospitality during progress on this work. Some of the numerical calculations in this paper were performed on the Odyssey cluster supported by the FAS Research Group at Harvard University. Feynman diagrams were drawn using JaxoDraw \cite{Binosi:2008ig}. This work is supported by NSF grant PHY-0855591 and the Harvard Center for the Fundamental Laws of Nature. YC is also supported in part by NSF grant PHY-0801323 and the Maryland Center for Fundamental Physics.


\begin{thebibliography}{99}


\bibitem{Nussinov:1985xr}
  S.~Nussinov,
  Phys.\ Lett.\  {\bf B165}, 55 (1985);
  S.~Dodelson, L.~M.~Widrow,
  Phys.\ Rev.\ Lett.\  {\bf 64}, 340-343 (1990).
  S.~M.~Barr, R.~S.~Chivukula, E.~Farhi,
  Phys.\ Lett.\  {\bf B241}, 387-391 (1990);
  S.~M.~Barr,
  Phys.\ Rev.\  {\bf D44}, 3062-3066 (1991);
  D.~B.~Kaplan,
  Phys.\ Rev.\ Lett.\  {\bf 68}, 741-743 (1992);
  R.~Foot, R.~R.~Volkas,
  Phys.\ Rev.\  {\bf D69}, 123510 (2004).
  [hep-ph/0402267];
  D.~Hooper, J.~March-Russell, S.~M.~West,
  Phys.\ Lett.\  {\bf B605}, 228-236 (2005).
  [hep-ph/0410114];
  S.~B.~Gudnason, C.~Kouvaris, F.~Sannino,
  Phys.\ Rev.\  {\bf D74}, 095008 (2006).
  [hep-ph/0608055];
  D.~E.~Kaplan, M.~A.~Luty, K.~M.~Zurek,
  Phys.\ Rev.\  {\bf D79}, 115016 (2009).
  [arXiv:0901.4117 [hep-ph]];
  T.~Cohen, K.~M.~Zurek,
  Phys.\ Rev.\ Lett.\  {\bf 104}, 101301 (2010).
  [arXiv:0909.2035 [hep-ph]];
  Y.~Cai, M.~A.~Luty, D.~E.~Kaplan,
  [arXiv:0909.5499 [hep-ph]];
  H.~An, S.~-L.~Chen, R.~N.~Mohapatra, Y.~Zhang,
  JHEP {\bf 1003}, 124 (2010).
  [arXiv:0911.4463 [hep-ph]];
  P.~-H.~Gu,
  Phys.\ Rev.\  {\bf D81}, 095002 (2010).
  [arXiv:1001.1341 [hep-ph]];
  T.~Cohen, D.~J.~Phalen, A.~Pierce, K.~M.~Zurek,
  Phys.\ Rev.\  {\bf D82}, 056001 (2010).
  [arXiv:1005.1655 [hep-ph]];
  J.~Shelton, K.~M.~Zurek,
  Phys.\ Rev.\  {\bf D82}, 123512 (2010).
  [arXiv:1008.1997 [hep-ph]];
  H.~Davoudiasl, D.~E.~Morrissey, K.~Sigurdson, S.~Tulin,
  Phys.\ Rev.\ Lett.\  {\bf 105}, 211304 (2010).
  [arXiv:1008.2399 [hep-ph]];
  N.~Haba, S.~Matsumoto,
  Prog.\ Theor.\ Phys.\  {\bf 125}, 1311-1316 (2011).
  [arXiv:1008.2487 [hep-ph]];
  E.~J.~Chun,
  Phys.\ Rev.\  {\bf D83}, 053004 (2011).
  [arXiv:1009.0983 [hep-ph]];
  P.~-H.~Gu, M.~Lindner, U.~Sarkar, X.~Zhang,
  Phys.\ Rev.\  {\bf D83}, 055008 (2011).
  [arXiv:1009.2690 [hep-ph]];
  M.~Blennow, B.~Dasgupta, E.~Fernandez-Martinez, N.~Rius,
  JHEP {\bf 1103}, 014 (2011).
  [arXiv:1009.3159 [hep-ph]];
  L.~J.~Hall, J.~March-Russell, S.~M.~West,
  [arXiv:1010.0245 [hep-ph]];
  R.~Allahverdi, B.~Dutta, K.~Sinha,
  Phys.\ Rev.\  {\bf D83}, 083502 (2011).
  [arXiv:1011.1286 [hep-ph]];
  B.~Dutta, J.~Kumar,
  Phys.\ Lett.\  {\bf B699}, 364-367 (2011).
  [arXiv:1012.1341 [hep-ph]];
  A.~Falkowski, J.~T.~Ruderman, T.~Volansky,
  JHEP {\bf 1105}, 106 (2011).
  [arXiv:1101.4936 [hep-ph]].
  Z.~Kang, J.~Li, T.~Li, T.~Liu, J.~Yang,
  [arXiv:1102.5644 [hep-ph]];
  M.~T.~Frandsen, S.~Sarkar, K.~Schmidt-Hoberg,
  [arXiv:1103.4350 [hep-ph]];
  D.~E.~Kaplan, G.~Z.~Krnjaic, K.~R.~Rehermann, C.~M.~Wells,
  [arXiv:1105.2073 [hep-ph]];
  N.~F.~Bell, K.~Petraki, I.~M.~Shoemaker, R.~R.~Volkas,
   [arXiv:1105.3730 [hep-ph]];
  C.~Cheung, K.~M.~Zurek,
  Phys.\ Rev.\  {\bf D84}, 035007 (2011).
  [arXiv:1105.4612 [hep-ph]];
  J.~March-Russell, M.~McCullough,
  [arXiv:1106.4319 [hep-ph]];
  C.~Arina, N.~Sahu,
  [arXiv:1108.3967 [hep-ph]];



  \bibitem{Kaplan:2009ag}
  D.~E.~Kaplan, M.~A.~Luty, K.~M.~Zurek,
  Phys.\ Rev.\  {\bf D79}, 115016 (2009).
  [arXiv:0901.4117 [hep-ph]].

\bibitem{Cohen:2009fz}
  T.~Cohen, K.~M.~Zurek,
  Phys.\ Rev.\ Lett.\  {\bf 104}, 101301 (2010).
  [arXiv:0909.2035 [hep-ph]].

\bibitem{Buckley:2010ui}
  M.~R.~Buckley, L.~Randall,
  JHEP {\bf 1109}, 009 (2011).
  [arXiv:1009.0270 [hep-ph]].



\bibitem{Cui:2011qe}
  Y.~Cui, L.~Randall, B.~Shuve,
    JHEP {\bf 1108}, 073 (2011).
  [arXiv:1106.4834 [hep-ph]].

\bibitem{Buckley:2011ye}
  M.~R.~Buckley, S.~Profumo,
  [arXiv:1109.2164 [hep-ph]],
  M.~Cirelli, P.~Panci, G.~Servant and G.~Zaharijas,
  arXiv:1110.3809 [hep-ph],
  S.~Tulin, H.~-B.~Yu and K.~M.~Zurek,
  arXiv:1202.0283 [hep-ph].



\bibitem{McDonald:2011zz}
  J.~McDonald,
  Phys.\ Rev.\  {\bf D83}, 083509 (2011),
  J.~McDonald,
  [arXiv:1108.4653 [hep-ph]].

\bibitem{Khlopov:1984pf}
  M.~Y.~.Khlopov, A.~D.~Linde,
  Phys.\ Lett.\  {\bf B138}, 265-268 (1984);
  J.~R.~Ellis, J.~E.~Kim, D.~V.~Nanopoulos,
  Phys.\ Lett.\  {\bf B145}, 181 (1984);
  M.~Kawasaki, T.~Moroi,
  Prog.\ Theor.\ Phys.\  {\bf 93}, 879-900 (1995).
  [hep-ph/9403364, hep-ph/9403061].

\bibitem{Kolb:1990vq}
  E.~W.~Kolb, M.~S.~Turner,
  Front.\ Phys.\  {\bf 69}, 1-547 (1990).

\bibitem{Kolb:1979qa}
  E.~W.~Kolb, S.~Wolfram,
  Nucl.\ Phys.\  {\bf B172}, 224 (1980).

\bibitem{Komatsu:2010fb}
  E.~Komatsu {\it et al.} [ WMAP Collaboration ],
  Astrophys.\ J.\ Suppl.\  {\bf 192}, 18 (2011).
  [arXiv:1001.4538 [astro-ph.CO]].

\bibitem{Bento:2001rc}
  L.~Bento, Z.~Berezhiani,
  Phys.\ Rev.\ Lett.\  {\bf 87}, 231304 (2001).
  [hep-ph/0107281];
  P.~-H.~Gu, U.~Sarkar,
  Phys.\ Lett.\  {\bf B679}, 118-121 (2009).
  [arXiv:0903.3473 [hep-ph]];
  C.~R.~Das, L.~V.~Laperashvili, H.~B.~Nielsen, A.~Tureanu,
  Phys.\ Lett.\  {\bf B696}, 138-144 (2011).
  [arXiv:1010.2744 [hep-ph]].


\bibitem{Pukhov:2004ca}
  A.~Pukhov,
  [hep-ph/0412191].

\bibitem{Harvey:1990qw}
  J.~A.~Harvey and M.~S.~Turner,
  Phys.\ Rev.\ D {\bf 42}, 3344 (1990).

\bibitem{Mohapatra:2009wp}
  R.~N.~Mohapatra,
  J.\ Phys.\ G {\bf G36}, 104006 (2009).
  [arXiv:0902.0834 [hep-ph]].

\bibitem{Raby:1987nb}
  S.~A.~Raby, G.~West,
  Nucl.\ Phys.\  {\bf B292}, 793 (1987).

\bibitem{Kopp:2009et}
  J.~Kopp, V.~Niro, T.~Schwetz, J.~Zupan,
  Phys.\ Rev.\  {\bf D80}, 083502 (2009).
  [arXiv:0907.3159 [hep-ph]].

\bibitem{xenon}
  E.~Aprile {\it et al.}  [XENON100 Collaboration],
  Phys.\ Rev.\ D {\bf 84}, 052003 (2011)
  [arXiv:1103.0303 [hep-ex]].

\bibitem{cdms}
  Z.~Ahmed {\it et al.}  [The CDMS-II Collaboration],
  Science {\bf 327}, 1619 (2010)
  [arXiv:0912.3592 [astro-ph.CO]].

  \bibitem{xenon1t}
  E.~Aprile, http://www.physics.ucla.edu/hep/dm10/talks/aprile.pdf.


\bibitem{Davoudiasl:2011fj}
  H.~Davoudiasl, D.~E.~Morrissey, K.~Sigurdson, S.~Tulin,
  [arXiv:1106.4320 [hep-ph]].

\bibitem{collaboration:2011wa}
  T.~F.~-L.~collaboration,
  [arXiv:1108.3546 [astro-ph.HE]].

\bibitem{Abazajian:2011ak}
  K.~N.~Abazajian and J.~P.~Harding,
  JCAP {\bf 1201}, 041 (2012)
  [arXiv:1110.6151 [hep-ph]].

\bibitem{Cui:2010ud}
  Y.~Cui, J.~D.~Mason, L.~Randall,
  JHEP {\bf 1011}, 017 (2010).
  [arXiv:1006.0983 [hep-ph]].

\bibitem{Ahlen:1994ct}
  S.~P.~Ahlen, V.~M.~Balebanov, R.~Battiston, U.~Becker, J.~Burger, M.~Capell, H.~F.~Chen, H.~S.~Chen {\it et al.},
  Nucl.\ Instrum.\ Meth.\  {\bf A350}, 351-367 (1994).





\bibitem{Barate:1999fs}
  R.~Barate {\it et al.} [ ALEPH Collaboration ],
  Eur.\ Phys.\ J.\  {\bf C11}, 193-216 (1999);
  D.~Fouchez,
  ``Search for charginos and neutralinos with the ALEPH experiment at LEP-2.''



\bibitem{Kribs:2009yh}
  G.~D.~Kribs, A.~Martin, T.~S.~Roy, M.~Spannowsky,
  Phys.\ Rev.\  {\bf D81}, 111501 (2010).
  [arXiv:0912.4731 [hep-ph]];
  G.~D.~Kribs, A.~Martin, T.~S.~Roy, M.~Spannowsky,
  Phys.\ Rev.\  {\bf D82}, 095012 (2010).
  [arXiv:1006.1656 [hep-ph]].

\bibitem{arXiv:1109.6572}
  G.~Aad {\it et al.} [ATLAS Collaboration],
  arXiv:1109.6572 [hep-ex].



\bibitem{Baer:2003wx}
  H.~Baer, C.~Balazs, A.~Belyaev, T.~Krupovnickas, X.~Tata,
  JHEP {\bf 0306}, 054 (2003).
  [hep-ph/0304303].

\bibitem{Abazov:2010wq}
  V.~M.~Abazov {\it et al.} [ D0 Collaboration ],
  Phys.\ Lett.\  {\bf B693}, 95-101 (2010).
  [arXiv:1005.2222 [hep-ex]].


\bibitem{Baker:2006ts}
  C.~A.~Baker, D.~D.~Doyle, P.~Geltenbort, K.~Green, M.~G.~D.~van der Grinten, P.~G.~Harris, P.~Iaydjiev, S.~N.~Ivanov {\it et al.},
  Phys.\ Rev.\ Lett.\  {\bf 97}, 131801 (2006).
  [hep-ex/0602020].

\bibitem{Hudson:2011zz}
  J.~J.~Hudson, D.~M.~Kara, I.~J.~Smallman, B.~E.~Sauer, M.~R.~Tarbutt, E.~A.~Hinds,
  Nature {\bf 473}, 493-496 (2011).



\bibitem{Binosi:2008ig}
  D.~Binosi, J.~Collins, C.~Kaufhold, L.~Theussl,
  Comput.\ Phys.\ Commun.\  {\bf 180}, 1709-1715 (2009).
  [arXiv:0811.4113 [hep-ph]].


\end{thebibliography}
\end{document}